\documentclass[a4paper,10pt]{article}%
\usepackage{amsfonts}
\usepackage{graphicx}
\usepackage{amsmath}
\usepackage{amssymb}%
\usepackage{cite}
\pdfoutput=1
\makeatletter

\@addtoreset{equation}{section}
\makeatother
\newcommand{\beq}{\begin{equation}}
\newcommand{\eeq}{\end{equation}}
\renewcommand{\a}{\alpha}
\renewcommand{\b}{{{\beta}}}

\newcommand{\be}{\begin{eqnarray}}
\newcommand{\ee}{\end{eqnarray}}

\usepackage[utf8]{inputenc}
\usepackage{amsmath}
\usepackage{amscd}
\usepackage{amsfonts}
\usepackage{amssymb}
\usepackage[english]{babel}
\usepackage{braket}
\usepackage{mathtools}
\usepackage{amsthm}
\usepackage{graphicx}
\usepackage{graphicx,color}
\usepackage{enumerate}
\usepackage{verbatim}
\usepackage[all]{xy}
\usepackage{xcolor}
\usepackage{comment}

\newcommand{\virgolette}{``}

\numberwithin{equation}{section}

\theoremstyle{plain}

\theoremstyle{definition}

\newtheorem{ex}{Example}
\newtheorem{obs}{Observation}
\theoremstyle{remark}
\newtheorem{rmk}{RMK}

\begin{document}
\baselineskip=18pt
\baselineskip 0.6cm

\begin{titlepage}
\setcounter{page}{0}
\renewcommand{\thefootnote}{\fnsymbol{footnote}}
\begin{flushright}
ARC-19-08
\\
\today
\end{flushright}

\vskip 2cm
\begin{center}
{\Huge \bf Pictures from Super } \\\vskip .2cm
{\Huge \bf  Chern-Simons Theory}
\vskip 1cm
{
\large {\bf C.~A.~Cremonini}$^{~a,b}$\footnote{carlo.alberto.cremonini@gmail.com} 
and
\large {\bf P.~A.~Grassi}$^{~c,d,e}$\footnote{pietro.grassi@uniupo.it}
}
\vskip .5cm {
\small
\medskip
\centerline{$^{(a)}$ \it Dipartimento di Scienze e Alta Tecnologia (DiSAT),}
\centerline{\it Universit\`a degli Studi dell'Insubria, via Valleggio 11, 22100 Como, Italy}
\medskip
\centerline{$^{(b)}$ \it INFN, Sezione di Milano, via G.~Celoria 16, 20133 Milano, Italy} 
\medskip
\centerline{$^{(c)}$
\it Dipartimento di Scienze e Innovazione Tecnologica (DiSIT),} \centerline{\it Universit\`a del Piemonte Orientale, viale T.~Michel, 11, 15121 Alessandria, Italy}
\medskip
\centerline{$^{(d)}$
\it INFN, Sezione di Torino, via P.~Giuria 1, 10125 Torino, Italy}
\medskip
\centerline{$^{(e)}$
\it Arnold-Regge Center, via P.~Giuria 1,  10125 Torino, Italy}
\medskip
}
\end{center}
\vskip 0.2cm
\centerline{{\bf Abstract}}
\medskip
We study super-Chern-Simons theory on a generic supermanifold. 
After a self-contained review of integration on supermanifolds, the complexes 
of forms (superforms, pseudo-forms and integral forms) and the extended 
Cartan calculus are discussed. We then introduce Picture Changing Operators. 
We provide several 
examples of computation of PCO's acting on different type of forms. We illustrate also the action of 
the $\eta$ operator, crucial ingredient to define the interactions of super Chern-Simons theory. 
Then, we discuss the action for super Chern-Simons theory on any supermanifold, first in the 
factorized form (3-form $\times$ PCO) and then, we consider the most general expression. The latter 
is written in term of psuedo-forms containing an infinite number of components. We show that the free equations 
of motion reduce to the usual Chern-Simons equations yielding the proof of the equivalence 
between the formulations at different pictures of the same theory. Finally, we discuss the interaction terms. They 
require a suitable definition in order to take into account the picture number. That implies the construction 
of a 2-product which is not associative that inherits an $A_\infty$ algebra structure. That shares several similarities 
with a recent construction of a super string field theory action by Erler, Konopka and Sachs.  


\end{titlepage}

\tableofcontents \noindent {}

\newpage
\setcounter{footnote}{0} \newpage\setcounter{footnote}{0}


\section{Introduction}

Our main motivation is to provide a general method for constructing classical actions for quantum field theories on supermanifolds with the powerful methods of supergeometry. As is known for general relativity, the powerful technique of differential forms on a given manifold permits the construction of physical interesting quantities (actions, observables, globally defined quantities). Here we would like to set up an equivalent framework for a supermanifold. Namely we would like to formulate quantum field theory models on supermanifolds as we use to do it in general relativity.

Given a supermanifold ${\cal SM}^{(n|m)}$  with $n$ bosonic dimensions and $m$ fermionic dimensions, we would like to construct an action of the form 
\begin{eqnarray}
\label{acc1}
S = \int_{{\cal SM}^{(n|m)}} {\cal L}^{(n|m)} \ ,
\end{eqnarray}
where ${\cal L}^{(n|m)}$ is an integral form \cite{Grassi:2016apf,CCGGeom,CCGhd,Witten:2012bg,Belopolsky:1997jz} 
with form degree $n$  and picture number $m$. $\displaystyle \mathcal{L}^{(n|m)}$ is a form which can be integrated on the supermanifold, i.e. it is a \emph{top form} and 
any super-diffeomorphism leaves the action invariant. 

 One strategy to build an action  ${\cal L}^{(n|m)}$ is to start from a conventional superform ${\cal L}^{(n|0)}$ in terms of the classical fields and their differentials and then complete it to an integral form as 
\begin{eqnarray}
\label{acc2}
{\cal L}^{(n|m)} = {\cal L}^{(n|0)} \wedge {\mathbb Y}^{(0|m)} \ ,
\end{eqnarray}
where ${\mathbb Y}^{(0|m)}$ is a Picture Changing Operator (PCO) mapping the superform ${\cal L}^{(n|0)}$ to an integral form ${\cal L}^{(n|m)}$. If ${\cal L}^{(n|0)}$ is closed, one can change $ {\mathbb Y}^{(0|m)}$ by exact pieces without changing the action $S$. The question is: is the factorized form (\ref{acc2}) always achievable or are there other possibilities? Namely, given the fields in a given picture, is there a way to build an action consistently producing meaningful results? 

For example, given a gauge field $A^{(1|0)}$ which is the usual 1-form connections at picture equal to zero, can one use a picture one field as $A^{(1|1)}$ instead? Then, we would replace the Lagrangian (\ref{acc2}) as
\begin{eqnarray}
\label{acc3}
 {\cal L}^{(n|0)}(A^{(1|0)})\wedge {\mathbb Y}^{(0|m)} \longrightarrow 
 {\cal L}^{(n|m)}(A^{(1|1)})  \ ,
\end{eqnarray}
such that the equations of motion are still dynamical equations. 

A similar issue is present in string theory  \cite{FMS} and 
string field theory \cite{Witten:1986qs}, where the ghost sector of RNS string theory model requires a choice of the vacuum due to the replicas of the same Hilbert space at different pictures. As is well known, the quantization 
of the $\beta-\gamma$ ghost sector leads to a Fock space filtered according to the ghost number and with respect to the picture number. That translates into the definition of the vertex operators representing the target space fields. 
Those vertex operators can be chosen in different pictures such that the total sum of pictures of the vertex operators inserted into a correlation function saturates the required picture charge (see also \cite{pol,DiVecchia:1997vef}) at given genus and number of punctures.  The result should be independent of the choice of the picture. In the case of string field theory, the situation is slightly different. In order to write a string field theory action, one needs to take into account the saturation of the picture on a disk (tree level classical action) and for that some alternatives were proposed (see \cite{Preitschopf:1989fc} and \cite{Witten:1986qs}). However, despite some 
interestring features for these models, 
they fail to give a complete interacting superstring field theory action. Only recently, 
by the work of \cite{Erler:2013xta}, 
a complete interacting superstring field theory action has been proposed. The string 
fields are taken into a picture one Hilbert space leading to a meaningful kinetic term. Nonetheless, the interactions are constructed in terms of a non-associative product which multiplies two string fields without increasing the picture and  which is the first non-trivial element of an $A_\infty$ 
algebra. That algebra has been built completely\cite{Erler:2013xta,Erler:2016ybs}. As shown in 
\cite{Catenacci:2018xsv,Catenacci-Grassi-Noja}, for any supermanifold, in terms of the PCO built in the complexes of forms, one can define a corresponding $A_\infty$-algebra \cite{Kajiura:2003ax,Keller,PenSch,Penkava,Sachs:2019gue} on the geometrical data and therefore we expect that we can follow the same pattern. 

In the same way, for the construction of quantum field theories on supermanifolds (we recall 
that the picture in string theory is related to the superghosts zero modes which are in relations 
with the supermoduli space of the underlying super-Riemann surface), one needs to fix the total picture 
of the action, but that does not select a given picture for the fields involved. This means that one can choose 
different set of fields, defined as forms in the complete complex, and construct the corresponding action 
(See \cite{CCGGeom,CCGhd}).  

To provide an illustration of this constructing procedure, we focus on a specific model, namely super Chern-Simons theory on a $(3|2)$ supermanifold. The classical action can be written in terms of the $(1|0)$ connection $A^{(1|0)}$. It is show that by using the factorized form ${\cal L}^{(3|0)} \wedge {\mathbb Y}^{(0|2)}$ all superspace formulations can be obtained. The choice of the 
PCO $ {\mathbb Y}^{(0|2)}$ ranging from the simplest example to more symmetric expressions 
(see \cite{Grassi:2016apf}) leads to different actions with manifest supersymmetry or in components. 

In the present work, we consider an action for super Chern-Simons theory (henceforth SCS) built in terms of the $A^{(1|1)}$ gauge fields, namely 
those at picture one. Their expansions in term of component fields are infinite dimensional, then the kinetic term is obtained 
by using repeated distributional properties and integrating on the supermanifold. The goal is to verify that the kinetic term 
yields the correct equations of motion, namely Chern-Simons flat connection and vanishing fermions. That is achieved by 
showing that, on-shell, all unphysical components can be gauged away except the Chern-Simons connection and its 
flat curvature condition. Here we consider those gauge transformations which are obtained by derivatives along the 
fermionic directions. Since the fermionic fibres do not have any topology, those gauge transformations can be reabsorbed (without any topological obstractions) 
leaving only usual gauge transformations along $x$-coordinates. 

Once we have verified that the free action leads to the correct equations of motion, we consider the interaction terms. 
Thus, following the string field theory construction \cite{Erler:2013xta}, we define a $2$-product which multiplies two 
$(1|1)$ gauge fields and decreases the picture by one unity \cite{Catenacci:2018xsv}. This
leads to the conventional interaction term plus additional interactions due to the infinite components of the 
picture one connections. The product used for the interaction is non-associative, but its associator 
is cohomologically trivial and therefore can be compensated by a $3$-product. This leads to an $A_\infty$ algebra 
which consistently provides the complete set of equations of motion. Together with the cyclicity of the 
inner product, we finally derive the equations of motion from a consistent variational principle. 

At the end, we discuss the supersymmetry in this framework. We found that even in the picture one 
setting for the gauge fields, the rheonomic conditions do show the same set of physical fields. In addition, 
since the construction is based on supermanifolds, the superdiffeomorphisms are those transformations of the 
action which preserve the entire structure. 

The paper is organized as follows: in sec.~2, we collect all mathematical tools needed for construction of the 
action and the derivation of the equations of motion. In particular, we describe the action of various operators on the 
space of forms for the supermanifold ${\cal SM}^{(3|2)}$. In sec.~3, we recall the action for super Chern-Simons theory 
in the factorized form. In sec.~4, we get to the main derivation of the equations of motion in the non-factorized 
form, interaction terms, gauge invariance and supersymmetry. In appendices, we collect some review material 
on $A_\infty$ algebras and their automorphisms and some explicit computations omitted in main text.


\section{Mathematical Tools}
\subsection{Supermanifolds and Superspaces}

Let us briefly recall the most basic definitions in supergeometry. For a rigorous and thorough mathematical treatment of the subject we suggest the reader to refer to\cite{Varadarajan:2004yz, VorGeom, Manin, DELIGNE}. The most basic and most important example of \emph{superspace} is given by $\mathbb{R}^{(p|q)}$, that is the pair given by the manifold $\mathbb{R}^p$ and the sheaf $C^\infty _{\mathbb{R}^p} \left[ \theta^1 , \ldots , \theta^q \right]$
\begin{equation}
	\mathbb{R}^{(p|q)} = \left( \mathbb{R}^p , C^\infty _{\mathbb{R}^p} \left[ \theta^1 , \ldots , \theta^q \right] \right) \ .
\end{equation}
This means that the superspace is constructed over the usual space $\mathbb{R}^p$ and the functions we consider are $C^\infty$ functions over the coordinates of the space $\mathbb{R}^p$ and they have polynomial dependence on the Grassmann coordinates $\left\lbrace \theta^i \right\rbrace_{i=1}^q$. Since the $\theta$ coordinates are anticommuting, this is equivalent to consider the exterior algebra generated by $q$ variables with value into $C^\infty$ functions:
\begin{equation}
	C^\infty _{\mathbb{R}^p} \left[ \theta^1 , \ldots , \theta^q \right] \cong \bigwedge^\bullet \mathbb{R}^q \otimes C^\infty_{\mathbb{R}^p} \ .
\end{equation}
This means that a general function can be expanded as
\begin{equation}
	f(x,\theta) = f_0 (x) + f_{i_1} (x) \theta^{i_1} + \ldots + f_q (x) \theta^1 \ldots \theta^q \ ,
\end{equation}
where the Einstein's summation convention is understood.

A \emph{(real) supermanifold} $\mathcal{SM}$ of \emph{dimension} $\mbox{dim} \, \mathcal{SM} = (p|q)$ is a superspace $\displaystyle \left( \left| \mathcal{SM} \right| , \mathcal{O}_{\mathcal{SM}} \right)$ where $|\mathcal{SM}|$ is a real manifold and $\mathcal{O}_{\mathcal{SM}}$ is a sheaf which is locally isomorphic to $\displaystyle C^\infty \left[ \theta^1 , \ldots , \theta^q \right]$. In other words a real supermanifold of dimension $(p|q)$ is a superspace which is locally isomorphic to $\mathbb{R}^{(p|q)}$. In this paper we will only deal with real supermanifolds: in particular this means that we will not be concerned by the subtleties related to non-projected and non-split supermanifolds which arise only in the context of complex supermanifolds 
\cite{Donagi:2013dua,1DCY,PiGeo,NojaG,CNR,CN,Witten:2012ga}

We consider the case of a real supermanifold $\mathcal{SM}^{(3|2)}$; in terms of the coordinates, we define the following differential operators 
\begin{equation}
	\partial_a = \frac{\partial}{\partial x^a} \ , \ D_\alpha = \frac{\partial}{\partial \theta^\alpha} - \left( \gamma^a \theta \right)_\alpha \partial_a \ , \ Q_\alpha = \frac{\partial}{\partial \theta^\alpha} + \left( \gamma^a \theta \right)_\alpha \partial_a \ ,
\end{equation}
where the second and the third are known as superderivative and supersymmetry generator, respectively. They satisfy the superalgebra relations
\begin{equation}
\left[ \partial_a , \partial_b \right] = 0 \, , \ \
\{D_{\a},D_{\b} \}=-2 \gamma^a_{\a\b} \partial_a\,, ~~~~
\{Q_{\a},Q_{\b} \}= 2 \gamma^a_{\a\b} \partial_a\,, ~~~~
\{D_{\a},Q_{\b} \}=0\,,
\left\lbrace \partial_a , D_\alpha \right\rbrace = 0 = \left\lbrace \partial_a , Q_\alpha \right\rbrace \ .
\label{susy3dB}
\end{equation}
In 3d, for the local subspace we use the Lorentzian metric $\eta_{ab} = (-,+,+)$, and the real and symmetric Dirac matrices $\gamma^a_{\a\b}$ given by
\begin{eqnarray}
\label{dicA}
&&\gamma^0_{\a\b} = (C \Gamma^0) = - {\mathbf 1}\,, ~~~
\gamma^1_{\a\b} = (C \Gamma^1) = \sigma^3\,,  \nonumber \\
&&\gamma^2_{\a\b} = (C \Gamma^2) = - \sigma^1\,, ~~
C_{\a\b} = i \sigma^2 = \epsilon_{\a\b}\,. 
\end{eqnarray}
Numerically, we have $\hat\gamma_a^{\a\b} = \gamma^a_{\a\b}$ and 
$\hat\gamma_a^{\a\b} = \eta_{ab} (C \gamma^b C)^{\a\b} = C^{\a\gamma} \gamma_{a, \gamma\delta} C^{\delta\beta}$. 
The conjugation matrix is
$\epsilon^{\a\b}$ and a bi-spinor is decomposed as follows $R_{\a\b} = R \epsilon_{\a\b}  + R_a \gamma^a_{\a\b}$ where
$R = - \frac12 \epsilon^{\a\b} R_{\a\b}$ and $R_a = {\rm Tr}(\gamma_a R)$ are
a scalar and a vector, respectively.
In addition, it is easy to show that
$\gamma^{ab}_{\a\b} \equiv \frac12 [\gamma^a, \gamma^b]_{\alpha \beta} = \epsilon^{abc} \gamma_{c \a\b}$.

The differential of $\Phi$ is expanded on a basis of forms as follows
\begin{equation}
d \Phi = dx^{a} \partial_{a} \Phi + d\theta^{\alpha}%
\partial_{\alpha}\Phi =
\end{equation}
\[
=\Big(dx^{a} + \theta\gamma^{a} d\theta\Big) \partial_{a} \Phi
+ d\theta^{\alpha}D_{\alpha}\Phi \equiv V^{a} \partial_{a}
\Phi + \psi^{\alpha}D_{\alpha}\Phi\,,
\]
where $V^a = dx^a + \theta \gamma^a d\theta$ and $\psi^\a = d \theta^\a$ which satisfy the Maurer-Cartan 
equations 
\begin{eqnarray}
\label{DVDPSI}
d V^a = \psi \gamma^a \psi\,, ~~~~~~ d \psi^\a = 0\,. 
\end{eqnarray}

Given a form $\Phi$, we can compute the supersymmetry variation as a Lie derivative $\mathcal{L}_{\epsilon}$ with $\epsilon=\epsilon^{\a}Q_{\a}+\epsilon^{a}\partial_{a}$ ($\epsilon^{a}$ are the infinitesimal parameters of the translations and $\epsilon^{\a}$ are the supersymmetry parameters) and by means of the Cartan formula we have
\begin{equation}
\delta_{\epsilon}\Phi=\mathcal{L}_{\epsilon}\Phi%
=\iota_{\epsilon}d\Phi + d \iota_{\epsilon} \Phi=\iota_{\epsilon}\Big(dx^{a}\partial_{a}%
\Phi+d\theta^{\a}\partial_{\a}\Phi\Big) + d \iota_{\epsilon} \Phi =
\end{equation}%
\[
=(\epsilon^{a}+\epsilon\gamma^{a}\theta)\partial_{a}\Phi
+\epsilon^{\a}\partial_{\a}\Phi + d \iota_{\epsilon} \Phi =\epsilon^{a}\partial_{a}%
\Phi+\epsilon^{\a}Q_{\a}\Phi + d \iota_{\epsilon} \Phi\,,%
\]
where the term $ d \iota_{\epsilon} \Phi$ is simply a gauge transformation. It follows easily that $\delta_{\epsilon}V^{a} = \delta_{\epsilon}\psi^{\alpha}=0$ and  $\delta_{\epsilon}d \Phi = d \delta_{\epsilon}\Phi$.


\subsection{Superforms, Integral forms and Pseudoforms}

As seen in 
\cite{Witten:2012bg,Belopolsky:1996cy,Belopolsky:1997bg,Belopolsky:1997jz,Voronov:1999mrk,VorGeom,CDGM, CCGir}, 
the space of differential forms has to be extended in order to define a meaningful integration theory. We define $\displaystyle \Omega^{(\bullet | \bullet)} \left( \mathcal{SM} \right) $ as the complete complex of forms; they are graded w.r.t. two gradings as
\begin{equation}
	\Omega^{(\bullet | \bullet)} = \oplus_{p,q} \Omega^{(p|q)} \ ,
\end{equation}
where $ q = 0 , \ldots , m $, $ p \leq n $ if $ q = m $ , $p \geq 0$ if $q=0$ and $p \in \mathbb{Z}$ if $q \neq 0,m$. The usual wedge product for form multiplication is defined as 
\begin{equation}\label{inNA}
\wedge: \Omega^{(p|r)} ({\cal SM}) \times \Omega^{(q|s)} ({\cal SM}) \longrightarrow \Omega^{(p+q|r+s)} ({\cal SM}) $$ $$ \left( \omega^{(p|r)} , \omega^{(q|s)} \right) \longrightarrow \omega^{(p|r)} \wedge \omega^{(q|s)} \,. 
\end{equation}
where $0\leq p,q \leq n$ and $0\leq r,s \leq m$ with $(n|m)$ are the bosonic and fermonic dimensions of the supermanifold ${\cal SM}$ \footnote{Notice that the wedge product is defined to be graded commuting.}.

Locally, a $(p|r)$-form $\omega$ formally reads 
\begin{equation}\label{inNBA}
\omega = \sum_{l,h,r} \omega_{[a_1 \dots a_l] (\a_{1} \dots \a_{h}) [\b_{1} \dots \b_{r}]}  dx^{a_1} \dots dx^{a_l} 
d\theta^{\a_1} \dots d\theta^{\a_h} \delta^{^{g(\beta_1)}}(d\theta^{\b_1}) \wedge   \dots_\wedge \delta^{^{g(\beta_r)}}(d\theta^{\b_r})
\end{equation}
where $g(x)$ denotes the differentiation degree of the Dirac delta function corresponding to the 1-form $d\theta^x$.  The 
three indices $l, h$ and $r$ satisfy the relation 
\begin{equation}\label{inNBB}
l + h - \sum_{k=1}^r g(\b_k) = p\,, ~~~~~~\a_l \neq \{\b_1, \dots, \b_r\} ~~~ \forall l=1,\dots,h\,,
\end{equation}
where the last equation means that each $\alpha_l$ in the above summation should be different from any $\beta_k$, otherwise the 
degree of the differentiation of the Dirac delta function can be reduced and the corresponding 1-form $d\theta^{\a_k}$ is removed from the basis. 
 The components  $\omega_{[i_1 \dots i_l] (\a_{1} \dots \a_{m}) [\b_{1} \dots \b_{r}]}$  
of $\omega$ are superfields. 

Due to the anticommuting properties of the $\delta$ forms, this product is by definition equal to zero if the forms to be multiplied contain $\delta$ localized in the same variables $d\theta$, since the $\delta$'s have to be considered as \emph{de Rham currents} \cite{Witten:2012ga}. In fig. 1, we display the complete complex of forms. We refer to the first line as the complex of \emph{superforms}, to the last line as the complex of \emph{integral forms} and to the middle lines as the complex of \emph{pseudoforms}.
We notice that the first line and the last line are bounded from below and from above, respectively. This is due to the fact that in the first line, being absent any delta functions, 
the form number cannot be negative, and in the last line, having saturated the number of delta functions we cannot admit any power of $d\theta$ (because of the 
distributional law $d\theta \delta(d\theta) =0$). In our case, we have $n=3$ and $m=2$, hence the complex has three lines.
 
 \def\de{\mathrm{d}}
\begin{figure}[t]
\begin{center}
\begin{tabular}{c@{\hskip -0mm}c@{\hskip 0mm}c}
\begin{tabular}{c@{\hskip 1.5mm}c@{\hskip 0.1mm}c}
&&\\[-0.34cm]
&\hspace{0.7cm}$0$&$\stackrel{\de}{\longrightarrow}$\\[0.06cm]
&\hspace{0.655cm}\tiny{$Z$}\hspace{0.3mm}\large{$\uparrow$}\phantom{\tiny{$Z$}}&\\[-0.05cm]
&\hspace{0.68cm}\vdots&\\[-0.05cm]
$\cdots$&$\Omega^{(-1|s)}$&$\stackrel{\de}{\longrightarrow}$\\[-0.15cm]
&\hspace{0.68cm}\vdots&\\[0.05cm]
&\hspace{0.655cm}\tiny{$Z$}\hspace{0.3mm}\large{$\uparrow$}\phantom{\tiny{$Z$}}&\\[0.05cm]
$\cdots$&$\Omega^{(-1|m)}$&$\stackrel{\de}{\longrightarrow}$\\[0.08cm]
\end{tabular}&
\begin{tabular}{|@{\hskip 1.5mm}c@{\hskip 1mm}c@{\hskip 2mm}c@{\hskip 1.5mm}c@{\hskip 1.5mm}c@{\hskip 1.5mm}c@{\hskip 1.5mm}c@{\hskip 1.5mm}|}
\hline
&&&&&&\\[-0.34cm]
$\Omega^{(0|0)}$&$\stackrel{\de}{\longrightarrow}$&$\cdots$&$\Omega^{(r|0)}$ &$\cdots$&$\stackrel{\de}{\longrightarrow}$&$\Omega^{(n|0)}$\\[0.06cm]
\tiny{$Z$}\hspace{0.3mm}\large{$\uparrow$}\large{$\downarrow$}\hspace{0.5mm}\tiny{$Y$}&&&\tiny{$Z$}\hspace{0.3mm}\large{$\uparrow$}\large{$\downarrow$}\hspace{0.5mm}\tiny{$Y$}&&&\tiny{$Z$}\hspace{0.3mm}\large{$\uparrow$}\large{$\downarrow$}\hspace{0.5mm}\tiny{$Y$}\\[-0.05cm]
\vdots&&&$\vdots$&&&$\vdots$\\[-0.05cm]
$\Omega^{(0|s)}$&$\stackrel{\de}{\longrightarrow}$&$\cdots$&$\Omega^{(r|s)}$ &$\cdots$&$\stackrel{\de}{\longrightarrow}$&$\Omega^{(n|s)}$\\[-0.15cm]
\vdots&&&$\vdots$&&&$\vdots$\\[0.05cm]
\tiny{$Z$}\hspace{0.3mm}\large{$\uparrow$}\large{$\downarrow$}\hspace{0.5mm}\tiny{$Y$}&&&\tiny{$Z$}\hspace{0.3mm}\large{$\uparrow$}\large{$\downarrow$}\hspace{0.5mm}\tiny{$Y$}&&&\tiny{$Z$}\hspace{0.3mm}\large{$\uparrow$}\large{$\downarrow$}\hspace{0.5mm}\tiny{$Y$}\\[0.05cm]
$\Omega^{(0|m)}$&$\stackrel{\de}{\longrightarrow}$&$\cdots$&$\Omega^{(r|m)}$&$\cdots$ &$\stackrel{\de}{\longrightarrow}$&$\Omega^{(n|m)}$\\[0.08cm]
\hline
\end{tabular}
&
\begin{tabular}{c@{\hskip 2mm}c@{\hskip 1.5mm}c}
&&\\[-0.34cm]
$\stackrel{\de}{\longrightarrow}$&$\Omega^{(n+1|0)}$&$\cdots$\\[0.06cm]
&\hspace{-1.15cm}\phantom{\tiny{$Y$}}\hspace{0.5mm}\large{$\downarrow$}\hspace{0.5mm}\tiny{$Y$}&\\[-0.05cm]
&\hspace{-1.15cm}\vdots&\\[-0.05cm]
$\stackrel{\de}{\longrightarrow}$&$\Omega^{(n+1|s)}$&$\cdots$\\[-0.15cm]
&\hspace{-1.15cm}\vdots&\\[0.05cm]
&\hspace{-1.15cm}\phantom{\tiny{$Y$}}\hspace{0.5mm}\large{$\downarrow$}\hspace{0.5mm}\tiny{$Y$}&\\[0.05cm]
$\stackrel{\de}{\longrightarrow}$&\hspace{-1.15cm}$0$&\\[0.08cm]
\end{tabular}
\end{tabular}
\end{center}
\vskip -0.2cm\caption{\rm \scriptsize Structure of the supercomplex of forms on a supermanifold of dimension $(m|n)\,$. The form degree $r$ increases going from left to right while the picture degree $s$ increases going from up to down. The rectangle contains the subset of the supercomplex where the various pictures are isomorphic in the cohomology of the $d$ differential.}\label{scomplex}
\end{figure}

The top form can be represented by the expression
\begin{equation}
\label{top3d}\omega^{(3|2)} = \omega(x,\theta) \epsilon_{abc} V^{a}\wedge V^{b} \wedge V^{c} \wedge\epsilon_{\alpha\beta} \delta(\psi^{\alpha}) \wedge
\delta(\psi^{\beta})\,,
\end{equation}
where $\omega(x,\theta)$ is a superfield 
which has the properties
\begin{equation}
\label{top3dB}d \omega^{(3|2)} = 0\,, ~~~~~ \mathcal{L}_{\epsilon}
\omega^{(3|2)} =d \Sigma^{(2|2)}\,.
\end{equation}

It is important to point out the transformation properties of $\omega^{(3|2)}$ under a Lorentz transformation of $SO(2,1)$.
Considering $V^a$, which transforms in the vector representation of $SO(2,1)$, the combination $\epsilon_{abc} V^{a}\wedge V^{b} \wedge V^{c}$ is clearly invariant. On the other hand, $d\theta^\a$ transform under the spinorial
representation of $SO(2,1)$, say $\Lambda_\a^{~\b} = (\gamma^{ab})_\a^{~\b}  \Lambda_{ab}$ with $\Lambda_{ab} \in so(2,1)$, and thus
an expression like $\delta(d\theta^\a)$ is not covariant. Nonetheless, the combination $\epsilon^{\a\b} \delta(d\theta^\a) \delta(d\theta^\b) = 2 \delta(d\theta^1) \delta(d\theta^2)$ is invariant using formal mathematical properties of 
distributions, for instance $d\theta \delta(d\theta) = 0$ and $d\theta \delta'(d\theta) = - \delta(d\theta)$. We  recall that 
$\delta(\psi^\a) \wedge \delta(\psi^\b) = - \delta(\psi^\b) \wedge \delta(\psi^\a)$. 
In addition, $\omega^{(3|2)}$ has a bigger symmetry group: we can transform the variables ($V^\a, \psi^\a)$ under an element of the supergroup $SL(3|2)$. The form $\omega^{(3|2)}$ is a representative of the Berezinian bundle, the
equivalent for supermanifolds of the canonical bundle on bosonic manifolds.

Let us consider the space $\Omega^{(1|1)}$ in the middle complex, spanned (in the sense of formal series) by the following 
psuedo-forms 
\begin{eqnarray}\label{midA}
\Omega^{(1|1)} &=& {\rm span} \Big\{(d\theta^\a)^{n+1} \delta^{(n)}(d\theta^\b), dx^{a} (d\theta^\a)^n \delta^{(n)}(d\theta^\b), \nonumber \\
&& \epsilon_{a b c }dx^{b} dx^{c} 
(d\theta^\a)^n \delta^{(n+1)}(d\theta^\b), \epsilon_{a b c } dx^a dx^{b} dx^{c}  (d\theta^\a)^n \delta^{(n+2)}(d\theta^\b)\Big\}_{n \geq 0} \ ,
\end{eqnarray}
where the number $n$ is not fixed and it must be a non-negative integer. For example, consider the pseudoform spanned by the second element in \eqref{midA} with $n=1$: 
\begin{equation}\label{midAA}
	dx^a A_{a \alpha \beta}^{(0)} d \theta^\alpha \delta' \left( d \theta^\beta \right) \ ;
\end{equation}
we have the implicit summation for the indices $\alpha$ and $\beta$, thus \eqref{midAA} becomes
\begin{equation}
	dx^a A_{a \alpha \beta}^{(0)} d \theta^\alpha \delta' \left( d \theta^\beta \right) = dx^a \left[ A_{a 1 1}^{0} d \theta^1 \delta' \left( d \theta^1 \right) + A_{a 1 2}^{0} d \theta^1 \delta' \left( d \theta^2 \right) + A_{a 2 1}^{0} d \theta^2 \delta' \left( d \theta^1 \right) + A_{a 2 2}^{0} d \theta^2 \delta' \left( d \theta^2 \right) \right] \ .
\end{equation}
We have to recall the distributional identity
\begin{equation}
	\left\langle x \delta^{(p)} \left( x \right) , \phi \right\rangle = - p \left\langle \delta^{(p-1)} \left( x \right) , \phi \right\rangle \ ,
\end{equation}
for any test function $\phi$, which extends to the same rule for the $1$-form form $d \theta$:
\begin{equation}
	d \theta \delta^{(p)} \left( d \theta \right) = - p \delta^{(p-1)} \left( d \theta \right) \ .
\end{equation}
Hence we get
\begin{equation}
	dx^a A_{a \alpha \beta}^{(0)} d \theta^\alpha \delta' \left( d \theta^\beta \right) = dx^a \left[ -A_{a 1 1}^{0} \delta \left( d \theta^1 \right) + A_{a 1 2}^{0} d \theta^1 \delta' \left( d \theta^2 \right) + A_{a 2 1}^{0} d \theta^2 \delta' \left( d \theta^1 \right) - A_{a 2 2}^{0} \delta \left( d \theta^2 \right) \right] \ .
\end{equation}
Notice that the first and the last terms are elements that can be spanned by $\displaystyle dx^{a} (d\theta^\a)^n \delta^{(n)}(d\theta^\b) $ for $n=0$; this means that by a redefinition of the fields $A_{a \alpha \beta}^{(p)}$ we can assume w.l.o.g. that $\alpha \neq \beta$ in the implicit sums. This reflects the property that elements spanned by $\left( d \theta^\alpha \right)^{n+1} \delta^{(n)} \left( d \theta^\beta \right)$ are exactly equal to 0 if $\alpha = \beta , \forall n \geq 0$.

Due to $1$-forms $dx^a$ and due to the fact that we are free to set $\a \neq \b$, the number of generators (monomial forms) at a given $n$ is $(8|8)$, but the total number of monomial generators in $\Omega^{(1|1)}$ is infinite.

\subsection{Integration}

Once the integral forms are defined, we have to clarify how the integration is performed. For that 
we consider an integral form given by  
\begin{eqnarray}
\label{IN_A}
\omega^{(3|2)} = \omega(x,\theta) \epsilon_{abc} dx^a dx^b dx^c \epsilon^{\a\b} \delta(d\theta^\a) \delta(d\theta^\b)  \ ,
\end{eqnarray}
where $\omega(x,\theta)$ is a superfield section of the Berezinian bundle $\Omega^{(3|2)}({\cal SM})$. 
Then, the integral on the supermanifold ${\cal SM}^{(3|2)}$ is 
\begin{eqnarray}
\label{IN}
\int_{{\cal SM}^{(3|2)}} \omega^{(3|2)}  = 
\int  \omega(x,\theta) [d^3xd^2\theta]  \ .
\end{eqnarray}
We obtain the last integral, by performing the integration over $dx$'s, viewed as anticommuting 
variables. Consequently we use the Berezin integral, and the integration over $d\theta$, viewed as 
algebraic bosonic variables \cite{CDGM,Witten:2012bg,CCGGeom} and the distributional properties of $\delta(d\theta)$. The final expression 
contains a usual Riemann/Lebesgue integral on $x$'s and 
the Berezin integral over $\theta$'s. The symbol $ [d^3xd^2\theta]$ is only a reminder on which variables 
the integral has to be performed. 

For example, in the case of ${\cal SM}^{(3|2)} = {\mathbb R}^{(3|2)}$ we have 
\begin{eqnarray}
\label{IN_C}
\int_{{\cal SM}^{(3|2)}} \omega^{(3|2)}  = \frac12 \int \left. 
\epsilon^{\a\b} D_\a D_\b  \omega(x,\theta) \right|_{\theta =0} [d^3x]  \ ,
\end{eqnarray}
where the Berezin integration has been performed and we are left with the Riemann/Lebesgue integral. 

We define a product (Serre's duality) between $\Omega^{(p|q)}$ and $\Omega^{(r|s)}$ 
forms as
\begin{equation}\label{midB}
\Big\langle \omega^{(p|r)}, \omega^{(q|s)}\Big\rangle = \int_{{\cal SM}^{(3|2)}} 
\omega^{(p|r)} \wedge \omega^{(q|s)}  \ ,
\end{equation}
which is non-vanishing only if $p+q = 3$ and $r+s =2$. Under these conditions, the spaces $\Omega^{(p|r)}$ and $\Omega^{(q|s)}$ are isomorphic and therefore 
there is a (super)form in $\Omega^{(p|0)}$ corresponding to an integral form in $\Omega^{(3-p|2)}$. By partially 
computing the 
form integral (leaving undone only the Berezin integral over the coordinates $\theta$ and the Riemann/Lebesgue integral over $x$),
we have 
\begin{equation}\label{midC}
\Big\langle\omega^{(p|0)}, \omega^{(n-p|m)}\Big\rangle = \sum_{{\cal J}=1}^{{\rm dim}(\Omega^{(p|0)})} \int   \omega_{\cal J} \widetilde\omega^{\cal J}(x,\theta) [dx^n d^m\theta]  \ ,
\end{equation}
where $\omega_{\cal J}(x,\theta)$ are the coefficients (the index ${\cal J}$ stands for the collection of indices needed to 
define the form) of the form $\omega^{(p|r)}$, while $\widetilde\omega^{\cal J}$ are the coefficients of the dual forms in $\Omega^{(3-p|2)}$. 
For the space $\mathbb{R}^{(3|2)}$, if we consider for example the spaces $\Omega^{(1|0)}$ and $\Omega^{(2|2)}$
we have: 
\begin{equation}\label{midD}
\omega^{(1|0)} = \omega_a dx^a + \omega_\a d\theta^\a\,, ~~~~ \omega_{\cal J} = \{\omega_a, \omega_\a\}  \ ,
\end{equation}
and 
\begin{equation}\label{midE}
\widetilde \omega^{(2|2)} = \widetilde\omega^a \epsilon_{a b c} dx^b dx^c \delta^2(d\theta) + \widetilde\omega^{\a} \epsilon_{a b c} dx^a dx^b dx^c \iota_\a \delta^2(d\theta)\,, ~~~~~
\widetilde \omega^{\cal J} = \{\widetilde\omega^a, \widetilde\omega^{~\a}\}  \ .
\end{equation}
Then, we can compute $\Big\langle\omega^{(1|0)},  \widetilde\omega^{(2|2)}\Big\rangle$ as
\begin{equation}\label{midF}
\Big\langle\omega^{(1|0)}, \widetilde\omega^{(2|2)}\Big\rangle = 
\int_{{\cal SM}^{(3|2)}} 
\omega^{(1|0)} \wedge \omega^{(2|2)} =
 \int \Big( \omega_a \widetilde\omega^a  -  \omega_\a \widetilde\omega^{\a} \Big)[dx^3 d^2\theta]  \ .
\end{equation}
Notice that the product is a pairing and it does not need to be positive definite.
 
If we use the same technique for $\Omega^{(1|1)}$ and $\Omega^{(2|1)}$, we have to recall that the dimension of these spaces is infinite and therefore the 
sum over ${\cal J}$ must be substituted with formal series. In the same way as described in the previous subsection, for a general supermanifold ${\cal SM}^{(3|2)}$ 
any form belonging to the middle complex $\Omega^{(p|1)}$ is decomposed into 
an infinite number of components as in (\ref{midA}).

If we use the following distributional relation
\begin{equation}\label{midG}
(d\theta_2)^p \delta^{(q)}(d\theta_1) \wedge (d\theta_1)^q \delta^{(p)}(d\theta_2) = 
(-1)^{p+q} p! q!  \,  \delta(d\theta_1)  \wedge \delta(d\theta_2)  \ ,
\end{equation}
where $p,q\geq 0$, we can 
parametrise the space $\Omega^{(1|1)}$ as 
\begin{eqnarray}\label{midH}
\omega^{(1|1)} &=& \sum_n \Big(
\phi^{12}_n (d\theta_1)^{n+1} \delta^{(n)}(d\theta_2) + \phi^{21}_n (d\theta_2)^{n+1} \delta^{(n)}(d\theta_1)
+ \nonumber \\
&+& 
H^{12}_{a, n} dx^a  (d\theta_1)^{n} \delta^{(n)}(d\theta_2) +
H^{21}_{a, n} dx^a  (d\theta_2)^{n} \delta^{(n)}(d\theta_1) +\nonumber \\
&+&  
K^{12}_{[ab], n} dx^a dx^b  (d\theta_1)^{n} \delta^{(n+1)}(d\theta_2) + 
K^{21}_{[ab], n} dx^a  dx^b (d\theta_2)^{n} \delta^{(n+1)}(d\theta_1) 
\nonumber \\
&+&
\psi^{12}_n d^3x (d\theta_1)^{n} \delta^{(n+2)}(d\theta_2) + 
\psi^{21}_n d^3x (d\theta_2)^{n} \delta^{(n+2)}(d\theta_1) \Big) \ ,
\end{eqnarray}
where again the various components $( \phi^{12}_n, \phi^{21}_n, \dots, \psi^{21}_n)$ are superfields. In the same way, 
we can parametrise the space $\Omega^{(2|1)}$ as
\begin{eqnarray}\label{midH}
\widetilde\omega^{(2|1)} &=& \sum_n \Big(
\rho^{12}_n (d\theta_1)^{n+2} \delta^{(n)}(d\theta_2) + 
\rho^{21}_n (d\theta_2)^{n+2} \delta^{(n)}(d\theta_1)
+ \nonumber \\
&+& 
L^{12}_{a, n} dx^a  (d\theta_1)^{n+1} \delta^{(n)}(d\theta_2) +
L^{21}_{a, n} dx^a  (d\theta_2)^{n+1} \delta^{(n)}(d\theta_1) +\nonumber \\
&+&  
M^{12}_{[ab], n} dx^a dx^b  (d\theta_1)^{n} \delta^{(n)}(d\theta_2) + 
M^{21}_{[ab], n} dx^a  dx^b (d\theta_2)^{n} \delta^{(n)}(d\theta_1) 
\nonumber \\
&+&
\tau^{12}_n d^3x (d\theta_1)^{n} \delta^{(n+1)}(d\theta_2) + 
\tau^{21}_n d^3x (d\theta_2)^{n} \delta^{(n+1)}(d\theta_1) \Big) \ ,
\end{eqnarray}
where the various components $( \rho^{12}_n, \rho^{21}_n, \dots, \tau^{21}_n)$ are superfields.
 
Now, we compute the product between two forms $\omega^{(1|1)}$ and $\omega^{(2|1)}$ as follows 
\begin{eqnarray}\label{midI}
\Big\langle\omega^{(1|1)}, \widetilde\omega^{(2|1)}\Big\rangle &=& \int_{{\cal SM}^{(3|2)}} \omega^{(1|1)} \wedge \widetilde\omega^{(2|1)} \nonumber \\ 
&=& \sum_{n=0}^\infty \int \Big( (\phi^{12}_n \tau^{21}_n - \phi^{21}_n \tau^{12}_n) + (\psi^{12}_n \rho^{21}_n - \psi^{21}_n \rho^{12}_n) + \nonumber \\
&+& (H^{12}_{a,n} M^{21}_{bc, n} - H^{21}_{a,n} M^{12}_{bc,n})\epsilon^{abc} + (K^{12}_{ab,n} L^{21}_{c, n} - K^{21}_{ab,n} L^{12}_{c,n})\epsilon^{abc}
\Big)[d^3x d^2\theta]\,. \nonumber 
\end{eqnarray}

Apparently, the previous expression does not seem to be covariant under Lorentz transformations. However, 
since the various superfields are independent, they can be re-organized into covariant expressions of the 
form 
\begin{eqnarray}\label{midL}
\Big\langle\omega^{(1|1)}, \widetilde\omega^{(2|1)}\Big\rangle &=& \int_{\cal SM} \omega^{(1|1)} \wedge \widetilde\omega^{(2|1)} \nonumber \\ 
&=& \sum_{n=0}^\infty \int \Big( ( \Phi^{\a\b}_n \Psi^{\gamma\delta}_n)\epsilon_{\a\gamma} \epsilon_{\b\delta}+ 
(R^{\a\b}_{ab,n} S^{\gamma\delta}_{c, n})\epsilon^{abc} \epsilon_{\a\gamma} \epsilon_{\b\delta}
\Big)[d^3x d^2\theta] \ , \nonumber 
\end{eqnarray}
where we have collected the superfields $\phi^{12}_n, \dots \tau^{21}_n$ into the two superfields $\Phi^{\a\b}_n, \Psi^{\gamma\delta}_n$, $H^{12}_{a,n}, \dots, L^{21}_{a, n}$ into  $S^{\gamma\delta}_{c, n}$ and $M^{21}_{ab, n}, \dots, K^{12}_{ab,n}$ 
into $R^{\a\b}_{ab,n}$. The important issue of the Lorentz covariance is discussed in the next subsection. 


\subsection{Covariance on $\Omega^{(p|r)}$}

In this subsection, we clarify how the Lorentz symmetry is implemented in the space of pseudo-forms. This is a crucial point in order to understand how the covariance is recovered at any picture number. 

We consider an infinitesimal Lorentz transformation $\delta^a_b + w^{a}_{~b} + {\cal O}(w^2)$ 
of $SO(2,1)$ (with $w_{ab} = - w_{ba}$). It acts on coordinates $x^a, \theta^\a$ according to vector and spinor representations
\begin{equation}\label{LOCA}
\delta  x^a = w^{a}_{~b} x^b\,, ~~~~~~
\delta  \theta^\a = \frac14 w_{ab} ( \gamma^{ab})^\a_{~\b} \theta^\b\,. 
\end{equation}
In the same way, the $(1|0)$-superforms 
$(dx^a, d\theta^\a)$ transform in the vector and spinor representations, respectively. Thus, all forms belonging to 
the complex with zero picture, namely $\Omega^{(p|0)}$, transform under the tensorial representations of each single monomial. For example, given 
$\omega_{[ab]( \a_1 \dots \a_n)} dx^a dx^b d\theta^{\a_1} \dots d\theta^{\a_n}$, the components $\omega_{[ab]( \a_1 \dots \a_n)}(x,\theta)$ 
transform under the anti-symmetrized product of the conjugated vector representation tensored with $n$-symmetrized conjugated spinor representation.  

If we consider the complex of integral forms $\Omega^{(p|2)}$, and we perform an infinitesimal Lorentz transformation. We have to 
use the distributional relation
\begin{equation}\label{LOCB}
\delta(\a d\theta^1 + \b d\theta^2) \delta(\gamma d\theta^1 + \delta d\theta^2) =
\frac{1}{\det
\left(
\begin{array}{cc}
 \a & \b   \\
 \gamma & \delta     
\end{array}
\right)
} 
\delta(d\theta^1) \delta(d\theta^2) 
 \end{equation}
to check that the product of $\delta(d\theta^1)\delta(d\theta^2)$ transforms as an inverse of a  density (we avoid the absolute value of the determinant since we are keeping track of the orientation of the integration) 
and therefore, each monomial of the 
complex $\Omega^{(p|2)}$ transforms according to a tensorial representation and the inverse of the determinant of a Lorentz transformation in 
the spinor representation (sections of the Berezinian bundle). This confirms the fact that the top form $d^3x \delta^2(d\theta)$ is indeed invariant under Lorentz transformations. In addition, when the derivatives of the product $\delta(d\theta^1)\delta(d\theta^2)$  are taken into account, for example as in the $\Omega^{(-2|2)}$ form 
\begin{eqnarray}
\label{LOCBA}
\omega^{(-2|2)} = \omega^{\a\b} \iota_\alpha \iota_\beta \delta(d\theta^1)\delta(d\theta^2) \ ,
\end{eqnarray}
the components $\omega^{\a\b}(x,\theta)$ transform as in a linear tensor representation of the spinorial representation. This means that the spinorial indices in (\ref{LOCBA}) are covariantly contracted. Therefore, 
for both the superforms $\Omega^{(p|0)}$ and the integral forms $\Omega^{(p|2)}$, the 
usual covariant techniques can be used. 

Let us now consider the infinite dimensional complex $\Omega^{(p|1)}$. As seen above, it is unbounded from above and from below and each space $\Omega^{(p|1)}$ is (double)-infinite dimensional. Double means that 
we have two ways to construct a pseudo form, along $\delta(d\theta^1)$ and along $\delta(d\theta^2)$. 
However, under any transformation which mixes $\theta^1$ with $\theta^2$ (for example Lorentz transformations) the two directions indeed mix and the following situation arises. 

If we consider a single Dirac delta function $\delta(d\theta^1)$, we cannot use the distributional identity 
(\ref{LOCB}), but we observe that, infinitesimally,
\begin{eqnarray}\label{LOCC}
\delta(d\theta^1) &\longrightarrow& \delta\Big(d\theta^1 + \frac14 w_{ab} (\gamma^{ab})^1_{~\b} d\theta^\b\Big) =  (1 - \frac14  (\gamma^{ab})^1_{~1}) \delta(d\theta^1) + \frac14 w_{ab} (\gamma^{ab})^1_{~2} d\theta^2 \delta^{(1)}(d\theta^1) + {\cal O}(w^2) \nonumber \\
\delta(d\theta^2) &\longrightarrow& \delta\Big(d\theta^2 + \frac14 w_{ab} (\gamma^{ab})^2_{~\b} d\theta^\b\Big) =  (1 - \frac14  (\gamma^{ab})^2_{~2}) \delta(d\theta^2) + \frac14 w_{ab} (\gamma^{ab})^2_{~1} d\theta^1 \delta^{(1)}(d\theta^2) + {\cal O}(w^2) 
\end{eqnarray}
where $\delta^{(1)}(d\theta^\a)$ is the first derivative of $\delta(d\theta^\a)$ and we have neglected higher order terms. The first and the second terms come from the Taylor expansion of the Delta distribution, with $d\theta^1$ and $d\theta^2$ respectively. This fact implies that in order to implement the Lorentz symmetry in the 
space of pseudo-forms $\Omega^{(p|1)}$, one necessarily needs an infinite dimensional space. Indeed, for a finite Lorentz transformation one needs all components in the 
$n$ expansion of a generic pseudoform in $\Omega^{(p|1)}$. 
For example, let us consider a $(0|1)$-pseudoform, it can be written as
\begin{eqnarray}
\label{LOCD}
\omega^{(0|1)} = \omega^{(0|1)}_0 + \omega^{(-1|1)}_1 + \omega^{(-2|1)}_2 + \omega^{(-3|1)}_3 \ ,
\end{eqnarray}
where we collected the pieces with different powers of $dx$'s (we use a little abuse of notation by omitting the $dx's$ and writing as superscripts only the fermionic form number and the picture number). Since the first term $\omega^{(0|1)}_0$ does not contains powers of $dx$, it can be written as
\begin{eqnarray}
\label{LOCE}
\omega^{(0|1)}_0 = \sum_{n=0}^\infty \left( 
\omega^{(n)}_{12}(x,
\theta) (d\theta^1)^n \delta^{(n)}(d\theta^2) + 
\omega^{(n)}_{21}(x,
\theta) (d\theta^2)^n \delta^{(n)}(d\theta^1)\right)  \ ,
\end{eqnarray}
where the coefficients $\omega^{(n)}_{12}(x,\theta), \omega^{(n)}_{21}(x,\theta)$ 
are superfields. Since we have distinguished the terms with $d \theta^1$ and $d \theta^2$, the covariance of the expression is not manifest. Indeed, it might be better to write \eqref{LOCE} as
\begin{eqnarray}
\label{LOCF}
\omega^{(0|1)}_0 = \sum_{n=0}^\infty \left( 
\omega^{(n), ~\beta}_{\a}(x,\theta) (d\theta^\alpha)^n \delta^{(n)}(d\theta^\beta) \right) \ ,
\end{eqnarray}
where the indices $\alpha$ and $\beta$ are summed, as conventionally. Notice that if $\alpha = \beta$, we have $(d\theta^\a)^n$ multiplying $\delta^{(n)}(d\theta^\a)$ 
and, by using the distributional property 
$(d\theta^\a)^n \delta^{(n)}(d\theta^\a) = (-1)^n n! \delta(d\theta^\a)$, the coefficient 
$\omega^{(n), ~\alpha}_{\a}(x,\theta)$ is reabsorbed into a redefinition of 
$\omega^{(0)} (x,\theta)$ which multiplies $\delta(d\theta^\a)$.   

If we perform an infinitesimal Lorentz transformation $w_{ab}$, 
we have that 
\begin{eqnarray}
\label{LOCG}
\omega^{(0|1)}_0 &\rightarrow&  
\sum_{n=0}^\infty \left( 
\omega^{(n), ~\beta}_{\a}(x,\theta) \Big(d\theta^\alpha + \frac14 w_{ab} 
(\gamma^{ab})^\a_{~\b} d\theta^\b\Big)^n 
\delta^{(n)}\Big(d\theta^\beta + \frac14 w_{ab} (\gamma^{ab})^\b_{~\gamma} d\theta^\gamma\Big) \right) \nonumber \\
&=& \sum_{n=0}^\infty 
	\omega^{(n), ~\beta}_{\a}(x,\theta) 
		\Big(
			( d\theta^\alpha)^n  + \frac{n}{4} w_{ab} 
			(\gamma^{ab})^\a_{~\b} ( d\theta^\alpha)^{n-1} d\theta^\b
		\Big) 
		\Big(
		\delta^{(n)}(d\theta^\beta) 
+ \frac14 w_{ab} (\gamma^{ab})^\b_{~\gamma} d\theta^\gamma 
\delta^{(n+1)}(d\theta^\beta)  \Big)\nonumber \\
&=& 
 \sum_{n=0}^\infty \left( 
	\widehat\omega^{(n), ~\beta}_{\a}(x,\theta)  
	(d\theta^\alpha)^n \delta^{(n)}(d\theta^\beta) \right) \ ,
\end{eqnarray}
where the coefficients $\widehat\omega^{(n), ~\beta}_{\a}(x,\theta)$ 
are suitably redefined using
\begin{eqnarray}
\label{LOCH}
&&\Big(
			( d\theta^\alpha)^n  + \frac{n}{4} w_{ab} 
			(\gamma^{ab})^\a_{~\b} ( d\theta^\alpha)^{n-1} d\theta^\b
		\Big) 
= \Big(1 + \frac{n}{4} w_{ab} 
			(\gamma^{ab})^\a_{~\a}\Big)  ( d\theta^\alpha)^n + 
			 \frac{n}{4} w_{ab} 
			\sum_{\b \neq \a} (\gamma^{ab})^\a_{~\b} ( d\theta^\alpha)^{n-1} d\theta^\b \ , \nonumber \\
&&\Big(
		\delta^{(n)}(d\theta^\beta) 
+ \frac14 w_{ab} (\gamma^{ab})^\b_{~\gamma} d\theta^\gamma 
\delta^{(n+1)}(d\theta^\beta)  \Big) = 
	\Big(1 - \frac{n+1}{4} w_{ab} 
			(\gamma^{ab})^\b_{~\b}\Big) 
\delta^{(n)}(d\theta^\beta) 
+ \frac14 w_{ab} \sum_{\gamma\neq \b} 
(\gamma^{ab})^\b_{~\gamma} d\theta^\gamma 
\delta^{(n+1)}(d\theta^\beta) \ . \nonumber \\	
 \end{eqnarray}
Then the coefficients $\omega^{(n),~\b}_\a$ are shifted as 
\begin{eqnarray}
\label{LOCK}
\delta \omega^{(n),~\b}_\a &=& \Big(\delta^\b_\a + \frac{n}{4} w_{ab} (\gamma^{ab})^\a_\a - \frac{n+1}{4} 
w_{ab} (\gamma^{ab})^\b_\b\Big)  \omega^{(n),~\b}_\a \nonumber \\
\delta \omega^{(n+1),~\b}_\a &=& -\frac{n+1}{4} w_{ab} (\gamma^{ab})^\b_{~\a} \omega^{(n),~\b}_\a \nonumber \\
\delta \omega^{(n-1),~\b}_\a &=& -\frac{n^2}{4} w_{ab} (\gamma^{ab})^\a_{~\b} \omega^{(n),~\b}_\a \,. 
\end{eqnarray}
This holds at the infinitesimal level, but for a finite transformation all the coefficients $\omega^{(n),~\a}_\b$ are 
involved. Therefore, the covariance of the expressions is maintained only if the complete series is taken into account. For other pieces $\omega^{(-p|1)}_p$ with $p=1,2,3$, we notice that the dependence upon $dx^a$ is 
polynomial and therefore they transform linearly as always, but in addition there is a complete reshuffling of 
the coefficients of the series. In the next sections we will adopt the notation of writing the Greek indices of the components fields of forms both below.

\subsection{Geometric Picture Changing Operators: some explicit results}

Having clarified the form complexes and having outlined how usual differential operators of Cartan calculus ($\displaystyle d , \iota_X , \mathcal{L}_X$) work on superspace, we point out that we can build a new set of differential operators\footnote{We use the words \emph{differential operator} in order to indicate any generalised function of usual differential operators.} acting on general forms such as $\displaystyle \delta \left( \iota_v \right) , \Theta \left( \iota_v \right) , Z_v , \mathbb{Y}_v , \eta_v $\cite{Belopolsky:1997jz,Catenacci:2018xsv,Catenacci-Grassi-Noja}. These operators are used to change the picture number of a given form (and eventually its form number as well) and are usually referred to as \emph{Picture Changing Operators} (PCO's). The specific form of those operators is suggested by String Theory analogy \cite{FMS,pol} and their geometric interpretation \cite{Belopolsky:1997bg}. In the present section we provide some results that will be used in the rest of the paper.

The first PCO we define is $\mathbb{Y}$: given a $(p|q)$-form $\displaystyle \omega^{(p|q)} \in \Omega^{(p|q)}$, we define the \emph{Picture Raising Operator} $\mathbb{Y}^{(0|s)}$ as a multiplicative operator s.t.
	\begin{eqnarray}
		\nonumber \mathbb{Y}^{(0|s)} : \Omega^{(p|q)} &\longrightarrow& \Omega^{(p|q+s)} \\
		\label{PCOY} \omega^{(p|q)} &\mapsto& \omega^{(p|q)} \wedge \mathbb{Y}^{(0|s)} \ .
	\end{eqnarray}
	Since it is a multiplicative operator that raises the picture number by $s$, it follows that locally 
\begin{equation*}
	\mathbb{Y}^{(0|s)} \propto \delta \left( d \theta^{\alpha_1} \right) \cdots \delta \left( d \theta^{\alpha_s} \right).
\end{equation*}

Again, given a $(p|q)$-form $\displaystyle \omega^{(p|q)} \in \Omega^{(p|q)}$, we define the \emph{Picture Lowering Operator} $Z_D$ as
	\begin{eqnarray}
		\nonumber Z_v : \Omega^{(p|q)} &\longrightarrow& \Omega^{(p|q-1)} \\
		\label{PCOZ} \omega^{(p|q)} &\mapsto& Z_v \left( \omega^{(p|q)} \right) = \left[ d , -i \Theta ( \iota_D ) \right] \omega^{(p|q)} \ ,
	\end{eqnarray}
	where $[ \cdot , \cdot ]$ denotes as usual a graded commutator and the action of the operator $\Theta ( \iota_v )$ is defined by the Fourier-like relation of the Heaviside step function
	\begin{equation} \label{PCOT}
		\Theta ( \iota_v ) \omega^{(p|q)} ( d \theta^\alpha ) = - i \lim_{\epsilon \to 0} \int_{- \infty} ^\infty \frac{dt}{t + i \epsilon} e^{i t \iota_v} \omega^{(p|q)} \left( d \theta^\alpha \right) = - i \lim_{\epsilon \to 0} \int_{- \infty} ^\infty \frac{dt}{t + i \epsilon} \omega^{(p|q)} \left( d \theta^\alpha + i t v^\alpha \right) \ ,
	\end{equation}
	where we have used the fact that $\displaystyle e^{i t \iota_v} $ is a translation operator. Hence the operator $\Theta (\iota_v)$ is of the form
	\begin{equation*}
		\Theta (\iota_v) : \Omega^{p|q} \to \Omega^{p-1|q-1} \ ,
	\end{equation*}
	i.e. it lowers either the form degree or the picture degree. As we will see in the following examples this operator does not give a pseudoform as a result, but rather an $inverse$ $form$, i.e. an expression containing negative powers of $d \theta$. We remark, as was discussed in \cite{Catenacci:2018xsv}, that the introduction of inverse form requires the definition of a new complex $\Omega^{(\bullet|\bullet)}_L$ corresponding to 
	the Large Hilbert Space (LHS) used in string theory.  In the following, we will denote simply by 
	$\Omega^{(\bullet|\bullet)}$ the space suitably enlarged. The relation between Large Hilbert Space and 
	Small Hilbert Space (SHS) was clarified in \cite{Catenacci:2018xsv} in the case of a single fermionic variable. 
		
Here we list some examples, not only in order to explain how to manipulate the $\Theta ( \iota_v )$ operator, but also in order to prepare some results that will be used in the next sections. In particular we have opted to highlight some of the following results to stress their particular significance and because they will be directly employed.
\begin{ex}
	Let us consider the case where $\displaystyle \omega^{(p|q)} = \delta ( d \theta^\alpha ) $, we have
	\begin{equation}
		\Theta ( \iota_v ) \delta ( d \theta^\alpha ) = - i \lim_{\epsilon \to 0} \int_{- \infty} ^\infty \frac{dt}{t + i \epsilon} \delta \left( d \theta^\alpha + i t v^\alpha \right) = \frac{-i}{i v^\alpha} \lim_{\epsilon \to 0} \int_{- \infty} ^\infty \frac{dt}{t + i \epsilon} \delta \left( t + \frac{d \theta^\alpha}{i v^\alpha} \right) = \frac{i}{d \theta^\alpha} \ .
	\end{equation}
	We can also obtain the previous result in a slightly different way:
	\begin{equation}
		\Theta ( \iota_v ) \delta ( d \theta^\alpha ) = - i \lim_{\epsilon \to 0} \int_{- \infty} ^\infty \frac{dt}{t + i \epsilon} \delta \left( d \theta^\alpha + i t v^\alpha \right) = \frac{-i}{i v^\alpha} \int \frac{dy}{\frac{y - d \theta^\alpha}{i v^\alpha}} \delta \left( y \right) = \frac{i}{d \theta^\alpha} \ ,
	\end{equation}
	where in the second passage we have performed the substitution $\displaystyle y = d \theta^\alpha + i v^\alpha t $ .\footnote{Since we are working with pseudoforms, the rules of the $\delta$ distributions are to be considered formal algebraic rules (for example, in the previous calculation $\displaystyle - \frac{d \theta^\alpha}{i v^\alpha} $ in not a $c$-number).}
\end{ex}
\begin{ex}
	We have the following result:
	\begin{equation} \fbox{$\displaystyle
		\Theta ( \iota_v ) \delta^{(p)} ( d \theta^\alpha ) = \frac{-i (-1)^{p+1} p!}{(d \theta^\beta )^{p+1}} \ .
		$}
	\end{equation}
		The result already stated follows after a direct calculation:
		\begin{equation*}
			\Theta ( \iota_v ) \delta^{(p)} ( d \theta^\alpha ) = -i \lim_{\epsilon \to 0} \int_{- \infty}^\infty \frac{dt}{t + i \epsilon} \frac{d^p}{d \left( d \theta^\alpha + i v^\alpha t \right)^p} \delta \left( d \theta^\alpha + i v^\alpha t \right) = \frac{-i}{i v^\alpha} \int \frac{dy}{\frac{y - d \theta^\alpha}{i v^\alpha}} \frac{d^p}{d y^p} \delta (y) = $$ $$ = -i (-1)^p \int dy \left[ \frac{d^p}{dy^p} \left( y - d \theta^\alpha \right)^{-1} \right] \delta (y) = -i (-1)^p (-1)^p p! \left. \left( y - d \theta^\alpha \right)^{- p - 1}\right|_{y = 0} = \frac{-i (-1)^{p + 1} p!}{\left( d \theta^\alpha \right)^{p+1}} \ .
		\end{equation*}
\end{ex}
In order to get more general formulas we consider other simple examples.
\begin{ex}
	Let us consider $\displaystyle \omega^{(p|q)} = d \theta^\alpha \delta ( d \theta^\beta ) $, we have
	\begin{equation}
		\Theta ( \iota_v ) d \theta^\alpha \delta ( d \theta^\beta ) = -i \lim_{\epsilon \to 0} \int_{- \infty}^\infty \frac{dt}{t + i \epsilon} \left( d \theta^\alpha + i v^\alpha t \right) \delta \left( d \theta^\beta + i v^\beta t \right) = $$ $$ = \frac{i}{v^\beta d \theta^\beta} \left( v^\beta d \theta^\alpha - v^\alpha d \theta^\beta \right) = \frac{-i}{v^\beta d \theta^\beta} v \cdot d \theta \epsilon^{\alpha \beta} \ ,
	\end{equation}
	where we have defined $\displaystyle - \epsilon^{\alpha \beta} v \cdot d \theta = v^\alpha d \theta^\beta - v^\beta d \theta^\alpha$ . Observe that we expect to get $0$ if $\alpha = \beta$, since $d \theta^\alpha \delta ( d \theta^\alpha ) = 0$; if we put $\alpha = \beta$ in the result we exactly get 0.
\end{ex}

\begin{ex}
	We have the following result:
	\begin{equation} \label{T1} \fbox{$\displaystyle
			\Theta ( \iota_v ) \left( d \theta^\alpha \right)^p \delta^{(q)} ( d \theta^\beta ) = i (-1)^q q! \frac{(d \theta^\alpha)^p}{( d \theta^\beta )^{q+1}} \ , \ \text{if} \ q \geq p \ .
		$}
	\end{equation}
		Again, the result follows from direct computation:
		\begin{equation}
			\Theta ( \iota_v ) \left( d \theta^\alpha \right)^p \delta^{(q)} ( d \theta^\beta ) = -i \lim_{\epsilon \to 0} \int_{- \infty}^\infty \frac{dt}{t + i \epsilon} \left( d \theta^\alpha + i v^\alpha t \right)^p \frac{d^q}{d \left( d \theta^\beta + i v^\beta t \right)^q} \delta \left( d \theta^\beta + i v^\beta t \right) = $$ $$ = - i (-1)^q \frac{d^q}{d y^q} \left[ \frac{1}{y - d \theta^\beta} \left( d \theta^\alpha + \frac{v^\alpha}{v^\beta} ( y - d \theta^\beta ) \right)^p \right]_{y = 0} = - i (-1)^q (d \theta^\alpha )^p \frac{d^q}{d y^q} \left[ \left( y - d \theta^\beta \right)^{-1} \right]_{y = 0} = $$ $$ = i (-1)^q q! \frac{(d \theta^\alpha )^p}{( d \theta^\beta )^{q + 1}} \ ,
		\end{equation}
		where we have made use of the assumption $q \geq p$ when expanding the binomial: the term with highest power of $y$ behaves like $y^p$, but since it is multiplied by a $y^{-1}$ term, we have a global $y^{p-1}$ which is annihilated by $\displaystyle \frac{d^q}{d y^q} $ if $q > p - 1$, i.e. $q \geq p$. The same  happens for all the other terms of the expansion except for the $\displaystyle \left( d \theta^\alpha \right)^p$ term which is multiplied by $\displaystyle \left( y - d \theta^\beta \right)^{-1} $ and does not give a trivial result after derivation.
\end{ex}
The following examples are studied because they are explicitly needed in the following section.
\begin{ex}
	Let us consider $\displaystyle \omega^{(p|q)} = \left( d \theta^\alpha \right)^{p+1} \delta^{(p)} ( d \theta^\beta ) $, we have
	\begin{equation} \label{T2}
		\Theta ( \iota_v ) \left( d \theta^\alpha \right)^{p+1} \delta^{(p)} ( d \theta^\beta ) =
		 - i (-1)^p p! \left[ \left( \frac{v^\alpha}{v^\beta} \right)^{p+1} - \left( \frac{d \theta^\alpha}{d \theta^\beta} \right)^{p+1} \right] \ ,
	\end{equation}
	where the result arises from a straightforward calculation.
\end{ex}
\begin{ex}
	Let us consider $\displaystyle \omega^{(p|q)} = \left( d \theta^\alpha \right)^{p+2} \delta^{(p)} ( d \theta^\beta ) $, we have
	\begin{equation} \label{T3}
		\Theta ( \iota_v ) \left( d \theta^\alpha \right)^{p+2} \delta^{(p)} ( d \theta^\beta ) =
		 - i (-1)^p p! \left[ (p+2) d \theta^\alpha \left( \frac{v^\alpha}{v^\beta} \right)^{p+1} - (p+1) d \theta^\beta \left( \frac{v^\alpha}{v^\beta} \right)^{p+2} - d \theta^\alpha \left( \frac{d \theta^\alpha}{d \theta^\beta} \right)^{p+1} \right] \ ,
	\end{equation}
	where again the result follows from direct calculation.
\end{ex}
As a final example we evaluate the application of $\Theta$ to $\delta ( d \theta^\alpha ) \delta ( d \theta^\beta )$:
\begin{ex}
	Let us consider $\displaystyle \omega^{(p|q)} = \delta ( d \theta^\alpha ) \delta ( d \theta^\beta ) $, we have
	\begin{equation} \label{T4}
		\Theta (\iota_v ) \delta ( d \theta^\alpha ) \delta ( d \theta^\beta ) = -i \lim_{\epsilon \to 0} \int_{-\infty}^\infty \frac{dt}{t + i \epsilon} \delta ( d \theta^\alpha + i v^\alpha t ) \delta ( d \theta^\beta + i v^\beta t ) = $$ $$ = \frac{-i}{i v^\alpha} \frac{1}{- \frac{d \theta^\alpha}{i v^\alpha}} \delta ( d \theta^\beta - i v^\beta \frac{d \theta^\alpha}{i v^\alpha} ) = \frac{i v^\alpha}{ d \theta^\alpha} \delta \left( v \cdot d \theta \epsilon^{\alpha \beta} \right) \ .
	\end{equation}
\end{ex}
Observe that $\delta ( v \cdot d \theta)$ allows us to rewrite the result in two other equivalent ways: 
\begin{eqnarray}
	\label{T5} \Theta (\iota_v ) \delta ( d \theta^\alpha ) \delta ( d \theta^\beta ) &=& \frac{i v^\beta}{ d \theta^\beta} \delta \left( v \cdot d \theta \epsilon^{\alpha \beta} \right) \ , \\
	\label{T6} \Theta (\iota_v ) \delta ( d \theta^\alpha ) \delta ( d \theta^\beta ) &=& \frac{1}{2} \left( \frac{i v^\alpha}{d \theta^\alpha} \delta \left( v \cdot d \theta \epsilon^{\alpha \beta} \right) + \frac{i v^\beta}{d \theta^\beta} \delta \left( v \cdot d \theta \epsilon^{\alpha \beta} \right) \right) \ .
\end{eqnarray}

Starting from the operator $\Theta \left( \iota_v \right)$ we directly define the PCO $\delta \left( \iota_v \right)$ as the formal derivative w.r.t. the argument of $\Theta$:
\begin{equation}\label{PCODELTA}
	\delta \left( \iota_v \right) := \Theta' \left( \iota_v \right) \ ,
\end{equation}
such that it acts on a general $(p|q)$-form by using the Fourier representation
\begin{equation}
	\delta \left( \iota_v \right) \omega^{(p|q)} ( d \theta^\alpha ) = \int_{-\infty}^\infty dt e^{i v \iota_v} \omega^{(p|q)} ( d \theta^\alpha ) = \int_{-\infty}^\infty dt \omega \left( d \theta^\alpha + i t v^\alpha \right) \ .
\end{equation}

We define now the operator $\eta$ as the geometric partner of $\eta$ of String Theory\cite{Berkovits:1994vy} in terms of its action on forms: given a $(p|q)$-form $\displaystyle \omega^{(p|q)} \in \Omega^{(p|q)}$, we define the operator $\eta$ as
	\begin{eqnarray}
	\nonumber	\eta ( \iota_v ) : \left( \Omega^{(p|q)} \right) &\to& \Omega^{(p+1|q+1)} \ , \\
		\label{PCOETA} \omega^{(p|q)} &\mapsto& \eta_{v} \omega^{(p|q)} = -2 \Pi \lim_{\epsilon \to 0} \sin \left( \epsilon \iota_v \right) \omega^{(p|q)} = i \Pi \lim_{\epsilon \to 0} \left( e^{i \epsilon \iota_v} - e^{-i \epsilon \iota_v} \right) \omega^{(p|q)} \ .
	\end{eqnarray}
	where the action of $e^{i \epsilon \iota_v}$ is defined as a translation operator acting on generalised functions of $d \theta$ and $\Pi$ is the \emph{parity changing functor} which allows us to convert bosonic/fermionic quantities into fermionic/bosonic ones.\footnote{Notice that in String Theory \cite{Berkovits:1994vy} there exists only one operator $\eta$ associated to the zero-mode of the $\eta_z$ field emerging from bosonisation of $\beta , \gamma$ system. In our case we define an $\eta_v$ operator for each fermionic direction.}\footnote{The limit in \ref{PCOETA} has to be intended as a distributional limit.}

Let us consider a few examples in order to understand better the action of this operator.
\begin{ex} \label{ETA1}
	Let us now consider the action of $\eta$ on a generic fermionic p-form with picture number 0:
	\begin{equation}
		\eta d \theta^p = i \Pi \lim_{\epsilon \to 0} \left( e^{i \epsilon \iota_v} - e^{-i \epsilon \iota_v} \right) d \theta^p = i \Pi \lim_{\epsilon \to 0} \left( \left( d \theta +i v \epsilon \right)^p - \left( d \theta -i v \epsilon \right)^p \right) = 0 \ ,
	\end{equation}
	thanks to the $\epsilon$ limit.
\end{ex}
\begin{ex} \label{ETA2}
	Let us now consider the action of $\eta$ on a Dirac delta form:
	\begin{equation}
		\eta \delta \left( d \theta \right) = i \Pi \lim_{\epsilon \to 0} \left( e^{i \epsilon \iota_v} - e^{-i \epsilon \iota_v} \right) \delta \left( d \theta \right) = i \Pi \lim_{\epsilon \to 0} \left( \delta \left( d \theta +i \epsilon v \right) - \delta \left( d \theta -i \epsilon v \right) \right) \ ;
	\end{equation}
	this result should be considered distributionally, i.e.
	\begin{equation}
		\lim_{\epsilon \to 0} \left\langle \delta \left( x +i \epsilon v \right) - \delta \left( x -i \epsilon v \right) , f \left( x \right) \right\rangle = \lim_{\epsilon \to 0} \left( f \left( -i \epsilon v \right) - f \left( i \epsilon v \right) \right) = 0 \ ,
	\end{equation}
	since being $f$ a test function, it is certainly continuous in 0. This result is then extended for $x \equiv d \theta$.
\end{ex}
\begin{ex}
	We have that the $\eta$ operator acting on a general pseudoform with picture number 1 gives 0:
	\begin{equation} \label{ETAPSEUDO} \fbox{$
			\eta \left( d \theta^\alpha \right)^p \delta^{(q)} \left( d \theta^\beta \right) = 0 \ .
		$}
	\end{equation}
	The result follows after a direct calculation in the distributional sense, i.e. where it is involved the application to a generic $C^\infty$ test function.
\end{ex}
\begin{ex}
	Let us now consider the action of $\eta$ on a $(-1|0)$-inverse form:
	\begin{equation}
		\eta \frac{1}{d \theta} = i \Pi \lim_{\epsilon \to 0} \left( e^{i \epsilon \iota_v} - e^{-i \epsilon \iota_v} \right) \frac{1}{d \theta} = $$ $$ = i \Pi \lim_{\epsilon \to 0} \left( \frac{1}{d \theta +i \epsilon v} - \frac{1}{d \theta -i \epsilon v} \right) = \Pi \lim_{\epsilon \to 0} \frac{2 \epsilon v}{d \theta^2 + \epsilon ^2 v^2} = \delta ( d \theta) \ ,
	\end{equation}
	where we have used the normalization of the Dirac delta distribution without the $\displaystyle \frac{1}{2 \pi}$ factor.
\end{ex}
\begin{ex}\label{ETA3}
	Let us now consider the action of $\eta$ on a general inverse form with picture number 0:
	\begin{equation}
		\eta \left( \frac{1}{d \theta} \right)^p = i \Pi \lim_{\epsilon \to 0} \left( e^{i \epsilon \iota_v} - e^{-i \epsilon \iota_v} \right) \left( \frac{1}{d \theta} \right)^p = $$ $$ = i \Pi \lim_{\epsilon \to 0} \left( \left( \frac{1}{d \theta +i \epsilon v} \right)^p - \left( \frac{1}{d \theta -i \epsilon v} \right)^p \right) = i \Pi \lim_{\epsilon \to 0} \frac{\left( d \theta -i \epsilon v \right)^p - \left( d \theta +i \epsilon v \right)^p }{\left( d \theta^2 +\epsilon^2 v^2 \right)^p} = $$ $$ = \frac{(-1)^{p-1}}{(p-1)!} \delta^{(p-1)} \left( d \theta \right) \ ,
	\end{equation}
	where in the last passage we have left only the linear terms in $\epsilon v$ since they are the only ones contributing.
\end{ex}
\begin{rmk}
	The operator $\eta$ is, modulo the multiplicative constant $i$, the left inverse of the operator $\Theta$ acting on pseudoforms, i.e.
	\begin{equation} \label{ETATHETA} \fbox{$
			\eta \left( \iota_v \right) \Theta \left( \iota_v \right) = i \ .
		$}
	\end{equation}
	We can apply $\eta$ to the definition of the operator $\displaystyle \Theta \left( \iota_v \right)$ (\ref{PCOT}) in  order to find
		\begin{equation}
			\eta \left( \iota_v \right) \Theta ( \iota_v ) \omega^{(p|q)} ( d \theta^\alpha ) = \lim_{\epsilon , \epsilon' \to 0} \int_{- \infty} ^\infty \frac{dt}{t + i \epsilon} \left[ \omega^{(p|q)} \left( d \theta^\alpha + i \epsilon' v^\alpha + i t v^\alpha \right) - \omega^{(p|q)} \left( d \theta^\alpha - i \epsilon' v^\alpha + i t v^\alpha \right) \right] = $$ $$ = \lim_{ \epsilon' \to 0} \int \frac{dy}{y -  d \theta^\alpha - i \epsilon' v^\alpha } \omega^{(p|q)} \left( y \right) - \lim_{ \epsilon' \to 0} \int \frac{dy}{y -  d \theta^\alpha + i \epsilon' v^\alpha  } \omega^{(p|q)} \left( y \right) = $$ $$ = \lim_{ \epsilon' \to 0} \int \frac{2 i \epsilon' v dy}{\left( y -  d \theta^\alpha \right)^2 + \epsilon'^2 v^2 } \omega^{(p|q)} \left( y \right) \ ,
		\end{equation}
		and by passing the limit under the integral sign we get
		\begin{equation}
			\int dy i \delta ( y - d \theta^\alpha ) \Omega^{(p|q)} \left( y \right) = i \Omega^{(p|q)} \left( d \theta^\alpha \right) \ ,
		\end{equation}
		i.e.
		\begin{equation}
			\eta \left( \iota_v \right) \Theta \left( \iota_v \right) = i \ .
		\end{equation}
\end{rmk}
\begin{ex}
	We have the following results for general inverse forms of picture degree 0 and 1:
	\begin{eqnarray}
		\label{ETA4} \eta \frac{\left( d \theta^\alpha \right)^p}{\left( d \theta^\beta \right)^q} &=& \frac{(-1)^{q-1}}{(q-1)!} \left( d \theta^\alpha \right)^p \delta^{(q-1)} \left( d \theta^\beta \right) \ , \\
		\label{ETA5} \eta \frac{1}{\left( d \theta^\alpha \right)^p} \delta^{(q)} \left( d \theta^\beta \right) &=& \frac{(-1)^{p-1}}{(p-1)!} \delta^{(p-1)} \left( d \theta^\alpha \right) \delta^{(q)} \left( d \theta^\beta \right) \ .
	\end{eqnarray}

		The proof of (\ref{ETA4}) is a direct consequence of Ex.\ref{ETA1}, having used the result of Ex.\ref{ETA3} , i.e. the operator $\eta$ passes through the numerator without any contribution.
		
		The proof of (\ref{ETA5}) is again a direct consequence of Ex.\ref{ETA2}, having used the result of Ex.\ref{ETA3} .
\end{ex}
\begin{rmk}
	The operator $\Theta$ is, modulo the multiplicative constant $i$, the left inverse of the operator $\eta$ on inverse forms of picture degree 0 and negative form degree, i.e.
	\begin{equation} \label{THETAETA} \fbox{$
			\Theta \left( \iota_v \right) \eta \left( \iota_v \right) = i \ .
		$}
	\end{equation}
		The proof is a direct consequence of the previous proposition and of (\ref{T1}):
		\begin{equation}
			\Theta \left( \iota_v \right) \eta \left( \iota_v \right) \left( \frac{\left( d \theta^\alpha \right)^p}{\left( d \theta^\beta \right)^q} \right) = \Theta \left( \iota_v \right) \left( \frac{(-1)^{q-1}}{(q-1)!} \left( d \theta^\alpha \right)^p \delta^{(q-1)} \left( d \theta^\beta \right) \right) = $$ $$ = \frac{(-1)^{q-1}}{(q-1)!} i (-1)^{q-1} (q-1)! \frac{(d \theta^\alpha)^p}{( d \theta^\beta )^{q}} = i \left( \frac{\left( d \theta^\alpha \right)^p}{\left( d \theta^\beta \right)^q} \right) \ \implies $$ $$ \implies \ \Theta \left( \iota_v \right) \eta \left( \iota_v \right) = i \ .
		\end{equation}
\end{rmk}
By using the results from the previous propositions we want now to investigate the commutation relation between the operator $\eta$ and the operator $Z_v$. Before doing this, let us study the commutation relation between the operator $\eta$ and the exterior derivative $d$:
\begin{rmk}
	The operator $\eta$ and the operator $d$ anticommute:
	\begin{equation} \label{ETAD}
		\left\lbrace \eta , d \right\rbrace = 0 \ .
	\end{equation}
		The proof follows after direct calculation on different types of inverse forms and pseudoforms.
\end{rmk}

\begin{rmk}
	The successive application of the operators $\eta$ and $Z_v$ gives 0:
	\begin{equation}
		\eta Z_v = Z_v \eta = 0 \ .
	\end{equation}
		The proof is simply an application of the definition of $Z_D$ and of the results (\ref{ETATHETA}) , (\ref{THETAETA}) and (\ref{ETAD}):
		\begin{equation}
			\eta Z_v = \eta \left( d \Theta ( \iota_v ) + \Theta (\iota_v) d \right) = -d \eta \Theta ( \iota_v ) + \eta \Theta ( \iota_v ) d = 0 \ ;
		\end{equation}
		\begin{equation}
			Z_v \eta = \left( d \Theta ( \iota_v ) + \Theta ( \iota_v ) d \right) \eta = d \Theta ( \iota_v ) \eta + \Theta ( \iota_v ) d \eta = i d - \Theta ( \iota_v ) \eta d = 0 \ .
		\end{equation}
\end{rmk}

\section{Super Chern-Simons Actions (SCS)}

In the present section, we review some of the ingredients needed for the construction of the main body of 
the paper. We first review D=3 N=1 super Chern-Simons theory in its classical derivation \cite{Nishino-Gates}. We start 
from the superspace construction, but we provide also the component action. Then, we reformulate the theory 
using the geometrical methods discussed in the previous section and we give the rules for a Chern-Simons theory on any supermanifold. 
We show that it leads to a very complicate non-factorized form, to be the basis for a theory on 
any supermanifold.

\subsection{SCS in Components and in Superspace}

We start from a $(1|0)$-superform $A^{(1|0)} = A_a V^a  + A_\a \psi^\a$, (where the superfields $A_a(x,\theta)$ and $A_\alpha(x,\theta)$ take values in the adjoint representation of the gauge group) and we define the field strength
\begin{eqnarray}\label{SM-A}
F^{(2|0)} = d A^{(1|0)} +  A^{(1|0)} \wedge A^{(1|0)} = F_{[ab]} V^a \wedge V^b + F_{a\alpha} V^a \wedge \psi^\a + F_{(\a\b)} \psi^\a\wedge \psi^\b \,,
\end{eqnarray}
where 
\begin{eqnarray}
\label{SM-B}
F_{[ab]} &=& \partial_{[a} A_{b]} + [A_a, A_b]\,, ~~ \nonumber \\
F_{a\a} &=& \partial_a A_\alpha - D_\alpha A_a + [A_\alpha, A_b]\,, ~~ \nonumber \\
F_{(\a\b)} &=& D_{(\a} A_{\b)} + \gamma^a_{\a\b} A_a + \{A_\alpha, A_\beta\}\,. 
\end{eqnarray}
In order to reduce the redundancy of degrees of freedom of $A_a$ and $A_\a$ of the $(1|0)$-form $A^{(1|0)}$, one imposes (by hand) the \emph{conventional constraint}
\begin{eqnarray}
\label{SM-C}
\iota_\a \iota_\b F^{(2|0)} =0\,   ~~~ \Longleftrightarrow ~~~ F_{(\a\b)} = D_{(\a} A_{\b)} + \gamma^a_{\a\b} A_a +\{A_\alpha, A_\beta\} =0\,, 
\end{eqnarray}
from which it follows that $F_{a\alpha} = \gamma_{a, \alpha\beta} W^\beta$ with 
$W^\a = \nabla^\b \nabla^\a A_\b$ and $\nabla_\a W^\a =0$. 
The gaugino field strength $W^\a$ is gauge invariant 
under the non-abelian transformations $\delta A_\a = \nabla_\a \Lambda$. 
These gauge transformations descend from the gauge transformations of $A^{(1|0)}$, $\delta A^{(1|0)} = \nabla \Lambda$ where $\Lambda$ is 
a $(0|0)$-form. 

The field strengths $F_{[ab]} , F_{a \alpha} , F_{(\alpha \beta)}$ satisfy the following Bianchi's identities  
\begin{eqnarray}\label{SM-V}
&&\nabla_{[a} F_{bc]} =0 \,, ~~~~~\nabla_\a F_{[ab]} + (\gamma_{[a} \nabla_{b]} W)_\a =0\,, \nonumber\\
&&F_{[ab]} + \frac12 (\gamma_{ab})^\a_{~\beta} \nabla_\a W^\beta =0 \,, ~~~~~
\nabla_\a W^\a =0\,, 
\end{eqnarray}
\newcommand{\sint}{\int\!\!\!\!\!\!-}
and by expanding the superfields $A_a, A_\a$ and $W^\a$ in components we 
have 
\begin{eqnarray}
\label{SM-VA}
A_\a = (\gamma^a \theta)_\alpha a_a + \lambda_\a \frac{\theta^2}{2}\,, ~~~~~~~
A_a = a_a + \lambda \gamma_a \theta + \dots \,, ~~~~~~
W^\a = \lambda^\a + f^\a_{~\b} \theta^\b + \dots\,, 
\end{eqnarray}
where $a_a(x)$ is the gauge field, $\lambda_\a(x)$ is the gaugino and $f_{(\a\b)} = \gamma_{\a\b}^{ab} f_{[ab]}$ 
is the gauge field strength with $f_{[ab]} = \partial_{[a} a_{b]}$. 
(The Wess-Zumino gauge $\theta^\a A_\a =0$ has been used.)

In terms of those fields, the super-Chern-Simons Lagrangian becomes 
\begin{eqnarray}
\label{SM-D}
S_{SCS} &=& \int {Tr} A_\a \left(W^\a - \frac{1}{6} [A_\b, F^{(\a\b)}]\right) [d^3x d^2\theta]\,, 
\end{eqnarray}
(we denote by  $[d^2\theta]$ the Berezin integral over the $\theta$'s variables), 
which in component reads
\begin{eqnarray}
\label{SM-DA}
S_{SCS}  = \int  {Tr} \Big( \epsilon^{abc} (a_a \partial_b a_c + 
 \frac23 a_a a_b a_c) + \lambda_{\a} \epsilon^{\a\b} \lambda_\b\Big) [d^3x] \,. 
\end{eqnarray}
That coincides with  the bosonic Chern-Simons action  with free non-propagating fermions.  

\subsection{SCS on Supermanifold}

In order to obtain the same action by integration on supermanifolds we consider the rheonomic action and 
the corresponding action principle \cite{CDAF}. It requires 
the choice of a bosonic submanifold ${\cal M}^{(3)}$ immersed into a supermanifold 
${\cal SM}^{(3|2)}$ and  a $(3|0)$-form on it
\begin{eqnarray}
\label{reA}
S_{rheo}[A, {\cal M}^{(3)}] = \int_{{\cal M}^{(3)} \subset {\cal SM}^{(3|2)}} {\cal L}^{(3)}(A, dA) \ .
\end{eqnarray}
Here the choice of the $(3|0)$-form  ${\cal L}^{(3)}$ is a three-form Lagrangian constructed with the superform $A$, 
and its derivatives $dA$, without using the Hodge dual operator (that is, without any reference to a metric on the 
supermanifold ${\cal SM}^{(3|2)}$). The action $S_{rheo}[A, {\cal M}^{(3)}]$ is a functional of the superfields and of the embedding of ${\cal M}^{(3)}$ into ${\cal SM}^{(3|2)}$. We can then consider the classical equations of motion by minimizing the action both respect to the variation of the fields and of the embedding. However, the variation of the immersion can be compensated by  
diffeomorhisms on the fields if the action ${\cal L}^{(3)}$ is a differential form. This implies that the complete set of 
equations associated to action (\ref{reA}) are the usual equations obtained by varying the fields on a fixed surface 
${\cal M}^{(3)}$ with the proviso that these equations hold not only on ${\cal M}^{(3)}$, but on the whole supermanifold ${\cal SM}^{(3|2)}$. 

The rules to build the action (\ref{reA}) are listed and discussed in \cite{CDAF} in  details. 
An important ingredient is the fact that for the action to be supersymmetric invariant,  
the Lagrangian must be invariant up to a $d$-exact term and, in addition, if the algebra of supersymmetry closes off-shell 
(either because there is no need of auxiliary fields or because it exists a formulation with auxiliary fields), the 
Lagrangian must be closed: 
\begin{eqnarray}
\label{reAA}
d {\cal L}^{(3)}(A) =0\,,
\end{eqnarray}
upon using the {\it rheonomic parametrization}. 
This amounts to set $F_{\a\b} =0$, which is an algebraic equation and it can be directly used in the 
action. One of the
rules of the geometrical construction for supersymmetric theories given in \cite{CDAF} is 
that by setting to zero the coordinates $\theta^\a$ and its differential $\psi^\a = d\theta^\a$, the Lagrangian  
\begin{eqnarray}
\label{reB}
\left. {\cal L}^{(3)}(A, dA) \right|_{\theta =0, d\theta =0} = {Tr} \Big( \epsilon^{abc} (a_a \partial_b a_c + 
 \frac23 a_a a_b a_c) + \lambda^{\a} \epsilon_{\a\b} \lambda^\b\Big)\,, 
\end{eqnarray}
 reduces to the component Lagrangian invariant under supersymmetry (up to a total derivative). Furthermore, 
 the equations of motion in the full-fledged superspace implies the rheonomic constraints (which coincide with the conventional constraints of superspace formalism). 

In order to express the action (\ref{reA}) in a more geometrical way by including the dependence upon the 
embedding into the integrand, we use the Poincar\'e dual form (already named PCO) $\mathbb{Y}^{(0|2)}$ dual to the immersion 
of ${\cal M}^{(3)}$ into ${\cal SM}^{(3|2)}$.   The Poincar\'e dual form $\mathbb{Y}^{(0|2)}$ is closed, 
it is not exact and any of its variation is $d$-exact. 
The action can now be written on the full supermanifold as
\begin{eqnarray}
\label{reC}
S[A] = \int_{{\cal SM}^{(3|2)} } {\cal L}^{(3|0)}(A, dA)\wedge  \mathbb{Y}^{(0|2)}\,.
\end{eqnarray}
Therefore, by choosing the PCO ${\mathbb Y}^{(0|2)} = \theta^2 \delta^2(d\theta)$, 
its factor $\theta^2$ projects the Lagrangian $ {\cal L}^{(3|0)}(A, dA)$ to ${\cal L}^{(3)}(A, dA)_{\theta=0}$
while the factor $\delta^2(d\theta)$ projects the latter to  ${\cal L}^{(3)}(A, dA)_{\theta =0, d\theta=0}$ reducing 
${\cal L}^{(3)}(A, dA)$  to the component Lagrangian (\ref{SM-DA}).  

Any variation of the embedding yields $\delta \mathbb{Y}^{(0|2)} = d \Lambda^{(-1|2)}$ and leaves the action invariant if the Lagragian is closed. The rheonomic Lagrangian ${\cal L}^{(3|0)}(A, dA)$ reads
\begin{eqnarray}
\label{reD}
{\cal L}^{(3|0)}(A, dA) = {Tr}\Big( A^{(1|0)}\wedge dA^{(1|0)} + \frac23 A^{(1|0)}\wedge A^{(1|0)}\wedge A^{(1|0)} + W^{(0|0)\a} \epsilon_{\a\b} W^{(0|0)\b} V^3  \Big)\wedge {\mathbb Y}^{(0|2)}\,, 
\end{eqnarray}
which is a $(3|2)$ form, $V^3 = \frac{1}{3!}\epsilon_{abc} V^a \wedge V^b\wedge V^c$.\footnote{This $(3|0)$ Lagrangian in (\ref{reD}) already appeared in \cite{Fabbri:1999ay} by reducing their formula from $N=2$ to $N=1$.} 
Again, by choosing  the PCO ${\mathbb Y}^{(0|2)} = \theta^2 \delta^2(d\theta)$ we get the 
component action (\ref{SM-DA}) and the third term in the action is fundamental to get the mass term for the non-dynamical fermions. 

This is the most general action and the closure of ${\cal L}^{(3|0)}$ implies that any gauge invariant 
and supersymmetric action can be built by choosing a PCO $\mathbb{Y}^{(0|2)}$ inside the same cohomology class. Therefore, 
starting from the rheonomic action, one can choose a different ``gauge" -- or better said a different immersion 
of the submanifold ${\cal M}^{(3)}$ inside the supermanifold ${\cal SM}^{(3|2)}$ -- leading to different forms of the action with the same physical content. 
It should be stressed, however, that the choice of ${\mathbb Y}^{(0|2)}_{new}$ (defined in the following subsection), is a preferred ``gauge'' choice, which allows us to derive the conventional constraint  by varying the action without using the rheonomic parametrization.


\subsection{SCS in Superspace Revised}

The choice of the PCO could be done observing that there are representatives 
respecting some isometries. 
For example the new operator
\begin{eqnarray}
\label{SM-G}
\mathbb{Y}_{new}^{(0|2)} = V^a \wedge V^b (\gamma_{ab})^{\a\b} \iota_\a \iota_\b \delta^2(\psi)\,,
\end{eqnarray}
is manifestly supersymmetric. Computing the expression in the integral \eqref{reC}, we see that 
${\mathbb Y}^{(0|2)}_{new}$ picks up al least two powers of $\psi$'s and one power of $V^a$ and that forces us 
to expand ${\cal L}^{(3|0)}$ as 3-form selecting the monomial $\psi \gamma_a \psi V^a$ dual to 
${\mathbb Y}^{(0|2)}_{new}$. That finally gives the supersymmetric action described in (\ref{SM-D}). 

The equations of motion derived from the new action
are 
\begin{eqnarray}
\label{SM-IA}
&&\mathbb{Y}_{new}^{(0|2)}  \Big(d A^{(1|0)} + A^{(1|0)} \wedge A^{(1|0)}\big) =0 ~~~~\Longrightarrow ~~~~ \nonumber \\
&&~~~~~~~~~V^3 (\gamma^a \iota)^\a \delta^2(\psi) F_{a\a} + 
(V^a \wedge V^b) \epsilon_{abc} (\gamma^c)^{\a\b} F_{(\a\b)} =0\,. 
\end{eqnarray}
The equations of motion correctly imply $F_{(\a\b)} =0$ (which is the conventional constraint) and $W^\a =0$ 
which are the super-Chern-Simons equation of motions. The second condition follows from $F_{\a\b} =0$  and by the Bianchi identities which implies that $F_{a\a} = \gamma_{a\a\b} W^\b$. 

Notice that this formulation allows us to get the conventional constraint as an equation of motion. In particular we find that
the equations of motion, together with the Bianchi identity,
imply the vanishing of the full field-strength. 
\begin{eqnarray}
\left\{
\begin{array}{l}
\mathbb{Y}_{new}^{(0|2)} F^{(2|0)}=0,   \\
  ~   \\
 dF^{(2|0)}+[A^{(1|0)},F^{(2|0)}]=0,     
\end{array}
\right.  \quad\quad  \Longrightarrow \quad\quad F^{(2|0)}=0\,. 
\end{eqnarray}


\subsection{SCS with Semi-supersymmetric PCO}
\newcommand{\sdot}{\,\mbox{\tiny $\bullet$}\,}

The choice of the PCO implies the form of the action and we present here another possibility. 
We consider the following expression
\begin{eqnarray}
\label{semiA}
{\mathbb Y}_{half}^{(0|2)} = V^a \theta^\a \epsilon_{\a\b} \gamma_a^{\b\gamma} \iota_\gamma \delta^2(\psi)\,. 
\end{eqnarray}
It is closed because of $\delta^2(\psi)$ and by using gamma matrices algebra. The presence of the 
explicit $\theta$ implies that it is not manifestly supersymmetric, but 
its variation is $d$-exact
\begin{eqnarray}
\label{semiB}
\delta_\epsilon {\mathbb Y}_{half}^{(0|2)}  = d \left( \frac32 \epsilon^\a \iota_\a \theta^2 \delta^2(\psi) \right) = 
{\cal L}_{\epsilon}  \left( \frac32  {\mathbb Y}^{(0|2)}_{half}  \right) \,. 
\end{eqnarray}
It is easy to show that this PCO is also not exact. 

Before computing the action, we discuss some other aspect of the geometry of the PCO (\ref{semiB}). 
Consider the 
expression
\begin{eqnarray}
\label{semiC}
\omega^{(3|0)} = \epsilon_{abc} V^a \wedge V^b \theta \gamma^c \psi \ ,
\end{eqnarray}
this expression is the Hodge dual to the PCO (\ref{semiA}) since it satisfies  
\begin{eqnarray}
\label{semiD}
\omega^{(3|0)} \wedge {\mathbb Y}_{half}^{(0|2)} = \theta^2 {\rm Vol}^{(3|2)}\,. 
\end{eqnarray}
Since the right hand side is closed (since it is a top integral form) and since ${\mathbb Y}^{(0|2)}_{half}$ 
is also closed, $\omega^{(3|0)}$ has to be closed or its variation is the kernel of 
${\mathbb Y}^{(0|2)}_{half}$. Let us verify the first possibility. Computing the variation 
of $\omega^{(3|0)}$, we have 
\begin{eqnarray}
\label{semiE}
d \omega^{(3|0)} = 2 V^a \psi \gamma_a \psi \psi \cdot \theta + V^a V^b \epsilon_{abc} \psi \gamma^c \psi\,, 
 \end{eqnarray}  
which does not vanish. Nevertheless, we can add two new terms and get 
\begin{eqnarray}
\label{semiF}
  \omega^{(3|0)} =  \left( 
  \epsilon_{abc} V^a \wedge V^b \theta \gamma^c \psi + V^a \psi \gamma_a \psi \theta^2 + \frac13 
 \epsilon_{abc} V^a V^b V^c\right) \ .
\end{eqnarray}
 The additional terms are needed to make $\omega^{(3|0)}$ closed, 
 but it does not affect the relation (\ref{semiD}) because of the powers of $\theta$'s and the powers of 
 $V$'s. 

 \subsection{SCS with Pseudoforms}

We consider now a new PCO. This is motivated by string theory, but we do not discuss here its origin,
since it can be also described in terms of the supermanifold structure. The fermionic space 
spanned by the coordinates $\theta^\a$ can be decomposed in terms of two commuting spinors $v^\a$ 
and $w^\a$ with the property that $v^\a \epsilon_{\a\b} w^\b \equiv \det (v, w) \equiv v \cdot w =1$ where $(v,w)$ is the $2\times 2$ matrix built with the spinors. Notice that any spinor $\theta^\a$ can be decomposed on 
that basis $\theta^\a =  - v^\a (w\cdot \theta) + w^\a (v \cdot \theta)$. Notice also that 
$\theta^\a \epsilon_{\a\b} \theta^\b = 2 (v \cdot \theta) (w \cdot \theta)$.

Any PCO $\mathbb{Y}^{(0|2)}$ can be decomposed into the product of two PCO's $\mathbb{Y}^{(0|1)}$ as follows 
 \begin{eqnarray}
\label{SM-NA}
\mathbb{Y}^{(0|2)}  = {\mathbb Y}^{(0|1)}_v  \wedge {\mathbb Y}^{(0|1)}_w  + d \Omega\,.
\end{eqnarray}

The piece $\Omega$ is a $(-1|2)$ form which depends on $v$ and $w$. The two PCO's are equivalent in 
the sense that they belong to the same cohomology class and they increase the picture by one unity. 
One can check by direct inspection that the product of the two PCO's inserted in the action 
does not lead to the conventional constraint $F_{\a\b} =0$ and therefore 
the exact term in (\ref{SM-NA}) relating the two actions is important to get the full-fledged action principle.  

Let us analyse the action with the new choice of PCO:
\begin{eqnarray}
\label{SM-NB}
S_{SCS} = \int_{{\cal SM}^{(3|2)}} Tr \Big( A \wedge d A  + \frac23 A\wedge A\wedge A + W^\a W_\a V^3\Big)\wedge 
  {\mathbb Y}^{(0|1)}_v \wedge {\mathbb Y}^{(0|1)}_w \ ,
  \end{eqnarray}
where the $\Omega$-term is dropped. Let us put aside the interaction term for the moment -- 
interaction terms will be discussed in the forthcoming sections --  and let us distribute the two $\mathbb{Y}$'s on the two pieces 
of the action as follows
\begin{eqnarray}
\label{SM-NC}
S^{quad}_{SCS} = \int_{{\cal SM}^{(3|2)}} {Tr} \Big( A \wedge d A \wedge {\mathbb Y}^{(0|1)}_v  {\mathbb Y}^{(0|1)}_w    + 
W^\a W_\a   {\mathbb Y}^{(0|1)}_v  {\mathbb Y}^{(0|1)}_w \wedge V^3\Big)\,. 
  \end{eqnarray}
Since the PCO's are closed, we can also bring them after each connection term $A^{(1|0)}$ and 
after the spinorial $W^{(0|0)}$ forms as 
\begin{eqnarray}
\label{SM-NC}
S^{quad}_{SCS} = \int_{{\cal SM}^{(3|2)}} {Tr} \Big( (A {}_\wedge {\mathbb Y}^{(0|1)}_v)  \wedge d (A {}_\wedge {\mathbb Y}^{(0|1)}_w)   + 
(W^\a {}_\wedge {\mathbb Y}^{(0|1)}_v) \wedge (W_\a {}_\wedge {\mathbb Y}^{(0|1)}_w) \wedge V^3\Big) \ ,
\end{eqnarray}
converting the gauge connection to a $(1|1)$ form as 
\begin{eqnarray}
\label{SM-NCA}
A^{(1|0)} \rightarrow  A^{(1|1)}  \equiv A^{(1|0)} {}_\wedge {\mathbb Y}^{(0|1)}_v \,. 
\end{eqnarray}
In the same way, the $(0|0)$-form $W^\a$ is converted into a $(0|1)$-pseudoform. 
Notice that, if we change the basis by linearly composing $v$ and $w$, we can write the same formula, up to an unessential overall 
factor. We can now forget that the connection pseudoform $A^{(1|1)}$ was originated by the factorised form \eqref{SM-NCA}, and consider an action built starting from a general $(1|1)$-pseudoform, the same applies for the pseudoform $W^{(0|1)}$.
Finally and most importantly, passing from $A^{(1|0)}$, which has a finite number 
of components, to  $A^{(1|1)}$, which has an infinite number of them, we have made an important 
assumption: we have moved to an infinite dimensional space. 

Therefore, we conclude that we have to take into account the 
generic action  
\begin{equation}\label{CSA}
S_{SCS} = \int_{{\cal SM}^{(3|2)}} {Tr} \Big( A^{(1|1)} \wedge d A^{(1|1)}  + 
W^{(0|1), \alpha} \epsilon_{\a\b} \wedge W^{(0|1), \beta} \wedge V^3  \Big)   \,.
\end{equation}
The wedge product is taken in the space of pseudoforms, therefore we have used the convention 
that two $(0|1)$-forms must be multiplied with the wedge product. 

In this way, we have succeeded to find the natural 
geometrical formulation of super Chern-Simons theory on a three-dimensional supermanifold. According to this 
observation, we notice that we have only explored the case $N=1$. This means two $\theta$'s and therefore two 
PCO's for the factorization ${\mathbb Y}^{(0|2)}_{new} = Y_v {}_\wedge Y_w + d \Omega$. However, the same conclusion 
can be achieved in any supermanifold with an even number of $\theta$'s 
and, if the even dimension of the fermionic sector is $m = 2 r$, we have the action 
\begin{equation}\label{CSA}
S_{SCS} = \int_{{\cal SM}^{(3|2 r)}} {Tr} \Big( A^{(1|r)} \wedge d A^{(1|r)}  + 
W^{(0|r), \alpha} \epsilon_{\a\b} \wedge W^{(0|r), \beta} \wedge V^3  \Big)   \,,
\end{equation}
 where $A^{(1|r)}$ is a $(1|r)$-pseudoform and $W^{(0|r), \alpha}$ a $(0|r)$-pseudoform. One can wonder whether 
 the Lagrangian is closed. For that, we need the rheonomic conditions and the observation that 
 they are left unchanged by multiplying them by the factor $Y_w$. 

We remark that in two previous works of one of the author, see \cite{DelMonte:2016czb,Castellani:2016ibp}, 
a non-factorized form of the action has been used. That has led, in the first case to new results and in the second case to a complete D=3 N=1 supergravity action in all possible superspace representations.

\section{General Super Chern-Simons Theory}

The following section is the main core of the present work. We demonstrate that the 
non-factorized action (\ref{CSA}) yields the same non-interacting equations of motion of super-Chern-Simons 
theory. We first write the explicit action by expanding the $A^{(1|1)}$ form in components, and,
by integrating according to the previous discussion on the various variables, we arrive at the action 
principle from which we compute the equations of motion. By an iterative procedure we verify that 
the physical content of these equations is exactly the same as free Chern-Simons theory. Then we 
introduce the interactions. Because of the picture number, the interaction term has to be introduced in a non-trivial way leading to a non-associative product. This product will be the starting point of an $A_\infty$-algebra generated by the gauge-invariance requirement. Finally we discuss the closure of the gauge algebra and the supersymmetric properties of the obtained Lagrangian.

\subsection{The Lagrangian}

Let us start from the pseudoform $\displaystyle A^{(1|1)} = A_0 + A_1 + A_2 + A_3 $ , where the subscript denotes the number of $dx$'s in the expression. We then have the decomposition:
\begin{eqnarray}
	\label{TLAAA} A_0 &=& \sum_{p=0}^{\infty} A_{\alpha \beta}^{(p)} ( d \theta^\alpha )^{p+1} \delta^{(p)} ( d \theta^\beta ) \ ,
\end{eqnarray}
\begin{eqnarray}
	\label{TLAAB} A_1 &=& \sum_{p=0}^{\infty} dx^m A_{m \alpha \beta}^{(p)} ( d \theta^\alpha )^{p} \delta^{(p)} ( d \theta^\beta ) \ , \\
	\label{TLAAC} A_2 &=& \sum_{p=0}^{\infty} dx^m dx^n A_{[m n] \alpha \beta}^{(p)} ( d \theta^\alpha )^{p} \delta^{(p + 1)} ( d \theta^\beta ) \ , \\
	\label{TLAAD} A_3 &=& \sum_{p=0}^{\infty} dx^m dx^n dx^r A_{[m n r] \alpha \beta}^{(p)} ( d \theta^\alpha )^{p} \delta^{(p + 2)} ( d \theta^\beta ) \ .
\end{eqnarray}
Let us clarify the notation: we have to fix a convention for the field $A^{1|1}$, i.e. we want to decide whether it is an even or odd field. However we are not free to choose a convention for the following reason: the field $A$ and the field $dA$ have opposite parity, since the exterior derivative $d$ is an odd operator, i.e. they commute $A \wedge dA = dA \wedge A$. This implies that
	\begin{equation*}
		d ( A \wedge A ) = dA \wedge A + A \wedge dA \ \text{if A is even} \ , \ d ( A \wedge A ) = dA \wedge A - A \wedge dA \ \text{if A is odd.}
	\end{equation*}
	In the even case we get that the Lagrangian $AdA$ is trivial, since it is equal to $\displaystyle \frac{1}{2} d (A \wedge A)$, while in the odd case we don't get a trivial Lagrangian, as it happens in usual Chern-Simons theory; therefore we must chose $A^{(1|1)}$ to be odd. This implies the following parities for the fields appearing in $A_i$, $i=0,1,2,3$:
\begin{equation}\label{TLAA}
	A_{\alpha \beta}^{(p)} , A_{[m n] \alpha \beta}^{(p)} \ \text{ are even fields and } \ A_{m \alpha \beta}^{(p)} , A_{[m n r] \alpha \beta}^{(p)} \ \text{ are odd fields } \ \forall p \in \mathbb{N}.
\end{equation}
Having these parities fixed we can track the signs when moving a field across another one.

Now let us derive the expressions for $dA_i$ , $i=0,1,2,3$:
\begin{eqnarray}
	\nonumber d A_0 &=&  \sum_{p=0}^{\infty} \left[ dx^m ( \partial_m A_{\alpha \beta}^{(p)} ) ( d \theta^\alpha )^{p+1} \delta^{(p)} (d \theta^\beta ) + d \theta^\gamma ( \partial_\gamma A_{\alpha \beta}^{(p)} ) ( d \theta^\alpha )^{p+1} \delta^{(p)} (d \theta^\beta ) \right] = \\ \nonumber &=& \sum_{p=0}^{\infty} \left[ dx^m \left( \partial_m A_{\alpha \beta}^{(p)} \right) ( d \theta^\alpha )^{p+1} \delta^{(p)} ( d \theta^\beta ) + \left( \partial_\alpha A_{\alpha \beta}^{(p)} \right) ( d \theta^\alpha )^{p+2} \delta^{(p)} ( d \theta^\beta ) + \right. \\  &&- p \left( \partial_\beta A_{\alpha \beta}^{(p)} \right) ( d \theta^\alpha )^{p+1} \delta^{(p - 1)} ( d \theta^\beta ) \Big]  = dA_0^{(0)} + dA_0^{(1)} \ .
\end{eqnarray}
Notice that we have decomposed the sum in $\gamma$ in a part with $\gamma = \alpha$ and in a part with $\gamma = \beta$.
Let us enumerate the other $dA_i$'s:
\begin{eqnarray}
	\nonumber dA_1 &=& - \sum_{p=0}^{\infty} \left[ dx^m dx^n \left( \partial_{[n} A_{m] \alpha \beta}^{(p)} \right) ( d \theta^\alpha )^{p} \delta^{(p)} ( d \theta^\beta ) + dx^m \left( \partial_\alpha A_{m \alpha \beta}^{(p)} \right) ( d \theta^\alpha )^{p+1} \delta^{(p)} ( d \theta^\beta ) + \right. \\ &&- p dx^m \left( \partial_\beta A_{m \alpha \beta}^{(p)} \right) ( d \theta^\alpha )^{p} \delta^{(p - 1)} ( d \theta^\beta ) \Big] =  dA_1^{(1)} + dA_1^{(2)} \ ; \\
	\nonumber dA_2 &=& \sum_{p=0}^{\infty} \left[ dx^m  dx^n   dx^r \left( \partial_{[r} A_{m n] \alpha \beta}^{(p)} \right) ( d \theta^\alpha )^{p} \delta^{(p + 1)} ( d \theta^\beta ) + \right. \\ \nonumber &&+ dx^m   dx^n \left( \partial_\alpha A_{[m n] \alpha \beta}^{(p)} \right) ( d \theta^\alpha )^{p+1} \delta^{(p + 1)} ( d \theta^\beta ) - (p+1) d x^m   dx^n \left( \partial_\beta A_{[m n] \alpha \beta}^{(p)} \right) ( d \theta^\alpha )^{p} \delta^{(p)} ( d \theta^\beta ) \Big] = \\ &=& dA_2^{(2)} + dA_2^{(3)} \ ; \\
	\nonumber dA_3 &=& - \sum_{p=0}^{\infty} \left[ dx^m   dx^n   dx^r \left( \partial_\alpha A_{[m n r] \alpha \beta}^{(p)} \right) ( d \theta^\alpha )^{p+1} \delta^{(p + 2)} ( d \theta^\beta ) + \right. \\ &&- (p+2) dx^m   dx^n   dx^r \left( \partial_\beta A_{[m n r] \alpha \beta}^{(p)} \right) ( d \theta^\alpha )^{p} \delta^{(p +1)} ( d \theta^\beta ) \Big] = dA_3^{(3)} \ .
\end{eqnarray}
In the final expression for the four terms we denote with a superscript the number of $dx$'s appearing. In the Lagrangian $A \wedge dA$, not every combination of the factors in the decompositions of $A$ and $dA$ is allowed, indeed we only need the terms where $d^3 x$ appears, in order to obtain the correct top-form for the bosonic integration. Thus the allowed combinations are
\begin{equation}\label{TLA}
	A dA = A_0 \left( d A_2^{(3)} + d A_3^{(3)} \right) + A_1 \left( d A_1^{(2)} + d A_2^{(2)} \right) + A_2 \left( d A_0^{(1)} + d A_1^{(1)} \right) + A_3 d A_0^{(0)} \ .
\end{equation}
We now want to evaluate the terms of (\ref{TLA}) in order to obtain the explicit form of the Lagrangian. Let us start from the last term, it has the form
\begin{align}
	\nonumber A_3 dA_0^{(0)} = & \left[ \sum_{p=0}^{\infty} dx^m   dx^n   dx^r A_{[m n r] \alpha \beta}^{(p)} ( d \theta^\alpha )^{p} \delta^{(p + 2)} ( d \theta^\beta ) \right] \wedge \\
	& \wedge \left[ \sum_{q=0}^{\infty} \left( \partial_\mu A_{\mu \nu}^{(q)} \right) ( d \theta^\mu )^{q+2} \delta^{(q)} ( d \theta^\nu ) - q \left( \partial_\nu A_{\mu \nu}^{(q)} \right) ( d \theta^\mu )^{q+1} \delta^{(q - 1)} ( d \theta^\nu ) \right] \ .
\end{align}
Recall that by definition we have $\displaystyle \delta ( d \theta^1 ) \delta ( d \theta^1 ) = 0 = \delta ( d \theta^2 ) \delta ( d \theta^2 ) $, this implies that in the previous product we have $\nu = \alpha$ and $\mu = \beta$. Moreover, in order to avoid vanishing terms, we need the power of $d \theta^\alpha$ in the first term to be smaller or equal to the derivation order of $\delta ( d \theta^\alpha )$ in the second term and the same holds true for $d \theta^\beta$ and $\delta ( d \theta^\beta )$ as well. This fixes uniquely $q$ in terms of $p$ and therefore we have the reduction to a single sum as
\begin{equation}\label{TLB}
	A_3 dA_0^{(0)} = \left[ \sum_{p=0}^{\infty} dx^m   dx^n   dx^r A_{[m n r] \alpha \beta}^{(p)} \left( - p! (p+2)! \partial_\beta A_{\beta \alpha}^{(p)} + p! (p+2)! (p+1) \partial_\alpha A_{\beta \alpha}^{(p+1)} \right) \right] \delta ( d \theta^\beta ) \delta ( d \theta^\alpha ) \ .
\end{equation}
The factors in (\ref{TLB}) are obtained by integrations by parts and by fixing $q=q(p)$ as discussed above.

In an analogous way we can calculate the other terms so that finally we get the Lagrangian:

\begin{eqnarray}
	\nonumber \mathcal{L}^{(3|2)} = {Tr} A^{(1|1)}dA^{(1|1)} &=& {Tr} \sum_{p=0}^\infty \left[ - p! (p+1)! A_{\alpha \beta}^{(p)} \left( \partial_{[r} A_{m n] \beta \alpha}^{(p)} - \partial_\beta A_{[m n r] \beta \alpha}^{(p - 1)} + (p+2) \partial_\alpha A_{[m n r] \beta \alpha}^{(p)} \right) + \right. \\
	\nonumber & - & p! p! A_{[m \alpha \beta}^{(p)} \left( \partial_r A_{n] \beta \alpha}^{(p)} - \partial_\beta A_{n r] \beta \alpha}^{(p - 1)} + (p+1) \partial_\alpha A_{n r] \beta \alpha}^{(p)} \right) + \\
	\label{TLC} & - & p! (p+1)! A_{[m n \alpha \beta}^{(p)} \left( \partial_{r]} A_{\beta \alpha}^{(p)} - \partial_\beta A_{r] \beta \alpha}^{(p)} + (p+1) \partial_\alpha A_{r] \beta \alpha}^{(p + 1)} \right) + \\
	\nonumber & - & p! (p+2)! A_{[m n r] \alpha \beta}^{(p)} \left( - \partial_\beta A_{\beta \alpha}^{(p)} + (p+1) \partial_\alpha A_{\beta \alpha}^{(p+1)} \right) \Big] dx^m \wedge dx^n \wedge dx^r \delta( d \theta^\beta ) \delta ( d \theta^\alpha ) \  . 
\end{eqnarray}
Notice that we have manifestly collected the $dx$'s and $\delta ( d\theta)$'s to ease the computation of the action integration on $dx$'s and $d \theta$'s:
\begin{eqnarray}
	S = \int_{\mathcal{SM}^{(3|2)}} Tr \left( A^{(1|1)} d A^{(1|1)} \right) = \int \left[ d^3x d^2 \theta \right] Tr \mathcal{L}^{(0|0)} \ .
\end{eqnarray} 

\subsection{Equations of Motion from the Action}

We can now investigate the equations of motion by varying the action w.r.t. the various fields; let us consider for example all the terms with $A_{[m n r] \alpha \beta}{(p)}$ in the Lagrangian (\ref{TLC}):
\begin{equation}\label{EMAA}
	\left[ - p! (p+2)! A_{m n r \alpha \beta}^{(p)} \partial_\beta A_{\beta \alpha}^{(p)} + p! (p+2)! (p+1) A_{m n r \alpha \beta}^{(p)} \partial_\alpha A_{\beta \alpha}^{(p+1)} - (p+1)! (p+2)! A_{\alpha \beta}^{(p+1)} \partial_\beta A_{m n r \beta \alpha}^{(p)} + \right. $$ $$ \left. + p! (p+1)! (p+2) A_{\alpha \beta}^{(p)} \partial_\alpha A_{m n r \beta \alpha}^{(p)} \right] \epsilon^{m n r} \epsilon^{\beta \alpha} \ ,
\end{equation}
where we have inserted the totally antisymmetric symbols $\epsilon$ as reminders for the symmetry of bosonic and fermionic indices. We can recast the last two terms of the previous expression as follows
\begin{align}
	\nonumber & \left[ \partial_\alpha \left( A_{\alpha \beta}^{(p)} A_{m n r \beta \alpha}^{(p)} \right) \right] \epsilon^{m n r} \epsilon^{\beta \alpha} = \left[ A_{m n r \alpha \beta}^{(p)} \left( \partial_\beta A_{\beta \alpha}^{(p)} \right) + A_{\alpha \beta}^{(p)} \left( \partial_\alpha A_{m n r \beta \alpha}^{(p)} \right) \right] \epsilon^{m n r} \epsilon^{\beta \alpha} \ \implies \\
	\implies & \left[ A_{\alpha \beta}^{(p)} \left( \partial_\alpha A_{m n r \beta \alpha}^{(p)} \right) \right] \epsilon^{m n r} \epsilon^{\beta \alpha} = \left[ \partial_\alpha \left( A_{\alpha \beta}^{(p)} A_{m n r \beta \alpha}^{(p)} \right) - A_{m n r \alpha \beta}^{(p)} \left( \partial_\beta A_{\beta \alpha}^{(p)} \right) \right] \epsilon^{m n r} \epsilon^{\beta \alpha} \ ; \label{EMAB} \\
	&\left[ - A_{\alpha \beta}^{(p+1)} \left( \partial_\beta A_{m n r \beta \alpha}^{(p)} \right) \right] \epsilon^{m n r} \epsilon^{\beta \alpha} = \left[ - \partial_\beta \left( A_{\alpha \beta}^{(p+1)} A_{m n r \beta \alpha}^{(p)} \right) + A_{m n r \alpha \beta}^{(p)} \left( \partial_\alpha A_{\beta \alpha}^{(p+1)} \right) \right] \epsilon^{m n r} \epsilon^{\beta \alpha} \ . \label{EMAC}
\end{align}
where we have used the graded Leibniz's rule according to the parity of the fields described in \eqref{TLAA}. The total derivative terms may be neglected since it would lead to null terms after the Berezin's integration of the $\theta$'s.

We can now insert (\ref{EMAB}) and (\ref{EMAC}) in (\ref{EMAA}) and we get
\begin{equation}
	\left[ 2 A_{m n r \alpha \beta}^{(p)} \left( - p! (p+2)! \partial_\beta A_{\beta \alpha}^{(p)} + (p+1)! (p+2)! \partial_\alpha A_{\beta \alpha}^{(p+1)} \right) \right] \epsilon^{m n r} \epsilon^{\beta \alpha} \ .
\end{equation}
The variation of the action w.r.t. the field $A_{[m n r] \alpha \beta}{(p)}$ leads to:
\begin{equation}\label{EMAD}
	- \partial_\beta A_{\beta \alpha}^{(p)} + (p + 1 ) \partial_\alpha A_{\beta \alpha}^{(p+1)} = 0 \ , \ \forall \ p \in \mathbb{N} \ .
\end{equation}


In an analogous way we can obtain the other equations of motion by varying the action w.r.t. the fields  $A_{\alpha \beta}^{(p)},$ $A_{m \alpha \beta}^{(p)}$ and $A_{[m n] \alpha \beta}^{(p)}$; the resulting equations are

\begin{eqnarray}
	\label{EMAE} \partial_r A_{\beta \alpha}^{(p)} - \partial_\beta A_{r \beta \alpha}^{(p)} + (p+1) \partial_\alpha A_{r \beta \alpha}^{(p + 1)} &=& 0 \ \forall \ p \in \mathbb{N} \ ; \\
	\label{EMAF} \partial_{[ r} A_{n ] \beta \alpha}^{(p)} - \partial_\beta A_{[n r] \beta \alpha}^{(p - 1)} + (p+1) \partial_\alpha A_{[n r] \beta \alpha}^{(p)} &=& 0 \ \forall \ p \in \mathbb{N} \ ; \\
	\label{EMAG} \partial_{[r} A_{m n] \beta \alpha}^{(p)} - \partial_\beta A_{[m n r] \beta \alpha}^{(p-1)} + \left( p+2 \right) \partial_\alpha A_{[m n r] \beta \alpha}^{(p)} &=& 0 \ \forall \ p \in \mathbb{N} \ ,
\end{eqnarray}
where we stress that in (\ref{EMAF}) and in (\ref{EMAG}) if $p=0$ the fields $A_{[n r] \beta \alpha}^{(- 1)}$ and $A_{[m n r] \beta \alpha}^{(-1)}$ are both defined to be zero.

\subsection{Equations of Motion from the Curvature $\displaystyle \left(  F^{(2|1)} = d A^{(1|1)} = 0 \right)$}


In this subsection we briefly show that the equations of motion already derived from the variational principle are the same that emerge by the usual flatness condition of (non-interacting) Chern-Simons theory $F = dA = 0$ at picture 1. For the sake of clarity we recall the strategy used to get the EoM. We gather the equations by the number of $dx$'s appearing, in other words we consider the various parts which are homogeneous in $dx$'s; the four homogeneous parts are then formal series into the $d \theta$'s, therefore by power counting we can set each single term of the series equal to zero. This yields
\begin{eqnarray}
	dA_0^{(0)} &=& 0 \ \implies \ \partial_\beta A_{\beta \alpha}^{(p)} - (p+1) \partial_\alpha A_{\beta \alpha}^{(p+1)} = 0 \ ; \\
	dA_0^{(1)} + dA_1^{(1)} &=& 0 \ \implies \ \partial_r A_{\beta \alpha}^{(p)} -\partial_\beta A_{r \beta \alpha}^{(p)} + (p+1) \partial_\alpha A_{r \beta \alpha}^{(p+1)} = 0 \ ; \\
	dA_1^{(2)} +dA_2^{(2)} &=& 0 \ \implies \ \partial_{[r} A_{n] \beta \alpha}^{(p)} - \partial_\beta A_{[n r] \beta \alpha}^{(p-1)} + (p+1) \partial_\alpha A_{[n r] \beta \alpha}^{(p)} = 0 \ ; \\
	dA_2^{(3)} +dA_3^{(3)} &=& 0 \ \implies \ \partial_{[r} A_{m n] \beta \alpha}^{(p)} - \partial_\beta A_{[m n r] \beta \alpha}^{(p-1)} + (p+2) \partial_\alpha A_{[m n r] \beta \alpha}^{(p)} = 0 \ .
\end{eqnarray}
This makes clear that the EoM emerging from the variational principle are exactly the same as those obtained by requiring the flatness of the connection $A$.

\subsection{Reducing the Equations of Motion}

In the previous subsections we have explicitly obtained the equations of motion for super Chern-Simons theory with pseudoforms. Since the Lagrangian \eqref{TLC} we started with contains an infinite number of fields, we therefore have an infinite number of EoM. We now want to use these equations to determine which fields, when \emph{on shell}, can be expressed as $d$-exact terms, i.e.\ we want to find out explicitly the cohomology (w.r.t. those forms which are related to the $\theta$-dependent gauge parameters) representatives of the free theory at picture number 1. In this subsection we omit the calculations and present only the strategy to apply and its result. The interested reader is suggested to refer to Appendix B where the explicit calculations are presented.

We use the following strategy: first we expand the fields in powers of $\theta$'s (recall that the expansion is truncated to the term $\theta^1 \theta^2$ since they are odd variables), then we insert the expansions in the EoM \eqref{EMAD} $\div$ \eqref{EMAG}. We then use the resulting equations in order to find algebraic relations between the fields of the expansion. The results are then inserted back in \eqref{TLAAA} $\div$ \eqref{TLAAD} and then we look for the terms that give rise to $d$-exact terms. We find that a representative of the cohomology class is
\begin{equation} \label{CSFIELD}
	A^{1|1} = dx^m \theta^\beta \tilde{B}_{m \alpha \beta}^{(0)} (x) \delta^{(0)} ( d \theta^\beta ) \ ,
\end{equation}
and the relative equation of motion is
\begin{equation}
	\partial_{[n} \tilde{B}_{m] \alpha \beta}^{(0)} (x) = 0 \ .
\end{equation}
Remarkably, notice that even if we started from a SCS Lagrangian with an infinite number of fields, we have shown that there is only one physical field, indeed all the other fields are $d$-exact $\theta$-dependent terms.

Moreover we have shown that starting from the free SCS action with a general $A^{1|1}$ pseudoform we obtain the factorisation
\begin{equation}
	A^{1|1} = A^{1|0} \wedge \mathbb{Y}^{0|1} \ , \ \text{ s.t. } \mathbb{Y}^{0|1} = \theta^\beta \delta ( d \theta^\beta ) + d \Omega^{-1|1} \ .
\end{equation}
Thus we have recovered a factorised form from a non-factorised Lagrangian.

\subsection{Interactions and the $m_2^{(-1)}$ Product}

We now define an interaction term which can be integrated on a supermanifold. 
Apparently, a problem arises. Indeed in order to define an interaction term, we need three gauge fields $A^{(1|1)}$, but the wedge product of three fields vanishes by anticommutativity of three Dirac delta functions in $d\theta_1$ or 
$d\theta_2$\footnote{Care must be used in defining the product of two pseudoforms since it might lead to divergencies in the Feynman diagrams \cite{cremonini-grassi}.}. 

Let us start from the action (3.22).
If we factorise $\mathbb{Y}^{(0|2)} = \prod_{\a=1}^2 v_{(\alpha)} \cdot \theta \delta( v_{(\a)} \cdot d \theta)$ , where 
we have chosen any two linear independent vectors $v_{(\alpha)}$, such that $v_{(1)} \cdot v_{(2)} \neq 0$, 
then we can set $ \mathbb{Y}^{(0|2)} A^{(1|0)}\wedge A^{(1|0)} = (\mathbb{Y}^{(0|1)}_{v_{(1)}} A^{(1|0)})\wedge (\mathbb{Y}^{(0|1)}_{v_{(2)}} A^{(1|0)})$. In other words 
we have distributed the PCO $\mathbb{Y}^{(0|2)}$ on the gauge fields $A^{(1|0)}$ which have now picture 1 each. In order to have one more gauge field, one needs one more PCO. This can be done by inserting 
the combination $ Z_{w_{(1)}} \mathbb{Y}_{w_{(1)}} = 1$, thus obtaining 
\begin{eqnarray}
\label{new_carloB}
S_{new} &=& \int_{\cal SM} 
{Tr} \Big( (\mathbb{Y}_{v_{(1)}}A^{(1|0)}) \wedge d  (\mathbb{Y}_{v_{(2)}} A^{(1|0)}) \nonumber \\
&+&
\frac{2}{3} ( \mathbb{Y}_{v_{(1)}} A^{(1|0)}) \wedge ( \mathbb{Y}_ {v_{(2)}} A^{(1|0)}) \wedge  Z_{w_{(1)}} (\mathbb{Y}_{w_{(1)}} A^{(1|0)}) \Big) \ .
\end{eqnarray}
The interaction term, rewritten in terms 
of pseudo-forms, has the following structure
\begin{equation}\label{INTA}
S_{int} = \int  {Tr}\Big( A^{(1|1)} A^{(1|1)} \wedge Z_v A^{(1|1)} \Big) \ .
\end{equation}


In \eqref{new_carloB} we have inserted the PCO $Z_v$ in a generic place in the interaction term. However, a priori, we have to consider all the possible places where to put the PCO. Therefore, following \cite{Catenacci-Grassi-Noja,Erler:2013xta}, we are led to define the $2$-$product$ $with$ $picture$ $degree$ -$1$ as
	\begin{align} \label{APA}
		\nonumber m_2^{(-1)} : \Omega^{(1|1)} \times \Omega^{(1|1)} & \to \Omega^{(2|1)} \\
		(A,A) & \mapsto m_2^{(-1)} (A,A) = \frac{1}{3} \left[ Z_v ( A \wedge A ) + Z_v (A) \wedge A + A \wedge Z_v (A) \right] \ .
	\end{align}
	This definition encodes the prescription of an \virgolette equally-weighted" application of the PCO $Z_v$, thus reflects the generality discussed above.\footnote{In first quantised String Theory, the PCO is independent of worldsheet coordinates and therefore it can be placed at any point into a correlation function. However, in order to formulate a String Field Theory action, that arbitrariness can not be used since a given choice might break gauge invariance \cite{WittenPerturbation}. In \cite{Erler:2013xta} the authors avoid this problem by suitably smearing the PCO on the disc on which the correlation functions are computed. This democratic choice preserves gauge invariance an leads to the 2-product discussed in the text.} Notice that, after introducing
	\begin{align} 
		\nonumber m_2^{(0)} : \Omega^{(1|1)} \times \Omega^{(1|1)} & \to \Omega^{(2|2)} \\
		\label{APB} (A,A) & \mapsto A \wedge A \ ,
	\end{align}
	we can recast the definition of $m_2^{(-1)}$ as
	\begin{equation} \label{APC}
		m_2^{(-1)} = \frac{1}{3} \left[ Z_v m_2^{(0)} + m_2^{(0)} \left( Z_v (A) \otimes 1 + 1 \otimes Z_v \right) \right] \ .
	\end{equation}
	 In \eqref{APC} we have adopted the coproduct formulation \cite{Catenacci-Grassi-Noja}. Observe that this product has form degree 0. In an analogous way, we can define a product with form degree $-1$ as
	\begin{align} 
		\nonumber \tilde{m}_2^{(-1)} : \Omega^{(1|1)} \times \Omega^{(1|1)} & \to \Omega^{(1|1)} \\
		\label{APD} (A,A) & \mapsto \tilde{m}_2^{(-1)} (A,A) = \frac{1}{3} \left[ -i \Theta (\iota_v) ( A \wedge A ) -i \Theta (\iota_v) (A) \wedge A - (-1)^{|A|} A \wedge i \Theta (\iota_v) (A) \right] \ .
	\end{align}
	From the definition \eqref{APA}, it follows that
	\begin{equation} \label{APE}
		m_2^{(-1)} = [ d , \tilde{m}_2^{(-1)} ] \ ,
	\end{equation}
	where $ [ \cdot , \cdot ] $ denotes as usual the graded commutator. 

Starting from the definition \eqref{APA}, we now find an explicit expression for the interaction term. Details can be found in Appendix B. Let us start from the action term
\begin{equation} \label{ITA}
	\left\langle A , m_2^{(-1)} (A , A ) \right\rangle = \int Tr \left( A \wedge m_2^{(-1)} (A , A ) \right) \ ,
\end{equation}
where the trace is to be taken with respect to the group indices. From \eqref{ITA}, we can extract the two following terms:
\begin{equation}
	A \wedge Z_v (A) \wedge A + A \wedge A \wedge Z_v (A) \ .
\end{equation}
Due to the cyclicity properties of the trace with respect to the group indices and of the wedge product with respect to the form indices we have
\begin{equation}
	A \wedge Z_v (A) \wedge A = A \wedge A \wedge Z_v (A) \ ,
\end{equation}
therefore we can recast the interaction term as
\begin{equation} \label{ITB}
	\left\langle A , m_2^{(-1)} (A , A ) \right\rangle = \frac{1}{3} Tr \left[ A \wedge Z_v \left( A \wedge A \right) + 2 A \wedge A \wedge Z_v A \right] \ .
\end{equation}
We first calculate the action of the operator $Z_v$ on the $A_i\text{'s} , i = 0,1,2,3$ and on the product $A \wedge A$:
\begin{align}
	\label{ITBA} i Z_v A_0 = & \sum_{p=0}^\infty i (-1)^{p+1} (p+1)! \frac{1}{\left( v^\beta \right)^2} \left( \frac{v^\alpha}{v^\beta} \right)^p \epsilon^{\alpha \beta} v \cdot d \theta \left( v^\alpha \partial_\alpha A_{\alpha \beta}^{(p)} + v^\beta \partial_\beta A_{\alpha \beta}^{(p)} \right) \ , \\
	\label{ITBB} i Z_v A_1 = & \sum_{p=0}^\infty i (-1)^{p+1} p! dx^m \left( \frac{v^\alpha}{v^\beta} \right)^p \frac{1}{v^\beta} \left( \partial_\alpha A_{m \alpha \beta}^{(p)} v^\alpha + v^\beta \partial_\beta A_{m \alpha \beta}^{(p)} \right) \ , \\
	\label{ITBC} i Z_v A_2 = & 0 \ , \\
	\label{ITBD} i Z_v A_3 = & 0 \ ,
\end{align}
\begin{align}
	\nonumber i Z_v \left( A \wedge A \right) = & \sum_{p=0}^\infty \left\lbrace dx^m dx^n p! p! \left[ (p+1) \partial_\alpha \left[ A_{[m n] \alpha \beta}^{(p)} , A_{\beta \alpha}^{(p)} \right] + \partial_\alpha \left( A_{[m \alpha \beta}^{(p)} A_{n] \beta \alpha}^{(p)} \right) \right] i v^\alpha \delta \left( \epsilon^{\alpha \beta} v \cdot d \theta \right) + \right. \\
	\label{ITBE} & \left. + dx^m dx^n p! p! \left[ (p+1) \partial_\beta \left[ A_{[m n] \alpha \beta}^{(p)} , A_{\beta \alpha}^{(p)} \right] + \partial_\beta \left( A_{[m \alpha \beta}^{(p)} A_{n] \beta \alpha}^{(p)} \right) \right] i v^\beta \delta \left( \epsilon^{\alpha \beta} v \cdot d \theta \right) \right\rbrace \ .
\end{align}
Notice that \eqref{ITBC} and \eqref{ITBD} are consequence of the general property
\begin{equation}
	Z_v \left[ \left( d \theta^\alpha \right)^p \delta^{(q)} \left( d \theta^\beta \right) \right] = 0 \ , \ \forall q \geq p+1 \ .
\end{equation}
By using these results in \eqref{ITB} we get to the explicit interaction term:
\begin{align}
	\nonumber A \wedge Z_v \left( A \wedge A \right) + 2 A \wedge A \wedge Z_v A = & \sum_{p,q=0}^\infty (-1)^p p! q! q! dx^m dx^n dx^r \delta^2 \left( d \theta \right) \cdot \\
	\nonumber & 3 \left\lbrace (q+1) \left[ A_{m n \alpha \beta}^{(q)} , A_{\beta \alpha}^{(q)} \right] + A_{m \alpha \beta}^{(q)} A_{n \beta \alpha}^{(q)} \right\rbrace \cdot \\
	\label{ITC} & \left\lbrace \left( \frac{v^\alpha}{v^\beta} \right)^p \left( \frac{v^\alpha}{v^\beta} \partial_\alpha + \partial_\beta \right) A_{r \alpha \beta}^{(p)} + \left( \frac{v^\beta}{v^\alpha} \right)^p \left( \frac{v^\beta}{v^\alpha} \partial_\beta + \partial_\alpha \right) A_{r \beta \alpha}^{(p)} \right\rbrace \ .
\end{align}
Therefore the Lagrangian reads
\begin{align} 
	\mathcal{L}^{(0|0)}_{INT} = & 2 Tr \left\lbrace \sum_{p,q=0}^\infty (-1)^p p! q! q! \left\lbrace (q+1) \left[ A_{m n \alpha \beta}^{(q)} , A_{\beta \alpha}^{(q)} \right] + A_{m \alpha \beta}^{(q)} A_{n \beta \alpha}^{(q)} \right\rbrace \cdot \right. \\
	\label{ITC} & \left. \left\lbrace \left( \frac{v^\alpha}{v^\beta} \right)^p \left( \frac{v^\alpha}{v^\beta} \partial_\alpha + \partial_\beta \right) A_{r \alpha \beta}^{(p)} + \left( \frac{v^\beta}{v^\alpha} \right)^p \left( \frac{v^\beta}{v^\alpha} \partial_\beta + \partial_\alpha \right) A_{r \beta \alpha}^{(p)} \right\rbrace \right\rbrace \epsilon^{mnr} \epsilon^{\alpha \beta} \ .
\end{align}
Notice that the interaction term depends on the constant vector $v^\alpha$ through $\displaystyle \frac{v^1}{v^2}$, namely their relative phase. That resembles the usual frame dependence of Superstring Field Theory actions. That dependence is supposed to disappear whenever a calculation of a correlation function is performed. In our case, this is a consequence of the fact that any variation of the PCO $Z_v$, by means a variation of its reference vector $v^\alpha$, is $d$-exact. Therefore, the action might depend upon the reference vector $v$, but the correlation functions will turn out to be independent of $v$.\footnote{The same dependence appears also in the construction of EKS when they build the PCO by spreading it on the disk.}

The meaning of this interaction term can be understood if we consider the result \eqref{CSFIELD}. Indeed it is a straightforward calculation to verify that, if \eqref{CSFIELD} holds, then the $m_2^{(-1)}$ product reduces to the usual wedge product, hence the interaction term is the usual Chern-Simons one. Our result \eqref{ITC} is coherent with this observation as we can readily verify:
\begin{align}
	\nonumber \mathcal{L}^{(0|0)}_{INT} = & 2 Tr \left\lbrace \theta^\beta \tilde{B}_{m \alpha \beta}^{(0)} \theta^\alpha \tilde{B}_{n \beta \alpha}^{(0)} \left[ \left( \frac{v^\alpha}{v^\beta} \partial_\alpha + \partial_\beta \right) \theta^\beta \tilde{B}_{r \alpha \beta}^{(0)} + \left( \frac{v^\beta}{v^\alpha} \partial_\beta + \partial_\alpha \right) \theta^\alpha \tilde{B}_{r \beta \alpha}^{(p)} \right] \right\rbrace = \\
	\label{ITD} = & 2 Tr \left\lbrace \theta^\beta \theta^\alpha \tilde{B}_{m \alpha \beta}^{(0)} \tilde{B}_{n \beta \alpha}^{(0)} \left[ \tilde{B}_{r \alpha \beta}^{(0)} + \tilde{B}_{r \beta \alpha}^{(p)} \right] \right\rbrace \ .
\end{align}
\eqref{ITD} shows that for the cohomology representative field the interaction term reduces to the usual one. A few remarks are necessary: first, we see that the interaction term does not depend on the vector $v^\alpha$, as expected; second, we see that we have two copies of the interaction term corresponding to the two propagating fields obtained in \eqref{CSFIELD}, i.e. $\displaystyle \tilde{B}_{m \alpha \beta} $ and $\displaystyle \tilde{B}_{m \beta \alpha} $.

\subsection{Cyclicity of $\langle \cdot , \cdot \rangle$}

In order to derive the equations of motion we need the interior product to be cyclic. For the sake of completeness, let us verify it explicitly: let $A, B, C$ be three $(1|1)$-pseudoforms, we want to verify that
\begin{equation} \label{CA}
	\int Tr ( A \wedge m_2^{(-1)} ( B \wedge C ) ) = (-1)^{|C| ( |A| + |B| )} \int Tr ( C \wedge m_2^{(-1)} ( A \wedge B ) ) = \int Tr ( C \wedge m_2^{(-1)} ( A \wedge B ) ) \ ,
\end{equation}
since $|A| = |B| = |C| = 1$. In order to avoid a cumbersome notation, we omit the integration and trace symbols; we have
\begin{align}
	\label{CF} A \wedge m_2^{(-1)} ( B \wedge C ) &= A \wedge Z_v \left( B \wedge C \right) + C \wedge A \wedge Z_v \left( B \right) + A \wedge B \wedge Z_v \left( C \right) \ , \\
	\label{CG} C \wedge m_2^{(-1)} ( A \wedge B ) &= C \wedge Z_v \left( A \wedge B \right) + B \wedge C \wedge Z_v \left( A \right) + C \wedge A \wedge Z_v \left( B \right) \ .
\end{align}
We observe that the second term of \eqref{CF} and the third term of \eqref{CG} are the same. We now write the extended expressions for the other four terms:
\begin{align}
	\nonumber A \wedge Z_v \left( B \wedge C \right) = & \left[ \sum_{p=0}^{\infty} dx^m A_{m \alpha \beta}^{(p)} ( d \theta^\alpha )^{p} \delta^{(p)} ( d \theta^\beta ) \right] \wedge \left[ \sum_{q=0}^\infty \left[ dx^n dx^r q! q! \left[ (q+1) \partial_\mu \left( B_{[n r] \mu \nu}^{(q)} C_{\nu \mu}^{(q)} - B_{\mu \nu}^{(q)} C_{[nr] \nu \mu}^{(q)} \right) + \right. \right. \right. \\
	\nonumber & \left. \left. \left. + \partial_\mu \left( B_{[n \mu \nu}^{(q)} C_{r] \nu \mu}^{(q)} \right) \right] v^\mu \delta ( \epsilon^{\mu \nu} v \cdot d \theta ) + dx^n dx^r q! q! \left[ (q+1) \partial_\nu \left( B_{[n r] \mu \nu}^{(q)} C_{\nu \mu}^{(q)} - B_{\mu \nu}^{(q)} C_{[nr] \nu \mu}^{(q)} \right) + \right. \right. \right. \\
	\label{CH} & \left. \left. \left. + \partial_\nu \left( B_{[n \mu \nu}^{(q)} C_{r] \nu \mu}^{(q)} \right) \right] v^\nu \delta ( \epsilon^{\mu \nu} v \cdot d \theta ) \right] \right] \ ; \\
	\nonumber A \wedge B \wedge Z_v ( C ) = & - \left[ \sum_{p=0}^\infty dx^m dx^n p! p! \left[ (p+1) \left( A_{[m n] \alpha \beta}^{(p)} B_{\beta \alpha}^{(p)} - A_{\alpha \beta}^{(p)} B_{[m n] \beta \alpha}^{(p)} \right) + A_{[m \alpha \beta}^{(p)} B_{n] \beta \alpha}^{(p)} \right] \delta ( d \theta^\beta ) \delta ( d \theta^\alpha ) \right] \wedge \\
	\label{CI} & \wedge \left[ \sum_{q=0}^\infty (-1)^{q+1} q! dx^r \left( \frac{v^\mu}{v^\nu} \right)^q \left( \partial_\mu C_{r \mu \nu}^{(q)} \frac{v^\mu}{v^\nu} + \partial_\nu C_{r \mu \nu}^{(q)} \right) \right] \ ;
\end{align}
\begin{align}
	\nonumber C \wedge Z_v \left( A \wedge B \right) = & \left[ \sum_{p=0}^{\infty} dx^m C_{m \alpha \beta}^{(p)} ( d \theta^\alpha )^{p} \delta^{(p)} ( d \theta^\beta ) \right] \wedge \left[ \sum_{q=0}^\infty \left[ dx^n dx^r q! q! \left[ (q+1) \partial_\mu \left( A_{[n r] \mu \nu}^{(q)} B_{\nu \mu}^{(q)} - A_{\mu \nu}^{(q)} B_{[nr] \nu \mu}^{(q)} \right) + \right. \right. \right. \\
	\nonumber & \left. \left. \left. + \partial_\mu \left( A_{[n \mu \nu}^{(q)} B_{r] \nu \mu}^{(q)} \right) \right] v^\mu \delta ( \epsilon^{\mu \nu} v \cdot d \theta ) + dx^n dx^r q! q! \left[ (q+1) \partial_\nu \left( A_{[n r] \mu \nu}^{(q)} B_{\nu \mu}^{(q)} - A_{\mu \nu}^{(q)} B_{[nr] \nu \mu}^{(q)} \right) + \right. \right. \right. \\
	\label{CJ} & \left. \left. \left. + \partial_\nu \left( A_{[n \mu \nu}^{(q)} B_{r] \nu \mu}^{(q)} \right) \right] v^\nu \delta ( \epsilon^{\mu \nu} v \cdot d \theta ) \right] \right] \ ; \\
	\nonumber C \wedge A \wedge Z_v ( B ) = & - \left[ \sum_{p=0}^\infty dx^m dx^n p! p! \left[ (p+1) \left( C_{[m n] \alpha \beta}^{(p)} A_{\beta \alpha}^{(p)} - C_{\alpha \beta}^{(p)} A_{[m n] \beta \alpha}^{(p)} \right) + C_{[m \alpha \beta}^{(p)} A_{n] \beta \alpha}^{(p)} \right] \delta ( d \theta^\beta ) \delta ( d \theta^\alpha ) \right] \wedge \\
	\label{CK} & \wedge \left[ \sum_{q=0}^\infty (-1)^{q+1} q! dx^r \left( \frac{v^\mu}{v^\nu} \right)^q \left( \partial_\mu B_{r \mu \nu}^{(q)} \frac{v^\mu}{v^\nu} + \partial_\nu B_{r \mu \nu}^{(q)} \right) \right] \ .
\end{align}
It is now easy to observe that the terms from \eqref{CH} and \eqref{CI} arrange together with the terms of \eqref{CJ} and \eqref{CK} in order to for total derivative terms; for example if we subtract from the first term of \eqref{CI} the first from \eqref{CJ} we get an expression of the form
\begin{equation}
	A B \partial C - C \partial ( A B ) = - \partial ( A B C ) \ ,
\end{equation}
where attention should be paid when considering the correct minus signs. The same thing is easily verified for all the other terms as well.

The cyclicity of the inner product is crucial in what concerns the variational principle involved in order to derive the expected equations of motion; indeed, when varying the interaction term with respect to the field $A$ we get three contributions:
\begin{equation}
	\delta \left\langle A , m_2^{(-1)} ( A , A ) \right\rangle = \left\langle \delta A , m_2^{(-1)} ( A , A ) \right\rangle + \left\langle A , m_2^{(-1)} ( \delta A , A ) \right\rangle + \left\langle A , m_2^{(-1)} ( A , \delta A ) \right\rangle \ .
\end{equation}
Thanks to the cyclicity of $\langle \cdot , \cdot \rangle$, we have
\begin{equation}
	\left. \begin{aligned}
		\left\langle A , m_2^{(-1)} ( \delta A , A ) \right\rangle = \left\langle \delta A , m_2^{(-1)} ( A , A ) \right\rangle \\
		\left\langle A , m_2^{(-1)} ( A , \delta A ) \right\rangle = \left\langle \delta A , m_2^{(-1)} ( A , A ) \right\rangle
	\end{aligned}
	\right\}
	 \implies \delta \left\langle A , m_2^{(-1)} ( A , A ) \right\rangle = 3 \left\langle \delta A , m_2^{(-1)} ( A , A ) \right\rangle \ .
\end{equation}
This implies that the variational principle give rise to the following equations of motion:
\begin{equation}
	dA + m_2^{(-1)} (A , A) = 0 \ .
\end{equation}
The gauge invariant EoM are consistent at the present level of $m_2$ product. In the forthcoming subsections we will show that as a consequence of non-associativity of the $m_2$ product it is necessary to modify the Lagrangian (hence the EoM) and the definition of gauge variation.

\subsection{Gauge Invariance and the Emergence of the $A_\infty$ Algebra}

In this subsection we study the gauge invariance of the action. It is a well known result that a section of the bundle of Lie algebra-valued 1-forms under the action of $g$ transforms as
\begin{equation}
	A \ \rightarrow \ \tilde{A} = g^{-1} A g + g^{-1} d \left( g \right) \ ,
\end{equation}
which infinitesimally becomes
\begin{equation} \label{GIA}
	\tilde{A} = A - c A + A c + d c = A + \delta_c A \ ,
\end{equation}
so that we have $\delta_c A = A c - c A + d c $ . In our case, $A \in \Omega^{(1|1)}$ and thus a few remarks are mandatory: in order to have the right matching of form and picture degrees, we have that the gauge parameter $c$ is a $(0|1)-$pseudoform and the products $Ac$ and $cA$ must be considered as $m_2^{(-1)}$ products in order to respect the correct picture number; thus the gauge transformation reads
\begin{equation} \label{GIB}
	\delta_c A = m_{2}^{(-1)} ( A , c ) - m_{2}^{(-1)} ( c , A ) + dc \ .
\end{equation}


Bosonic Chern-Simons theory is invariant (up to boundary terms) under infinitesimal gauge transformations \eqref{GIA}, indeed
\begin{equation} \label{GIC}
	\delta_c \mathcal{L}_{CS} = d ( c A \wedge A + c d A ) \ .
\end{equation}
In this setting, we have that the algebra of gauge transformations closes with respect to the commutator $[ \cdot , \cdot ]$ operation, that is
\begin{equation} \label{GID}
	\left[ \delta_{c_1} , \delta_{c_2} \right] A = \delta_{\left[ c_1 , c_2 \right]} A \ .
\end{equation}
The closure of the algebra is a direct consequence of the Jacobi identity as it can be easily verified.

In our case, the non-associativity of the product $m_2^{(-1)}$ invalidate both the previous results: the action is no longer gauge invariant and the algebra of gauge transformations does not close any longer. Let us see these two facts explicitly:
\begin{equation}
	\frac{1}{2} \delta_c \left\langle A , d A + \frac{2}{3} m_2^{(-1)} ( A , A ) \right\rangle = \left\langle \delta_c A , d A + m_2^{(-1)} ( A , A ) \right\rangle \ ,
\end{equation}
having used the cyclicity of the internal product; thus we have
\begin{equation}
	\left\langle \delta_c A , d A + m_2^{(-1)} ( A , A ) \right\rangle = \left\langle m_{2}^{(-1)} ( A , c ) - m_{2}^{(-1)} ( c , A ) + dc , d A + m_2^{(-1)} ( A , A ) \right\rangle \ .
\end{equation}
By recalling that $|A| = 1 , |c| = 0$ and equation \eqref{CA} for the cyclicity, we have
\begin{align}
	\nonumber \left\langle m_{2}^{(-1)} ( A , c ) , dA \right\rangle = & (-1)^{(|A|+1)(|A|+|c|)} \left\langle dA , m_{2}^{(-1)} ( A , c ) \right\rangle = \left\langle dA , m_{2}^{(-1)} ( A , c ) \right\rangle = \\
	= & \left\langle c , m_{2}^{(-1)} ( dA , A ) \right\rangle \ ; \\
	\nonumber - \left\langle m_{2}^{(-1)} ( c , A ) , dA \right\rangle = & -(-1)^{(|A|+1)(|A|+|c|)} \left\langle dA , m_{2}^{(-1)} ( c , A ) \right\rangle = - \left\langle dA , m_{2}^{(-1)} ( A , c ) \right\rangle = \\
	= & - \left\langle c , m_{2}^{(-1)} ( A , dA ) \right\rangle \ .
\end{align}
These two terms, together with the term $\displaystyle \left\langle dc , m_{2}^{(-1)} ( A , A ) \right\rangle $ can be arranged as
\begin{equation}
	\int d \, Tr \left( c \, m_2^{(-1)} ( A , A ) \right) \ ,
\end{equation}
i.e. a boundary term; another boundary term is
\begin{equation}
	\left\langle dc , dA \right\rangle = \int d \, Tr \left( c dA \right) \ ,
\end{equation}
and these two terms together are exactly the analogous boundary term of \eqref{GIC}. However, we are left with the two terms
\begin{align}
	\nonumber \left\langle m_2^{(-1)} ( A , c ) , m_2^{(-1)} ( A , A ) \right\rangle - \left\langle m_2^{(-1)} ( c , A ) , m_2^{(-1)} ( A , A ) \right\rangle = \\
	\label{GIE} = \left\langle A , m_2^{(-1)} ( m_2^{(-1)} ( A , c ) , A ) \right\rangle - \left\langle A , m_2^{(-1)} ( A , m_2^{(-1)} ( c , A ) ) \right\rangle \ .
\end{align}
Since $m_2^{(-1)}$ is not associative, these two terms do not sum to zero and therefore the action is no longer gauge invariant. This is the reason why a non-associative product leads to the emergence a $A_\infty$-algebra structure: in order to have a gauge invariant theory, we need to add a piece with a 3-product, then a piece with a 4-product and so on.

Before doing this, let us analyse the closure of the gauge algebra. Let us rewrite \eqref{GIB} with the definition used in \eqref{APPI}:
\begin{equation} \label{GIEA}
	\delta_c A = m_{2}^{(-1)} ( A , c ) - m_{2}^{(-1)} ( c , A ) + dc = l_2^{(-1)} ( A , c ) + dc \ .
\end{equation}
With this convention we have
\begin{align}
	\nonumber \left[ \delta_{c_1} , \delta_{c_2} \right] A = & \delta_{c_1} \left( l_2^{(-1)} ( A , c_2 ) + dc_2 \right) - \delta_{c_2} \left( l_2^{(-1)} ( A , c_1 ) + dc_1 \right) = \\
	\nonumber = & l_2^{(-1)} \left( l_2^{(-1)} ( A , c_1 ) , c_2 \right) + l_2^{(-1)} \left( dc_1 , c_2 \right) - l_2^{(-1)} \left( l_2^{(-1)} ( A , c_2 ) , c_1 \right) - l_2^{(-1)} \left( dc_2 , c_1 \right) = \\
	\label{GIF} = & l_2^{(-1)} \left( l_2^{(-1)} ( A , c_1 ) , c_2 \right) + l_2^{(-1)} \left( l_2^{(-1)} ( c_2 , A ) , c_1 \right) + d l_2^{(-1)} \left( c_1 , c_2 \right) \ ,
\end{align}
being $c$ an even field. Since $l_2^{(-1)}$ does not satisfy the Jacobi identity, the algebra does not close. Notice, once again, that Jacobi identity plays a crucial role for the algebra to close. Indeed, if $l_2$ were to satisfy Jacobi identity, we would have had
\begin{align}
	\nonumber l_2^{-1} \left( l_2^{-1} ( A , c_1 ) , c_2 \right) + l_2^{-1} \left( l_2^{-1} ( c_2 , A ) , c_1 \right) + d l_2^{-1} \left( c_1 , c_2 \right) = & - l_2^{-1} \left( l_2^{-1} ( c_1 , c_2 ) , A \right) + d l_2^{-1} \left( c_1 , c_2 \right) = \\
	\nonumber = & l_2^{-1} \left( A , l_2^{-1} ( c_1 , c_2 ) \right) + d l_2^{-1} \left( c_1 , c_2 \right) = \\
	= & \delta_{l^{(-1)}_2 \left( c_1 , c_2 \right)} A \ ,
\end{align}
where in this case $l^{(-1)}_2 \left( c_1 , c_2 \right) = \left[ c_1 , c_2 \right]$.

The break down of gauge-invariance shown in \eqref{GIF} suggests that we should add to the Lagrangian other terms in order to have the cancellation of the terms arising from the gauge variation and, therefore, a gauge-invariant action. This translates mathematically into the introduction of an $A_\infty$-algebra as mentioned previously and as we are about to show explicitly.

We now proceed by constructing explicitly the first multiproduct of the $A_\infty$-algebra.
Let us consider the action discussed so far:
\begin{equation} \label{TPM3A}
	S = \left\langle A , \frac{1}{2} dA + \frac{1}{3} m_2^{(-1)} \left( A , A \right) \right\rangle \ .
\end{equation}
Another way to check the need to introduce other terms in the action is to study the \virgolette Bianchi identities", i.e. we have to check whether $\displaystyle dF = l_2^{(-1)} \left( F , A \right)$ . This is equivalent to verify the gauge invariance of the action, but this turns out to be useful for constructing explicitly the higher products. From \eqref{TPM3A} the field strength reads
\begin{equation} \label{TPM3B}
	F = dA + m_2^{(-1)} ( A , A ) \ .
\end{equation}
Upon applying the exterior derivative we get
\begin{equation} \label{TPM3C}
	d F = d m_2^{(-1)} ( A , A ) = m_2^{(-1)} ( dA , A ) - m_2^{(-1)} ( A , dA ) \ ,
\end{equation}
having used the fact that $d$ is a derivation with respect to $m_2^{(-1)}$, which follows as a consequence of $\left[ Z_v , d \right] = 0$. We now can use \eqref{TPM3B} in order to substitute in \eqref{TPM3C} the expression for $dA$ and we get
\begin{equation} \label{TPM3D}
	d F = m_2^{(-1)} ( F , A ) - m_2^{(-1)} ( A , F ) + m_2^{(-1)} \left( A , m_2^{(-1)} \left( A , A \right) \right) - m_2^{(-1)} \left( m_2^{(-1)} \left( A , A \right) , A \right) \ ,
\end{equation}
where, as expected, it appears the extra term given by the associator of $m_2^{(-1)}$. In order to get rid of this term, we introduce in the action an extra term, formally denoted by $\displaystyle \left\langle A , \frac{1}{4} m_3^{(-2)} \left( A , A , A \right) \right\rangle $, which we are about to construct. By requiring that
\begin{equation} \label{TPM3DA}
	d \left( F + m_3^{(-2)} ( A , A , A \right) = m_2^{(-1)} ( F , A ) - m_2^{(-1)} ( A , F ) \ ,
\end{equation}
we obtain the equation
\begin{equation} \label{TPM3E}
	d m_3^{(-2)} \left( A , A , A \right) + m_3^{(-2)} \left( dA , A , A \right) - m_3^{(-2)} \left( A , dA , A \right) + m_3^{(-2)} \left( A , A , dA \right) + $$ $$ - m_2^{(-1)} \left( m_2^{(-1)} \left( A , A \right) , A \right) + m_2^{(-1)} \left( A , m_2^{(-1)} \left( A , A \right) \right) = 0 \ .
\end{equation}
Notice that when applying the exterior derivative $d$ to the term $m_3^{(-2)}$, we have also the term $\displaystyle d m_3^{(-2)} \left( A , A , A \right) $; this in not equal to $\displaystyle m_3^{(-2)} \left( dA , A , A \right) - m_3^{(-2)} \left( A , dA , A \right) + m_3^{(-2)} \left( A , A , dA \right) $, since a priori $d$ is not a \virgolette derivation" with respect to $m_3^{(-2)}$; therefore, we must consider the action of $d$ either on the product or on each single term appearing as argument of the product.

Also, as we have anticipated, \eqref{TPM3E} is the equation that relates the product $m_2$ and the product $m_3$, once we have made the formal substitution $d \equiv m_1$. In particular it gives the non-associativity of $m_2$ in terms of the higher product $m_3$ and it is actually the first defining relations of an $A_\infty$-algebra that makes non-associativity manifest.

We are now ready to give an explicit expression for $m_3$. First of all some observations are in order. When we have introduced the 3-product, we have used the notation $m_3^{(-2)}$ because this product has to subtract 2 from the picture number, in order to have the correct picture number in the Lagrangian. Moreover, $m_3$ must have form number $-1$ in order to saturate the correct form number in the Lagrangian. These considerations lead to an ansatz: $m_3$ could be constructed as a combination of $m_2$ and $\tilde{m}_2$, since both have picture number $-1$ and have form number 0 and $-1$ respectively. An equally-weighted choice is
\begin{align}
	\nonumber m_3^{(-2)} ( A , B , C ) = & \frac{1}{2} \left[ m_2^{(-1)} \left( \tilde{m}_2^{(-1)} ( A , B ) , C \right) - (-1)^{|A|} m_2^{(-1)} \left( A , \tilde{m}_2^{(-1)} ( B , C ) \right) + \right. \\
	\label{TPM3F} & \left. + \tilde{m}_2^{(-1)} \left( m_2^{(-1)} ( A , B ) , C \right) - \tilde{m}_2^{(-1)} \left( A , m_2^{(-1)} ( B , C )\right) \right] + d \Lambda ( A , B , C ) \ ,
\end{align}
where the signs have been chosen such that \eqref{TPM3E} is respected and the last term does not appear in \eqref{TPM3E} since it is $d$-exact. Let us verify that \eqref{TPM3F} satisfies \eqref{TPM3E} on generic $(1|1)$-forms $A, B, C$:
\begin{equation}
	\frac{1}{2} \left[ m_2^{(-1)} \left( d \tilde{m_2}^{(-1)} \left( A , B \right) , C \right) - m_2^{(-1)} \left( \tilde{m_2}^{(-1)} \left( A , B \right) , dC \right) + m_2^{(-1)} \left( dA , \tilde{m_2}^{(-1)} \left( B , C \right) \right) + \right. $$ $$ \left. - m_2^{(-1)} \left( A , d \tilde{m_2}^{(-1)} \left( B , C \right) \right) + d \tilde{m_2}^{(-1)} \left( m_2^{(-1)} \left( A , B \right) , C \right) - d \tilde{m_2}^{(-1)} \left( A , m_2^{(-1)} \left( B , C \right) \right) + \right. $$ $$ \left. + m_2^{(-1)} \left( \tilde{m_2}^{(-1)} \left( dA , B \right) , C \right) - m_2^{(-1)} \left( dA , \tilde{m_2}^{(-1)} \left( B , C \right) \right) + \tilde{m_2}^{(-1)} \left( m_2^{(-1)} \left( dA , B \right) , C \right) + \right. $$ $$ \left. - \tilde{m_2}^{(-1)} \left( dA , m_2^{(-1)} \left( B , C \right) \right) - m_2^{(-1)} \left( \tilde{m_2}^{(-1)} \left( A , dB \right) , C \right) - m_2^{(-1)} \left( A , \tilde{m_2}^{(-1)} \left( dB , C \right) \right) + \right. $$ $$ \left. - \tilde{m_2}^{(-1)} \left( m_2^{(-1)} \left( A , dB \right) , C \right) + \tilde{m_2}^{(-1)} \left( A , m_2^{(-1)} \left( dB , C \right) \right) + m_2^{(-1)} \left( \tilde{m_2}^{(-1)} \left( A , B \right) , dC \right) + \right. $$ $$ \left. + m_2^{(-1)} \left( A , \tilde{m_2}^{(-1)} \left( B , dC \right) \right) + \tilde{m_2}^{(-1)} \left( m_2^{(-1)} \left( A , B \right) , dC \right) - \tilde{m_2}^{(-1)} \left( A , m_2^{(-1)} \left( B , dC \right) \right) \right] + $$ $$ - m_2^{(-1)} \left( m_2^{(-1)} \left( A , B \right) , C \right) + m_2^{(-1)} \left( A , m_2^{(-1)} \left( B , C \right) \right) = 0 \ .
\end{equation}


Since we have added a new term in the Lagrangian, we have to define a new field strength:
\begin{equation}
	F' = dA + m_2^{(-1)} ( A , A ) + m_3^{(-2)} ( A , A , A ) = F + m_3^{(-2)} ( A , A , A ) \ .
\end{equation}
Clearly the Bianchi identity for $F'$ does not hold, in particular from \eqref{TPM3DA} we have
\begin{align}
	\nonumber d F' = & m_2^{(-1)} ( F , A ) - m_2^{(-1)} ( A , F ) =  m_2^{(-1)} ( F' - m_3^{(-2)} ( A , A , A ) , A ) - m_2^{(-1)} ( A , F' - m_3^{(-2)} ( A , A , A ) ) = \\
	= & m_2^{(-1)} ( F' , A ) - m_2^{(-1)} ( A , F' ) - m_2^{(-1)} ( m_3^{(-2)} ( A , A , A ) , A ) + m_2^{(-1)} ( A , m_3^{(-2)} ( A , A , A ) ) \ .
\end{align}
Again, we have an extra term breaking the Bianchi identity. By following the prescription described above, we add to the action an extra term, which we denote by $\displaystyle \left\langle A , \frac{1}{5} m_4^{(-3)} \left( A , A , A , A \right) \right\rangle $, in order to restore the identity. As we get so far, one can easily infer the algorithm to be used in order to construct the whole $A_\infty$ Lagrangian. Before we go on to the closure of the gauge transformations, another issue is to be addressed: is $m_3$ in the Small Hilbert Space?


In order to answer this question, let us now look back at the definition of the 3-product given into the equation \eqref{TPM3F}. If we neglect the $d$-exact term $d \Lambda$ we have that the product $m_3$ is defined by a certain combination of $m_2$ and $\tilde{m_2}$. Now, since the product $m_2$ is defined via the application of the operator $Z$, it maps pseudoforms into pseudoforms, as discussed in sections 2 and 4. This means that $m_2$ maps the SHS into itself. On the other hand, $\tilde{m}_2$ is defined via the operator $\Theta$, that maps pseudoforms into inverse forms. Therefore, by contrast, $\tilde{m}_2$ maps the SHS into the LHS. This means, a priori, that $m_3$ gives values in the LHS. Here is where the $d$-exact term becomes relevant: it can be defined as a term that annihilates the LHS part resulting from the $\tilde{m}_2$ part of $m_3$.

Superstring Theory suggests a simple way to establish whether an objects lies in the SHS, this is based on the operator $\virgolette \eta$'', whose definition was given in subsection 2.6.

If we want $m_3^{(-2)}$ to have image in the SHS, we have to require that
\begin{equation} \label{M3SHSA}
	\eta \left[ m_3^{(-2)} (A , B , C ) \right] = 0 \ ;
\end{equation}
this equation translates to an equation for $d \Lambda$. Let us see the explicit expression of the previous relation term by term:
\begin{align}
	\nonumber \eta \left[ m_2^{(-1)} \left( \tilde{m_2}^{(-1)} \left( A , B \right) , C \right) \right] = & -\frac{i}{9} \eta \left[ Z_v \left( \Theta \left( A \wedge B \right) \wedge C \right) + Z_v \left( \Theta A \wedge B \wedge C \right) + (-1)^{|A|} Z_v \left( A \wedge \Theta B \wedge C \right) + \right. \\
	\nonumber & \left. + Z_v \left( \Theta \left( A \wedge B \right) \right) \wedge C + Z_v \left( \Theta \left( A \wedge B \right) \right) \wedge C + (-1)^{|A|} Z_v \left( A \wedge \Theta B \right) \wedge C + \right. \\
	\nonumber & \left. + \Theta \left( A \wedge B \right) \wedge Z_v C + \Theta A \wedge B \wedge Z_v C + (-1)^{|A|} A \wedge \Theta B \wedge Z_v C \right] = \\
	\label{M3SHSB} & = \frac{1}{3} A \wedge B \wedge Z_v C \ ,
\end{align}
having used extensively the properties of the operator $\eta$ described in Section 2. The second term is
\begin{align}
	\nonumber - (-1)^{|A|} \eta \left[ m_2^{(-1)} \left( A , \tilde{m_2}^{(-1)} \left( B , C \right) \right) \right] = & - (-1)^{|A|} (-1)^{|A| ( |B| + |C| - 1} \eta \left[ m_2^{(-1)} \left( \tilde{m_2}^{(-1)} \left( B , C \right) , A \right) \right] = \\
	\label{M3SHSC} = & - \frac{1}{3} Z_v A \wedge B \wedge C \ ,
\end{align}
having used \eqref{M3SHSB}. The third term is
\begin{align}
	\nonumber \eta \left[ \tilde{m_2}^{(-1)} \left( m_2^{(-1)} \left( A , B \right) , C \right) \right] = & -\frac{i}{9} \eta \left[ \Theta \left( Z_v \left( A \wedge B \right) \wedge C \right) + \Theta \left( Z_v A \wedge B \wedge C \right) + \Theta \left( A \wedge Z_v B \wedge C \right) + \right. \\
	\nonumber & \left. + \Theta \left( Z_v \left( A \wedge B \right) \right) \wedge C + \Theta \left( Z_v A \wedge B \right) \wedge C + \Theta \left( A \wedge Z_v B \right) \wedge C + \right. \\
	\nonumber & \left. + (-1)^{|A| + |B|} Z_v \left( A \wedge B \right) \wedge \Theta C + (-1)^{|A| + |B|} Z_v A \wedge B \wedge \Theta C + \right. \\
	\nonumber & \left. + (-1)^{|A| + |B|} A \wedge Z_v B \wedge \Theta C \right] = \\
	\label{M3SHSD} = & \frac{1}{3} Z_v \left( A \wedge B \right) \wedge C + \frac{1}{3} Z_v A  \wedge B \wedge C + \frac{1}{3} A \wedge Z_v B \wedge C \ .
\end{align}
Finally, the fourth term reads
\begin{align}
	\nonumber \eta \left[ - \tilde{m_2}^{(-1)} \left( A , m_2^{(-1)} \left( B , C \right) \right) \right] = & \frac{i}{9} \eta \left[ \Theta \left( A \wedge Z_v \left( B \wedge C \right) \right) + \Theta \left( A \wedge Z_v B \wedge C \right) + \Theta \left( A \wedge B \wedge Z_v C \right) + \right. \\
	\nonumber & \left. + \left( \Theta A \right) \wedge Z_v \left( B \wedge C \right) + \left( \Theta A \right) \wedge Z_v B \wedge C + \right. \\
	\nonumber & \left. + \left( \Theta A \right) \wedge B \wedge Z_v C + (-1)^{|A|} A \wedge \Theta \left( Z_v \left( B \wedge C \right) \right) + (-1)^{|A|} A \wedge \Theta \left( Z_v B \wedge C \right) + \right. \\
	\nonumber & \left. + (-1)^{|A|} A \wedge \Theta \left( B \wedge Z_v C \right) \right] = \\
	\label{M3SHSE} = & -\frac{1}{3} A \wedge Z_v \left( B \wedge C \right) - \frac{1}{3} A \wedge Z_v B \wedge C - \frac{1}{3} A \wedge B \wedge Z_v C \ .
\end{align}
By putting the four terms together we get
\begin{equation} \label{M3SHSF}
	\eta \left[ m_3^{(-2)} (A , B , C ) \right] = \frac{1}{3} \left[ Z_v \left( A \wedge B \right) \wedge C - A \wedge Z_v \left( B \wedge C \right) \right] + \eta d \Lambda \left( A , B , C \right) \ .
\end{equation}
We can now manipulate this expression in order to find an explicit formulation for the multiproduct $\Lambda$, indeed we have:
\begin{equation} \label{M3SHSG} 
	Z_v \left( A \wedge B \right) \wedge C - A \wedge Z_v \left( B \wedge C \right) = d \left[ \Theta \left( A \wedge B \right) \wedge C + A \wedge \Theta \left( B \wedge C \right) \right] + \Theta \left( dA \wedge B \right) \wedge C + $$ $$ - \Theta \left( A \wedge dB \right) \wedge C + \Theta \left( A \wedge B \right) \wedge d C - d A \wedge \Theta \left( B \wedge C \right)  - A \wedge \Theta \left( dB \wedge C \right) + A \wedge \Theta \left( B \wedge dC \right) \ .
\end{equation}
We can now define the formal expression
\begin{equation} \label{M3SHSH}
	\hat{m_3}^{(-1)} \left( A , B , C \right) = \Theta \left( A \wedge B \right) \wedge C - (-1)^{|A|} A \wedge \Theta \left( B \wedge C \right) \ ,
\end{equation}
so that the terms in \eqref{M3SHSF} become
\begin{equation} \label{M3SHSI}
	d \hat{m_3}^{(-1)} \left( A , B , C \right) + \hat{m_3}^{(-1)} \left( dA , B , C \right) - \hat{m_3}^{(-1)} \left( A , dB , C \right) + \hat{m_3}^{(-1)} \left( A , B , dC \right) = d \left[ \hat{m_3}^{(-1)} \left( A , B , C \right) \right] \ .
\end{equation}
By inserting \eqref{M3SHSI} in \eqref{M3SHSF} we obtain
\begin{equation} \label{M3SHSJ}
	\eta \Lambda ( A , B , C ) = - \frac{1}{3} \hat{m_3}^{(-1)} \left( A , B , C \right) \ .
\end{equation}
An immediate (and equally weighted) solution to this equation is suggested by the fact that the operator $\eta$ is the left-inverse of the operator $\Theta$ as seen in \eqref{ETATHETA}:
\begin{equation} \label{M3SHSK}
	\Lambda ( A , B , C ) = - \frac{1}{12} \left[ \Theta \left( \iota_v \right) \hat{m_3}^{(-1)} \left( A , B , C \right) - \hat{m_3}^{(-1)} \left( \Theta \left( \iota_v \right) A , B , C \right) + \right. $$ $$ \left. - (-1)^{|A|} \hat{m_3}^{(-1)} \left( A , \Theta \left( \iota_v \right) B , C \right) - (-1)^{|A| + |B|} \hat{m_3}^{(-1)} \left( A , B , \Theta \left( \iota_v \right) C \right) \right] \ .
\end{equation}
We have therefore that if the $d$-exact term appearing in \eqref{TPM3F} is set to be equal to \eqref{M3SHSK}, the product $m_3^{(-1)}$ lives in the small Hilbert space.


Let us now study the problem of the closure of the gauge algebra. Previously we have seen that, since the product $m_2^{(-1)}$ is not associative, the gauge algebra does not close. We now show that, in order the algebra to close, we have to modify the gauge transformation law \eqref{GIEA} by introducing multiproducts induced by the $A_\infty$ algebra discussed so far. We have already seen that the transformation law for the pseudoform $A$ is given by
\begin{equation}
	\delta_c A =  l_2^{(-1)} ( A , c ) + dc \ ,
\end{equation}
and that the commutator of two gauge transformations gives
\begin{equation}
	\left[ \delta_{c_1} , \delta_{c_2} \right] A = l_2^{-1} \left( l_2^{-1} ( A , c_1 ) , c_2 \right) + l_2^{-1} \left( l_2^{-1} ( c_2 , A ) , c_1 \right) + d l_2^{-1} \left( c_1 , c_2 \right) \ .
\end{equation}
We can now proceed by using a method analogous to the one for the $A_\infty$-algebra, i.e. we consider the gauge transformation of the field strength $F$; as a starting point, let us consider the case where the interaction term is given only by $m_2^{(-1)}$:
\begin{align}
	\nonumber \delta_c F = & \delta_c \left( d A + m_2^{(-1)} \left( A , A \right) \right) = l_2^{(-1)} \left( dA + m_2^{(-1)} \left( A , A \right) , c \right) + m_2^{(-1)} \left( m_2^{(-1)} \left( A , c \right) , A \right) + \\
	\nonumber & - m_2^{(-1)} \left( A , m_2^{(-1)} \left( c , A \right) \right) + m_2^{(-1)} \left( A , m_2^{(-1)} \left( A , c \right) \right) - m_2^{(-1)} \left( m_2^{(-1)} \left( A , A \right) , c \right) + \\
	\label{GTLA} & + m_2^{(-1)} \left( c , m_2^{(-1)} \left( A , A \right) \right) - m_2^{(-1)} \left( m_2^{(-1)} \left( c , A \right) , A \right) \ ,
\end{align}
where, except for the first term, we have used the explicit definition of the antisymmetrised product $l_2^{(-1)}$ in terms of the product $m_2^{(-1)}$. The first term of \eqref{GTLA} is exactly the generalisation of the commutator of a usual gauge transformation $\displaystyle \delta_c F = \left[ F , c \right] $; the other terms are arranged as three associators of the product $m_2^{(-1)}$. In \eqref{TPM3E} we have already discussed the relation between the associator of $m_2^{(-1)}$ and the product $m_3^{(-2)}$, thus we can easily infer how the gauge transformation law should be modified: let us define the \emph{modified infinitesimal gauge transformation} as
\begin{equation} \label{GTLB}
	\delta_c A = \frac{1}{2!} l_3^{(-2)} ( A , A , c ) + l_2^{(-1)} ( A , c ) + dc \ ,
\end{equation}
where $l_3^{(-2)}$ is the antisymmetrisation of $m_3^{(-2)}$ as defined in appendix A, and the factor $\displaystyle \frac{1}{2!} $ is necessary since in the definition of $l_3^{(-2)}$ we have a double counting when two fields are equal. With this definition we have that \eqref{GTLA} becomes
\begin{align}
	\nonumber \delta_c F = & l_2^{(-1)} \left( dA + m_2^{(-1)} \left( A , A \right) , c \right) + m_2^{(-1)} \left( m_2^{(-1)} \left( A , c \right) , A \right) - m_2^{(-1)} \left( A , m_2^{(-1)} \left( c , A \right) \right) + \\
	\nonumber & + m_2^{(-1)} \left( A , m_2^{(-1)} \left( A , c \right) \right) - m_2^{(-1)} \left( m_2^{(-1)} \left( A , A \right) , c \right) + m_2^{(-1)} \left( c , m_2^{(-1)} \left( A , A \right) \right) + \\
	\label{GTLC} & - m_2^{(-1)} \left( m_2^{(-1)} \left( c , A \right) , A \right) + d \left[ l_3^{(-2)} \left( A , A , c \right) \right] + m_2^{(-1)} \left( l_3^{(-2)} \left( A , A , c \right) , A \right) + m_2^{(-1)} \left( A , l_3^{(-2)} \left( A , A , c \right) \right) \ .
\end{align}
The term $\displaystyle d \left[ l_3^{(-2)} \left( A , A , c \right) \right] $ cancels out the three associators exactly, because of the third $A_\infty$ relation \eqref{TPM3E} as expected. The other new terms are made of combinations of the 2-product $m_2^{(-1)}$ and the 3-product $m_3^{(-2)}$ and give rise to the necessity of a new modification of the gauge transformation law for the field $A$. The algorithm is now analogous to the one used for the construction of multiproducts: first we have to update the definition of the field strength $F$ by adding the term $m_3^{(-2)} \left( A , A , A \right)$, then we have to apply the gauge transformation \eqref{GTLB}. The superfluous terms will have to be reabsorbed, by means of the $A_\infty$ relations, by the insertion of a term $\frac{1}{3!} l_4^{(-3)} \left( A , A , A , c \right)$ which is given by the antisymmetrisation of the product $m_4^{(-3)}$ that should have been constructed as described in the previous subsections. We observe that the numerical factor $\displaystyle \frac{1}{3!} $ is a consequence of the possible permutations of the three $A$ fields as arguments. Then the process should be iterated. This will lead to the final correct gauge transformation law given by
\begin{equation} \label{GTLD}
	\delta_c A = \sum_{i=1}^{\infty} \frac{1}{(i-1)!} l_i^{(-i+1)} \left( A , \ldots , A , c \right) \ ,
\end{equation}
which, under the identification $l_1^{(0)} \equiv m_1^{(0)} \equiv d $, is exactly the one described in \cite{Jurco:2018sby}.

\subsection{Supersymmetry at Picture 1}

In sections 2 and 3 we have discussed the supersymmetric action of free SCS and the rheonomic equations when working at picture 0. When working at picture 1, things work differently, nonetheless yield the same results. The gauge field $A^{(1|1)}$ can be decomposed in powers of $V^a$ exactly as we showed for the expansion in $dx^a$. Hence the field strength will be decomposed consequently as
\begin{equation} \label{SSP1A}
	F^{(2|1)} = V^a \wedge V^b \wedge V^c F_{[a b c]}^{(-1|1)} + V^a \wedge V^b F_{[a b]}^{(0|1)} + V^a F_a^{(1|1)} + F_0^{(2|1)} \ .
\end{equation}
In \eqref{SSP1A} we have a slight abuse of notation: in the superscripts $(n|1)$, $n$ is to be intended as the fermionic form number, and the pseudoforms $\displaystyle F_I^{(n|1)} $ do not have a further decomposition in $V$'s. We now apply the Bianchi identity (recall formulas in \eqref{DVDPSI}):
\begin{align}
	\nonumber d F = 0 \ \implies \ & 3 \psi^\alpha \gamma^a_{\alpha \beta} \psi^\beta V^b \wedge V^c F_{[a b c]}^{(-1|1)} - V^a \wedge V^b \wedge V^c \psi^\alpha D_\alpha F_{[a b c]}^{(-1|1)} + 2 \psi^\alpha \gamma^a_{\alpha \beta} \psi^\beta V^b F_{[a b]}^{(0|1)} + \\
	\nonumber & + V^a \wedge V^b \wedge V^c \partial_{[c} F_{a b]}^{(0|1)} + V^a \wedge V^b \psi^\alpha D_\alpha F_{[a b]}^{(0|1)} + \psi^\alpha \gamma^a_{\alpha \beta} \psi^\beta F_a^{(1|1)} + \\
	& - V^a \wedge V^b \partial_{[b} F_{a]}^{(1|1)} - V^a \psi^\alpha D_\alpha F_a^{(1|1)} + V^a \partial_a F_0^{(2|1)} + \psi^\alpha D_\alpha F_0^{(2|1)} = 0 \ .
\end{align}
By looking at the homogeneous parts in $V$, we get the system
\begin{equation} \label{SSP1B}
	\begin{cases}
		- D_\alpha F_{[a b c]}^{(-1|1)} + \partial_{[c} F_{a b]}^{(0|1)} = 0 \\
		3 \psi^\alpha \gamma^c_{\alpha \beta} \psi^\beta F_{[c a b]}^{(-1|1)} + \psi^\alpha D_\alpha F_{[a b]}^{(0|1)} - \partial_{[b} F_{a]}^{(1|1)} = 0 \\
		2 \psi^\alpha \gamma^a_{\alpha \beta} \psi^\beta F_{[a b]}^{(0|1)} - \psi^\alpha D_\alpha F_b^{(1|1)} + \partial_b F_0^{(2|1)} = 0 \\
		\psi^\alpha \gamma^a_{\alpha \beta} \psi^\beta F_a^{(1|1)} + \psi^\alpha D_\alpha F_0^{(2|1)} = 0
	\end{cases} \ .
\end{equation}
Now we can apply the \emph{conventional constraint} prescription: in \eqref{SSP1A} we put equal to 0 the term with no $V$, i.e. $F_0 = 0$. By means of this prescription we can solve the previous system as follows: the last equation becomes
\begin{equation} \label{SSP1C}
	\psi^\alpha \gamma^a_{\alpha \beta} \psi^\beta F_a^{(1|1)} = 0 \ \implies \ F_a^{(1|1)} = \psi^\alpha \gamma_{a \alpha \beta} W^{\beta (0|1)} \ ,
\end{equation}
where $W$ is any function, because of the \emph{Fierz identity}. This result gives us the correct way to identify the gaugino field strength $W$ in terms of the gauge field $A$. We can substitute this result in the third equation of \eqref{SSP1B} in order to get
\begin{align}
	\nonumber & 2 \psi^\alpha \gamma^a_{\alpha \beta} \psi^\beta F_{[a b]}^{(0|1)} - \psi^\alpha \psi^\mu \gamma_{b \mu \nu} D_\alpha W^{\nu (0|1)} = 0 \ \implies \\
	& \implies F_{[a b]}^{(0|1)} = \frac{1}{2} \gamma^{\alpha \beta}_{[a} \gamma_{b] \beta \nu} D_\alpha W^{\nu (0|1)} = \frac{1}{4} \gamma_{[a b] \nu}^{\alpha} D_\alpha W^{\nu (0|1)} \ ,
\end{align}
where $\gamma^{[ab]}_{\alpha \beta}$ was defined in section 2. Because of the trace properties of $\gamma$ matrices in three dimensions, it follows that $D_\alpha W^{\alpha (0|1)} = 0 $. Notice that, up to now, we have the same results as in the picture 0 case. From the known results at picture 0, we can infer that in the second equation of \eqref{SSP1B} we have
\begin{equation}
	\psi^\alpha D_\alpha F_{[a b]}^{(0|1)} - \partial_{[b} F_{a]}^{(1|1)} = 0 \ ,
\end{equation}
and then the equation becomes
\begin{equation} \label{SSP1D}
	\psi^\alpha \gamma^c_{\alpha \beta} \psi^\beta F_{[c a b]}^{(-1|1)} = 0 \ .
\end{equation}
By contrast to what we obtained in \eqref{SSP1C}, in this case we get exactly
\begin{equation} \label{SSP1E}
	F_{[c a b]}^{(-1|1)} = 0 \ ,
\end{equation}
and this is a consequence of the possibility of reordering the indices $a,b,c$. Consider indeed
\begin{equation}
	F_{[c a b]}^{(-1|1)} = \sum_{p=0}^\infty F_{[c a b], 1 2}^{(p)} \left( \psi^1 \right)^p \delta^{(p+1)} \left( \psi^2 \right) \ ,
\end{equation}
having made, without loss of generality, a choice for the direction $\psi^2$ of the $\delta$ term; we have
\begin{equation}
	\psi^\alpha \gamma^c_{\alpha \beta} \psi^\beta = \left( ( \psi^1 )^2 + ( \psi^2 )^2 , (\psi^1 )^2 - (\psi^2 )^2 , 2 \psi^1 \psi^2 \right) \ ,
\end{equation}
and then, for $p=0$ we have that \eqref{SSP1D} becomes
\begin{equation}
	( \psi^1 )^2 F_{1 a b, 1 2}^{(0)} + ( \psi^1 )^2 F_{2 a b, 1 2}^{(0)} + 2 \psi^1 F_{3 a b, 1 2}^{(0)} = 0 \ .
\end{equation}
This implies $\displaystyle F_{3 a b, 1 2}^{(0)} = 0 $, and by reshuffling $\displaystyle F_{3 a b, 1 2}^{(0)} \to F_{2 a b, 1 2}^{(0)} $ and $\displaystyle F_{3 a b, 1 2}^{(0)} \to F_{1 a b, 1 2}^{(0)} $ we get $\displaystyle F_{[c a b], 1 2}^{(0)} = 0 $. By iteration it follows that $\displaystyle F_{[c a b], 1 2}^{(p)} = 0 \ , \forall p \in \mathbb{N} $, hence $\displaystyle F_{[c a b]}^{(-1|1)} = 0 $.

Finally, the first equation of \eqref{SSP1B} is the usual Bianchi identity
\begin{equation}
	\partial_{[c} F_{a b]}^{(0|1)} = 0 \ .
\end{equation}
We have therefore proved that the Bianchi identities at picture 1 are the same as the Bianchi identities at picture 0. Moreover we have an explicit way to find the expression of the \emph{gaugino field strength} $W$ at picture 1, i.e. \eqref{SSP1C}.

\subsection{Supersymmetry at Picture 2}

In this section we study the prescriptions that supersymmetry imposes at picture 2. Our analysis is meant to be compared with its analogous at pictures 0 and 1, as to find analogies and differences.

\noindent Let us start from the decomposition of the field strength:
\begin{equation} \label{SSP2A}
	F^{(2|2)} = V^a \wedge V^b F^{(0|2)}_{[a b]} + V^a \wedge V^b \wedge V^c F^{(-1|2)}_{[a b c]} \ .
\end{equation}
Notice that the field strength components $F^{(0|2)}_{[a b]}$ and $F^{(-1|2)}_{[a b c]}$ are the only possible because of the presence of the two $\delta$'s. In particular we have the decomposition:
\begin{equation} \label{SSP2B}
	F^{(0|2)}_{[a b]} = F_{[a b]} \delta^2 \left( d \theta \right) \ , \ F^{(-1|2)}_{[a b c]} = F^\mu_{[a b c]} i_\mu \delta^2 \left( d \theta \right) \ ,
\end{equation}
where $\iota_\mu$ is the usual compact notation to indicate a fermionic derivation on (one of) the two $\delta$'s. We can now apply the Bianchi identity:
\begin{align}
	\nonumber d F^{(2|2)} = 0 \ \implies \ & 2 \psi^\alpha \gamma^a_{\alpha \beta} \psi^\beta V^b F^{(0|2)}_{[a b]} + V^a \wedge V^b \psi^\alpha D_\alpha F^{(0|2)}_{[a b]} + V^a \wedge V^b \wedge V^c \partial_{[c} F^{(0|2)}_{a b]} + \\
	\label{SSP2C} & + 3 \psi^\alpha \gamma^a_{\alpha \beta} \psi^\beta V^b \wedge V^c F^{(-1|2)}_{[a b c]} - V^a \wedge V^b \wedge V^c \psi^\alpha D_\alpha F^{(-1|2)}_{[a b c]} = 0 \ .
\end{align}
It is immediate to see that many terms in the expression are trivially 0 because of the explicit decomposition \eqref{SSP2B}, and the Bianchi identity reduces to
\begin{equation} \label{SSP2D}
	\partial_{[c} F^{(0|2)}_{a b]} - \psi^\alpha D_\alpha F^{(-1|2)}_{[a b c]} = 0 \ .
\end{equation}
In particular we have
\begin{equation} \label{SSP2E}
	\psi^\alpha D_\alpha F^{(-1|2)}_{[a b c]} = \psi^\alpha D_\alpha F^\mu_{[a b c]} i_\mu \delta^2 \left( d \theta \right) = D_\alpha F^\alpha_{[a b c]} \delta^2 \left( d \theta \right) = \epsilon_{a b c} D_\alpha W^\alpha \delta^2 \left( d \theta \right) \ ,
\end{equation}
where we have factorised the dependence on the bosonic indices with a totally antisymmetric tensor by writing $F^\alpha_{[a b c]} = \epsilon_{a b c} W^\alpha$.

As can be directly seen, in the picture 2 case it is not necessary to invoke a \emph{conventional constraint} prescription in order to solve the abstract Bianchi identities, since there is not a term with no $V$'s from the beginning; however, by imposing this condition one recovers the usual form of the theory, i.e.
\begin{equation}
	D_\alpha W^\alpha = 0 \ .
\end{equation}
Also, notice that under this constraint, we have the same field content of the previous two cases, i.e.\ a field with two antisymmetric bosonic indices satisfying the standard Bianchi identity and a field with a fermionic index satisfying the null superdivergence condition.

\subsection{Passing from a Picture to Another Picture}

In this section we want to discuss how the informations described above are recovered in term of PCO's. To this end, let us rewrite the field strength contents in a \virgolette diagrammatic'' fashion. With some abuse of notation, we omit the $V$'s and indicate only the fermionic form number and the picture number as to get:
\begin{align*}
	p = 0 \ , \ F^{(2|0)} = & F_0^{(2|0)} + F_1^{(1|0)} + F_2^{(0|0)} + \ 0 \ \ \ \ , \\
	p = 1 \ , \ F^{(2|1)} = & F_0^{(2|1)} + F_1^{(1|1)} + F_2^{(0|1)} + F_3^{(-1|1)} \ , \\
	p = 2 \ , \ F^{(2|2)} = & \ \ \ 0 \ \ \ + \ \ \ 0 \ \ \ + F_2^{(0|2)} + F_3^{(-1|2)} \ .
\end{align*}
The PCO $Z_v$ described in the previous sections acts vertically, from the last line to the first. A priori, one might expect that the first and the second 0's of the last line get mapped to 0's in the first and second line, but there is a subtlety to be considered. Indeed, we can modify the PCO $\Theta$ as follows
\begin{equation} \label{PPAPA}
	\Theta \left( \iota_v \right) \rightarrow \Theta \left( \iota_v \right) + i_X \delta \left( \iota_v \right) \ ,
\end{equation}
where $X$ is an even vector field. This modification allows to \virgolette move diagonally" in the previous diagram when reducing the picture. In the following we will provide a justification for this modification, showing that it corresponds to a sort of \virgolette gauge transformation" of $\Theta$.

The operator $Z_v$ is said to be a Picture Lowering Operator because it is the left inverse of the Picture Raising Operator $\mathbb{Y} = \theta^\alpha \delta \left( d \theta^\alpha \right)$, which is a representative of the cohomology as discussed in the previous sections:
\begin{equation} \label{PPAPB}
	Z_v \left( \theta^\alpha \delta \left( d \theta^\alpha \right) \right) = -i d \Theta \left( \iota_v \right) \left( \theta^\alpha \delta \left( d \theta^\alpha \right) \right) = -i d \left( \theta^\alpha \frac{i}{d \theta^\alpha} \right) = 1 \ ,
\end{equation}
having used the closure of $\mathbb{Y}$. In sections 2 and 3 we have shown that we can define the PCO $\mathbb{Y}$ modulo exact terms, i.e. we can consider any new PCO upon adding $d$-exact terms,
\begin{equation} \label{PPAPC}
	\tilde{\mathbb{Y}} = \mathbb{Y} + d \Lambda \ .
\end{equation}
We now show that it is also possible to make an analogous choice for the operator $\Theta$, and in particular the transformation \eqref{PPAPA} is allowed. We begin with the general identity
\begin{equation} \label{PPAPD}
	\left( Z_v + U \right) \left( \mathbb{Y} + d \Lambda \right) = 1 \ \implies \ Z_v d \Lambda + U \mathbb{Y} + U d \Lambda = 0 \ .
\end{equation}
A general form for $\Lambda$ is given by
\begin{equation} \label{PPAPE}
	\sum_{p=0}^{\infty} \theta^\alpha \left( d \theta^\alpha \right)^p \left( \iota_\beta \right)^{p+1} \delta \left( d \theta^\beta \right) \ \implies \ d \Lambda = \sum_{p=0}^{\infty} \left( d \theta^\alpha \right)^{p+1} \left( \iota_\beta \right)^{p+1} \delta \left( d \theta^\alpha \right) \ .
\end{equation}
Clearly the application of the operator $Z_v$ reduces to
\begin{equation} \label{PPAPF}
	-i d \Theta \left( \iota_v \right) d \Lambda = \sum_{p=0}^{\infty} -i d \left[ i (-1)^p p! \frac{(d \theta^\alpha)^p}{( d \theta^\beta )^{p+1}} \right] = 0 \ ,
\end{equation}
that is, $Z_v d \Lambda = 0 $. Therefore the consistency relation \eqref{PPAPD} reduces to
\begin{equation} \label{PPAPG}
	U \mathbb{Y} + U d \Lambda = 0 \ .
\end{equation}
This equation means that any modification $\Lambda$ as in \eqref{PPAPC} that satisfies \eqref{PPAPG} amounts to the same modification as adding a term $U$ to the PCO $Z_v$ compatible with \eqref{PPAPC}. It is easy to see that, in particular, the additional piece described in \eqref{PPAPA} works:
\begin{equation}
	U = d \delta \left( \iota_v \right) \iota_E + \delta \left( \iota_v \right) \iota_E d \ \implies \ U \mathbb{Y} = 0 \ , \ U d \Lambda = 0 \ ,
\end{equation}
because neither $Y$ nor $d \Lambda$ contain $dx$ pieces, thus the contraction $\iota_E$ gives automatically 0. This shows that it is possible to add additional pieces in $\Theta$ that do not change the equivalence class of the cohomology and in particular that $\delta \left( \iota_v \right) \iota_E$ does this.

This re-definition of the operator $\Theta$ really allows to move diagonally as follows:
\begin{equation*}
	\xymatrix{
		p = 0 \ , & \ F^{(2|0)} = & F_0^{(2|0)} \ \ + & F_1^{(1|0)} \ \ + & F_2^{(0|0)} \ \ + & \ 0 \\
		p = 1 \ , & \ F^{(2|1)} = & F_0^{(2|1)} \ar[u] \ \ + & F_1^{(1|1)} \ar[u] \ar[ul]  \ \ + & F_2^{(0|1)} \ar[u] \ar[ul] \ \ + & F_3^{(-1|1)} \ar[u] \ar[ul] \\
		p = 2 \ , & \ F^{(2|2)} = & \ 0 \ar[u] \ \ + & \ 0 \ar[u] \ar[ul] \ \ + & F_2^{(0|2)} \ar[u] \ar[ul] \ \ + & F_3^{(-1|2)} \ar[u] \ar[ul] }
\end{equation*}
Thanks to the diagonal arrows, this diagram is meant to show that the contributions at lower picture may come from various terms. For example, we see that $F_1^{(1|0)}$ may receive contributions either from $F_2^{(0|2)}$ or from $F_3^{(-1|2)}$. This allows to better understand the nature of the \emph{conventional constraint}, which, for example, is already implemented at picture 2.

\section{Conclusions and Outlook}

We have discussed in detail the construction of the super-Cherns-Simons theory using the language of 
pseudoforms. We pointed out that the interaction term has to be built in terms of a non-associative product 
leading to a tower of interactions organized into a $A_\infty$ algebra. Finally, the compatibility with supersymmetry is studied. This is the starting point to several applications and follow-ups. Let us list some of them. 

\begin{enumerate}
\item As we have learnt from string theory, the introduction of PCO is due in RNS formalism \cite{FMS}, but 
also in pure spinor framework \cite{Berkovits:2004dt}. The present analysis is in part directly related to pure spinor formalism as was addressed some time ago, by one of the author and G. Policastro in \cite{Grassi:2004tv}, where it is shown how the two frameworks are related. Since, at the moment, we have understood some of the aspects of classical field theories on supermanifolds, it would be very interesting to explore again the relation between PCO's in string theory and in quantum field theory. 
\item One of the main motivation  to build the $2$-product in string field theory was to avoid the singularities 
emerging when two PCO's collide at the same point creating a potential divergence. 
We have seen from a preliminary work that the same phenomena is at work also in the present context. 
Feynman diagrams computations will be presented somewhere else \cite{cremonini-grassi}. 
\item One of the famous work by Witten on the relation between topological strings and Chern-Simons 
gauge theories \cite{Witten:1992fb} can be finally repeated in the context of supermanifolds. Until now, these aspects 
of string theory and topological strings on supermanifolds have never been explored and we hope that the present framework might be suitable to address these problems.  
\item One of the examples worked out in detail in \cite{Castellani:2016ibp} was the 
case of $D=3, N=1$ supergravity. It was shown that the non-factorized form of the action 
leads to the component action matching the superspace constructions. Nonetheless, it has not been 
explored the same situation for higher dimensional supergravity models. A crucial aspect for 
supergravity models is the fact that PCO depends upon the dynamical fields and therefore a deeper analysis 
must be completed. 
\item A long standing issue is the problem of auxiliary fields for higher dimensional and extended supersymmetry gauge theories and supergravity. We established a complete new framework 
to re-think to that old problem and 
might serve to build off-shell supersymmetric models with extended supersymmetry. 

\vspace{0.5cm}

What is rather striking is the comparison between the factorised form of the action and the non-factorised form. Apparently, all the complications arise from the infinite-dimensional nature of pseudoforms, therefore it is natural to wander whether a suitable field redefinition might immediately prove the equivalence of the two actions. Furthermore, we must investigate the theory at the quantum level where some potential singularities and divergences might jeopardise the classical equivalence relation. Nevertheless, we have shown that a very simple classical theory with some basic assumptions on the worldvolume supermanifold leads to a consistent algebraic structure of an $A_{\infty}$ Chern-Simons theory. That construction parallels the EKS construction without referring to any string theory, conformal field theory, Riemann surfaces or using other mathematical ingredients.
\end{enumerate}

\section*{Acknowledgements}
We thank S. Cacciatori,  L. Castellani, R. Catenacci,  A. Cattaneo,  and R. Re, for useful comments and 
discussions. We thank C. Maccaferri and T. Erler for important considerations about string field theory and 
string theory. We thank S. Noja for a careful reading of the manuscript and for mathematical reviews on supergeometry. 

\pagebreak
\appendix

\section{$A_\infty$ and $L_\infty$ Algebras}

Let us first recall the definitions of $A_\infty$ and $L_\infty$ algebras following \cite{Stasheff} but by using a modern language:

	\noindent an $A_{\infty}$\emph{-algebra}, or a \emph{strongly homotopy associative structure} on $V$, is a collection of linear maps $m_i: V^{\otimes^n} \to V$ such that they satisfy
	\begin{equation} \label{APPA}
		\sum_{\substack{1 \leq p \leq d \\ 0 \leq q \leq d - p}} (-1)^{|a_1| + \ldots + |a_q - q|} m_{d - p + 1} \left( a_d , \ldots , a_{p + q + 1} , m_{p} \left( a_{p+q} , \ldots , a_{q + 1} \right) , a_q , \ldots , a_1 \right) = 0 \ .
	\end{equation}

	\noindent An $L_{\infty}$\emph{-algebra}, or a \emph{strongly homotopy Lie structure} on $V$, is a collection of skew-symmetric linear maps $l_i: V^{\otimes^n} \to V$ such that they satisfy
	\begin{equation} \label{APPB}
	\sum_{\substack{j+k=i \\ 0 \sigma \in \text{Sh} (j;i)}} \chi ( \sigma ; a_1 , \ldots , a_i ) (-1)^k l_{k+1} \left( l_j \left( a_{\sigma (1)} , \ldots , a_{\sigma (j)} \right) , a_{\sigma (j+1)} , \ldots , a_{\sigma (i)} \right) = 0 \ ,
\end{equation}
where Sh$(j;i)$ are the permutations $\sigma$ of $\left\lbrace 1 , \ldots , i \right\rbrace$ such that $\sigma (1) < \ldots < \sigma (j)$ and $\sigma(j+1) < \ldots < \sigma(i)$ and $\chi ( \sigma ; a_1 , \ldots , a_i )$ is the $graded$ $Koszul$ $sign$ defined by
\begin{equation} \label{APPC}
	a_1 \wedge \ldots \wedge a_i = \chi ( \sigma ; a_1 , \ldots , a_i ) a_{\sigma (1)} \wedge \ldots \wedge a_{\sigma (i)} \ .
\end{equation}

In \cite{Stasheff} the author stated that it is possible to obtain a SHLS starting from a SHAS by a \virgolette commutation operation". This can be rephrased in a more modern language by following \cite{Loday-Vallette} where it is proved the following proposition:

	\noindent the functor
	\begin{equation}
		A_\infty \mathsf{alg} \to L_\infty \mathsf{alg} \ ,
	\end{equation}
	consists in antisymmetrising the $A_\infty$ operations $m_n$ to get the $L_\infty$ operations $l_m$.

This amounts to say that given an $A_\infty$ relation \eqref{APPA} and by antisymmetrising it, we get the $L_\infty$ relation \eqref{APPB}. Let us try to understand this result by considering the first steps of the relations. First of all, we can observe that an hint to this result is given by the fact that by antisymmetrising the associative property we get the Jacobi identity:
\begin{equation} \label{APPD}
	a ( b c ) - ( a b ) c = 0 \  \implies \ [ a , [ b , c ] ] + [ c , [ a , b ] ] + [ b , [ c , a ] ] = 0 \ ,
\end{equation}
as an easy exercise can show. Now let us consider the $A_\infty$ relations; the first one reads
\begin{equation} \label{APPE}
	m_1 ( m_1 ( a ) ) = 0 \ ,
\end{equation}
and simply turns into
\begin{equation} \label{APPF}
	l_1 ( l_1 ( a ) ) = 0 \ ,
\end{equation}
i.e. $l_1$ is a nilpotent operation. The second $A_\infty$ relation is
\begin{equation} \label{APPG}
	m_2 \left( a_2 , m_1 ( a_1 ) \right) + (-1)^{|a_1| + 1} m_2 \left( m_1 ( a_2 ) , a_1 \right) + m_1 \left( m_2 ( a_2 , a_1 ) \right) = 0 \ ,
\end{equation}
and by antisymmetrising it we get
\begin{equation} \label{APPH}
	m_2 \left( a_2 , m_1 ( a_1 ) \right) - m_2 \left( m_1 ( a_1 ) , a_2 \right) + (-1)^{|a_1| + 1} m_2 \left( m_1 ( a_2 ) , a_1 \right) - (-1)^{|a_1| + 1} m_2 \left( a_1 , m_1 ( a_2 ) \right) + $$ $$ + m_1 \left( m_2 ( a_2 , a_1 ) \right) - m_1 \left( m_2 ( a_1 , a_2 ) \right) = 0 \ \implies $$ $$ \implies \ l_2 \left( a_2 , l_1 ( a_1 ) \right) + (-1)^{|a_1| + 1} l_2 \left( l_1 ( a_2 ) , a_1 \right) + l_1 \left( l_2 ( a_2 , a_1 ) \right) = 0 \ ,
\end{equation}
having used the definition
\begin{equation} \label{APPI}
	l_2 ( A , B ) = m_2 ( A , B ) - m_2 ( B , A ) \ .
\end{equation}
Therefore eq. (A.9) shows that $l_1$ is a differential with respect to the product $l_2$. If we go on with the $A_\infty$ products to the 3-product, we find the law of "failure of associativity", i.e. the $associator$ $a ( b c ) - ( a b ) c$ is no longer equal to 0. By antisymmetrising the relation we find the law of "failure of Jacobi identity", i.e. the $Jacobiator$ $[ a, [ b,c]] + [b,[c,a]]+[c,[a,b]]$ is no longer 0. This may be guessed by looking at the easy relation shown in (A.5): as well as the antisymmetrisation of the associative property gives the Jacobi identity, the antisymmetrisation of the failure of associativity gives the failure of Jacobi identity. Let us see this explicitly, where, for the sake of clarity, we fix $|a_i|=0 \forall i \in \mathbb{N}$ in order to recover the expressions found in the previous sections. The third $A_\infty$ relation is
\begin{equation} \label{APPIA}
	m_3 \left( a_3 , a_2 , m_1 ( a_1 ) \right) - m_3 \left( a_3 , m_1 (a_2) , a_1 \right) + m_3 \left( m_1 (a_3) , a_2 , a_1 \right) + m_1 \left( m_3 \left( a_3 , a_2 , a_1 \right) \right) + $$ $$ + m_2 \left( a_3 , m_2 \left( a_2 , a_1 \right) \right) - m_2 \left( m_2 \left( a_3 , a_2 \right) , a_1 \right) = 0 \ ,
\end{equation}
and shows explicitly the non-vanishing of the associator (the second line). Let us define the 3-product of the $L_\infty$-algebra by antisymmetrising the $m_3^{(-2)}$ product:
\begin{align} \label{AAPJ}
	\nonumber l_3^{(-2)} \left( A , B , C \right) = & m_3^{(-2)} \left( A , B , C \right) - m_3^{(-2)} \left( A , C , B \right) + m_3^{(-2)} \left( B , C , A \right) + \\ & - m_3^{(-2)} \left( B , A , C \right) + m_3^{(-2)} \left( C , A , B \right) - m_3^{(-2)} \left( C , B , A \right) \ .
\end{align}
We can now antisymmetrise \eqref{APPIA}, and an easy calculation shows that we get
\begin{equation} \label{APPK}
	l_3 \left( a_3 , a_2 , l_1 ( a_1 ) \right) + - l_3 \left( a_3 , l_1 (a_2) , a_1 \right) + l_3 \left( l_1 (a_3) , a_2 , a_1 \right) + l_1 \left( l_3 \left( a_3 , a_2 , a_1 \right) \right) + $$ $$ + l_2 \left( a_3 , l_2 \left( a_2 , a_1 \right) \right) + l_2 \left( a_1 , l_2 \left( a_3 , a_2 \right) \right) + l_2 \left( a_2 , l_2 \left( a_1 , a_3 \right) \right) = 0 \ .
\end{equation}
\eqref{APPK} shows explicitly the non-vanishing of the Jacobiator (the second line) as a direct consequence of the non-vanishing of the associator of the relative $A_{\infty}$ relation.

\section{Explicit calculations}

In this appendix we present the explicit calculations relative to the reduction of the equations of motion and to the interaction term.

\subsection{Reducing the Equations of Motion}

In this subsection we explain in deep details how to reduce the equations of motion n order to determine the cohomology representative fields of the presudoform $A^{(1|1)}$. In subsection 4.5 we have announced the strategy to be used and the result obtained, here we show the useful passages. First of all let us consider the expansion in powers of $\theta$ of any field:
\begin{equation}\label{SEMA}
	A (x , \theta) = \tilde{A} (x) + \theta^\alpha \tilde{B} (x) + \theta^\beta \tilde{C} (x) + \theta^\alpha \theta^\beta \tilde{D} (x) \ .
\end{equation}
We start by applying this expansion to (\ref{EMAD}):
\begin{equation}
	- \tilde{C}^{(p)}_{\beta \alpha} (x) + \theta^\alpha \tilde{D}^{(p)}_{\beta \alpha} + (p + 1 ) \tilde{B}^{(p+1)}_{\beta \alpha} + (p+1) \theta^\beta \tilde{D}^{(p+1)}_{\beta \alpha} = 0 \ .
\end{equation}
We can separately equal to 0 the different coefficients of the monomials in $\theta$, thus obtaining
\begin{equation}\label{SEMB}
	\begin{cases}
		\tilde{D}^{(p)}_{\beta \alpha} = 0 \\
		(p + 1 ) \tilde{B}^{(p+1)}_{\beta \alpha} = \tilde{C}^{(p)}_{\beta \alpha} (x)
	\end{cases}
	 \ \forall p \in \mathbb{N} \ .
\end{equation}
By inserting this result back in (\ref{SEMA}) we get
\begin{equation}\label{SEMC}
	A^{(p)}_{\beta \alpha} (x , \theta) = \tilde{A}^{(p)}_{\beta \alpha} (x) + \theta^\alpha \tilde{B}^{(p)}_{\beta \alpha} (x) + \theta^\beta (p+1) \tilde{B}^{(p+1)}_{\beta \alpha} (x) \ .
\end{equation}
Now we insert (\ref{SEMA}) in (\ref{EMAE}):
\begin{equation}
	\partial_r \tilde{A}_{\beta \alpha}^{(p)} + \theta^\alpha \partial_r \tilde{B}_{\beta \alpha}^{(p)} + \theta^\beta \partial_r \tilde{C}_{\beta \alpha}^{(p)} + \theta^\alpha \theta^\beta \partial_r \tilde{D}_{\beta \alpha}^{(p)} - \tilde{C}_{r \beta \alpha}^{(p)} + \theta^\alpha \tilde{D}_{r \beta \alpha}^{(p)} + (p+1) \tilde{B}_{r \beta \alpha}^{(p+1)} + (p+1) \theta^\beta \tilde{D}_{r \beta \alpha}^{(p+1)} = 0 \ ,
\end{equation}
and by substituting what we got in (\ref{SEMB}) we obtain
\begin{equation}
	\partial_r \tilde{A}_{\beta \alpha}^{(p)} + \theta^\alpha \partial_r \tilde{B}_{\beta \alpha}^{(p)} + \theta^\beta (p+1) \partial_r \tilde{B}_{\beta \alpha}^{(p+1)} - \tilde{C}_{r \beta \alpha}^{(p)} + \theta^\alpha \tilde{D}_{r \beta \alpha}^{(p)} + (p+1) \tilde{B}_{r \beta \alpha}^{(p+1)} + (p+1) \theta^\beta \tilde{D}_{r \beta \alpha}^{(p+1)} = 0 \ .
\end{equation}
Again, we separate this equation in homogeneous polynomials in $\theta$:
\begin{equation}\label{SEMD}
	\begin{cases}
		\partial_r \tilde{A}_{\beta \alpha}^{(p)} - \tilde{C}_{r \beta \alpha}^{(p)} + (p+1) \tilde{B}_{r \beta \alpha}^{(p+1)} = 0 \\
		\partial_r \tilde{B}_{\beta \alpha}^{(p)} + \tilde{D}_{r \beta \alpha}^{(p)} = 0 \\
		(p+1) \partial_r \tilde{B}_{\beta \alpha}^{(p+1)} + (p+1) \tilde{D}_{r \beta \alpha}^{(p+1)} = 0 \\
	\end{cases} \ \forall p \in \mathbb{N} \ .
\end{equation}
Since these expressions are valid $\forall p \in \mathbb{N}$, the second and the third equations are the same. The second equation is used in order to find a formal expression of $\tilde{D}_{r \beta \alpha}^{(p)}$ in terms of $\tilde{B}_{r \beta \alpha}^{(p)}$, while the first gives us a formal recursive relation between $\tilde{C}_{r \beta \alpha}^{(p)}$ and $\partial_r \tilde{A}_{\beta \alpha}^{(p)} + (p+1) \tilde{B}_{r \beta \alpha}^{(p+1)}$. Therefore when inserting these results back in (\ref{SEMA}) we get
\begin{equation}\label{SEME}
	A_{r \beta \alpha}^{(p)} (x , \theta) = \tilde{A}_{r \beta \alpha}^{(p)} (x) + \theta^\alpha \tilde{B}_{r \beta \alpha}^{(p)} (x) + \theta^\beta \left( \partial_r \tilde{A}_{\beta \alpha}^{(p)} (x) + (p+1) \tilde{B}_{r \beta \alpha}^{(p+1)} (x) \right) - \theta^\alpha \theta^\beta \partial_r \tilde{B}_{\beta \alpha}^{(p)} (x) \ .
\end{equation}
Now we substitute (\ref{SEMA}) in (\ref{EMAF}):
\begin{equation}
	\partial_{[r} \tilde{A}_{n] \beta \alpha}^{(p)} + \theta^\alpha \partial_{[r} \tilde{B}_{n] \beta \alpha}^{(p)} + \theta^\beta \partial_{[r} \tilde{C}_{n] \beta \alpha}^{(p)} + \theta^\alpha \theta^\beta \partial_{[r} \tilde{D}_{n] \beta \alpha}^{(p)} - \tilde{C}_{[n r] \beta \alpha}^{(p-1)} + \theta^\alpha \tilde{D}_{[n r] \beta \alpha}^{(p-1)} + (p+1) \tilde{B}_{[n r] \beta \alpha}^{(p)} + (p+1) \theta^\beta \tilde{D}_{[n r] \beta \alpha}^{(p)} = 0 \ ,
\end{equation}
and substituting $\tilde{C}_{n \beta \alpha}^{(p)}$ and $\tilde{D}_{n \beta \alpha}^{(p)}$ from (\ref{SEMD}) we get
\begin{equation}
	\partial_{[r} \tilde{A}_{n] \beta \alpha}^{(p)} + \theta^\alpha \partial_{[r} \tilde{B}_{n] \beta \alpha}^{(p)} + \theta^\beta (p+1) \partial_{[r} \tilde{B}_{n] \beta \alpha}^{(p+1)} - \tilde{C}_{[n r] \beta \alpha}^{(p-1)} + \theta^\alpha \tilde{D}_{[n r] \beta \alpha}^{(p-1)} + (p+1) \tilde{B}_{[n r] \beta \alpha}^{(p)} + (p+1) \theta^\beta \tilde{D}_{[n r] \beta \alpha}^{(p)} = 0 \ ,
\end{equation}
By separating the homogeneous terms in $\theta$ we get
\begin{equation}\label{SEMF}
	\begin{cases}
		\partial_{[r} \tilde{A}_{n] \beta \alpha}^{(p)} - \tilde{C}_{[n r] \beta \alpha}^{(p-1)} + (p+1) \tilde{B}_{[n r] \beta \alpha}^{(p)} = 0 \\
		\partial_{[r} \tilde{B}_{n] \beta \alpha}^{(p)} + \tilde{D}_{[n r] \beta \alpha}^{(p-1)} = 0 \\
		(p+1) \partial_{[r} \tilde{B}_{n] \beta \alpha}^{(p+1)} + (p+1) \tilde{D}_{[n r] \beta \alpha}^{(p)} = 0
	\end{cases} \ \forall p \in \mathbb{N} \ .
\end{equation}
As in (\ref{SEMD}), the second and the third equations are almost the same; however in (\ref{SEMF}) there is a slight difference: the case $p=0$ decouples from the other cases as
\begin{equation}\label{SEMG}
	\begin{cases}
		\partial_{[r} \tilde{A}_{n] \beta \alpha}^{(0)} + \tilde{B}_{[n r] \beta \alpha}^{(0)} = 0 \\
		\partial_{[r} \tilde{B}_{n] \beta \alpha}^{(0)} = 0
	\end{cases} \ .
\end{equation}
This will prove of fundamental importance as we will see shortly. For $p \neq 0$ the second and third equations in (\ref{SEMF}) are the same. We can insert these results back in (\ref{SEMA}) but now we keep the $p=0$ case separated:
\begin{equation}\label{SEMH}
	A_{[m n] \beta \alpha}^{(0)} (x , \theta) = \tilde{A}_{[m n] \beta \alpha}^{(0)} (x) - \theta^\alpha \partial_{[n} \tilde{A}_{m] \beta \alpha}^{(0)} (x) + \theta^\beta \partial_{[n} \tilde{A}_{m] \beta \alpha}^{(1)} (x) + 2 \theta^\beta \tilde{B}_{[m n] \beta \alpha}^{(1)} (x) - \theta^\alpha \theta^\beta \partial_{[n} \tilde{B}_{m] \beta \alpha}^{(1)} (x) \ ;
\end{equation}
\begin{equation}\label{SEMI}
	A_{[m n] \beta \alpha}^{(p)} (x , \theta) = \tilde{A}_{[m n] \beta \alpha}^{(p)} (x) + \theta^\alpha \tilde{B}_{[m n] \beta \alpha}^{(p)} (x) + \theta^\beta \partial_{[n} \tilde{A}_{m] \beta \alpha}^{(p+1)} (x) + (p+2) \theta^\beta \tilde{B}_{[m n] \beta \alpha}^{(p+1)} (x) - \theta^\alpha \theta^\beta \partial_{[n} \tilde{B}_{m] \beta \alpha}^{(p+1)} (x) \ .
\end{equation}
We are left with the substitution of (\ref{SEMA}) in (\ref{EMAG}):
\begin{equation}\label{SEMJ}
	\partial_{[r} \tilde{A}_{m n] \beta \alpha}^{(p)} + \theta^\alpha \partial_{[r} \tilde{B}_{m n] \beta \alpha}^{(p)} + \theta^\beta \partial_{[r} \tilde{C}_{m n] \beta \alpha}^{(p)} + \theta^\alpha \theta^\beta \partial_{[r} \tilde{D}_{m n] \beta \alpha}^{(p)} - \tilde{C}_{[m n r] \beta \alpha}^{(p-1)} + \theta^\alpha \tilde{D}_{[m n r] \beta \alpha}^{(p-1)} + $$ $$ + (p+2) \tilde{B}_{[m n r] \beta \alpha}^{(p)} + \theta^\beta (p+2) \tilde{D}_{[m n r] \beta \alpha}^{(p)} = 0 \ .
\end{equation}
Instead the $p=0$ case reads:
\begin{equation}
	\partial_{[r} \tilde{A}_{m n] \beta \alpha}^{(0)} + \theta^\beta 2 \partial_{[r} \tilde{B}_{m n] \beta \alpha}^{(1)} + 2 \tilde{B}_{[m n r] \beta \alpha}^{(0)} + \theta^\beta 2 \tilde{D}_{[m n r] \beta \alpha}^{(0)} = 0 \ .
\end{equation}
By separating the homogeneous terms in $\theta$ we get
\begin{equation}\label{SEMK}
	\begin{cases}
		\partial_{[r} \tilde{A}_{m n] \beta \alpha}^{(0)} + 2 \tilde{B}_{[m n r] \beta \alpha}^{(0)} = 0 \\
		2 \partial_{[r} \tilde{B}_{m n] \beta \alpha}^{(1)} + 2 \tilde{D}_{[m n r] \beta \alpha}^{(0)} = 0
	\end{cases} \ .
\end{equation}
By substituting in (\ref{SEMA}) we get
\begin{equation}\label{SEML}
	A_{[m n r] \beta \alpha}^{(0)} (x , \theta) = \tilde{A}_{[m n r] \beta \alpha}^{(0)} (x) - \theta^\alpha \frac{1}{2} \partial_{[r} \tilde{A}_{m n] \beta \alpha}^{(0)} (x) + \theta^\beta \partial_{[r} \tilde{A}_{m n] \beta \alpha}^{(1)} (x) + \theta^\beta 3 \tilde{B}_{[m n r] \beta \alpha}^{(1)} (x) - \theta^\alpha \theta^\beta \partial_{[r} \tilde{B}_{m n] \beta \alpha}^{(1)} (x) \ .
\end{equation}
Finally let us consider the case $p \neq 0$, eq. (\ref{SEMJ}) becomes
\begin{equation}
	\partial_{[r} \tilde{A}_{m n] \beta \alpha}^{(p)} + \theta^\alpha \partial_{[r} \tilde{B}_{m n] \beta \alpha}^{(p)} + \theta^\beta (p+2) \partial_{[r} \tilde{B}_{m n] \beta \alpha}^{(p+1)} - \tilde{C}_{[m n r] \beta \alpha}^{(p-1)} + \theta^\alpha \tilde{D}_{[m n r] \beta \alpha}^{(p-1)} + (p+2) \tilde{B}_{[m n r] \beta \alpha}^{(p)} + \theta^\beta (p+2) \tilde{D}_{[m n r] \beta \alpha}^{(p)} = 0 \ .
\end{equation}
As usual we separate the homogeneous terms in $\theta$ and get
\begin{equation}\label{SEMM}
	\begin{cases}
		\partial_{[r} \tilde{A}_{m n] \beta \alpha}^{(p)} - \tilde{C}_{[m n r] \beta \alpha}^{(p-1)} + (p+2) \tilde{B}_{[m n r] \beta \alpha}^{(p)} = 0 \\
		\partial_{[r} \tilde{B}_{m n] \beta \alpha}^{(p)} + \tilde{D}_{[m n r] \beta \alpha}^{(p-1)} = 0 \\
		(p+2) \partial_{[r} \tilde{B}_{m n] \beta \alpha}^{(p+1)} + (p+2) \tilde{D}_{[m n r] \beta \alpha}^{(p)} = 0  \ .
	\end{cases}
\end{equation}
Since we have $p \neq 0$, the second and third equations are exactly the same. By substituting back in (\ref{SEMA}) we get
\begin{equation}\label{SEMN}
	A_{[m n r] \beta \alpha}^{(p)} (x , \theta) = \tilde{A}_{[m n r] \beta \alpha}^{(p)} (x) + \theta^\alpha \tilde{B}_{[m n r] \beta \alpha}^{(p)} (x) + \theta^\beta \partial_{[r} \tilde{A}_{m n] \beta \alpha}^{(p+1)} (x) + \theta^\beta (p+3) \tilde{B}_{[m n r] \beta \alpha}^{(p+1)} (x) - \theta^\alpha \theta^\beta \partial_{[r} \tilde{B}_{m n] \beta \alpha}^{(p+1)} (x) \ .
\end{equation}
We now have to insert all the expressions found for the fields in (\ref{TLAAA}) $\div$ (\ref{TLAAD}) in order to see the combinations that are $d-$exact.
Let us forget for a while the expressions for $\tilde{B}_{m n \alpha \beta}^{(0)}$ and $\tilde{B}_{m n r \alpha \beta}^{(0)}$ that led to the separations of the expressions for $A_{m n \alpha \beta}^{(0)}$ and $A_{m n r \alpha \beta}^{(0)}$ and insert (\ref{SEMC}), (\ref{SEME}), (\ref{SEMI}) and (\ref{SEMN}) in (\ref{TLAAA}) $\div$ (\ref{TLAAD}). We get:
\begin{eqnarray}
	\label{SEMO} A_0 &=& \sum_{p=0}^{\infty} \left( \tilde{A}^{(p)}_{\alpha \beta} (x) + \theta^\beta \tilde{B}^{(p)}_{\alpha \beta} (x) + \theta^\alpha (p+1) \tilde{B}^{(p+1)}_{\alpha \beta} (x) \right) ( d \theta^\alpha )^{p+1} \delta^{(p)} ( d \theta^\beta ) \ ; \\
	\nonumber A_1 &=& \sum_{p=0}^{\infty} dx^m \left( \tilde{A}_{m \alpha \beta}^{(p)} (x) + \theta^\beta \tilde{B}_{m \alpha \beta}^{(p)} (x) + \theta^\alpha \left( \partial_m \tilde{A}_{\alpha \beta}^{(p)} (x) + (p+1) \tilde{B}_{m \alpha \beta}^{(p+1)} (x) \right) + \right. \\ \label{SEMP} &&- \theta^\beta \theta^\alpha \partial_m \tilde{B}_{\alpha \beta}^{(p)} (x) \Big) ( d \theta^\alpha )^{p} \delta^{(p)} ( d \theta^\beta ) \ ; \\
	\nonumber A_2 &=& \sum_{p=0}^{\infty} dx^m dx^n \left( \tilde{A}_{[m n] \alpha \beta}^{(p)} (x) + \theta^\beta \tilde{B}_{[m n] \alpha \beta}^{(p)} (x) + \theta^\alpha \left( \partial_{[n} \tilde{A}_{m] \alpha \beta}^{(p+1)} (x) + (p+2) \tilde{B}_{[m n] \alpha \beta}^{(p+1)} (x) \right) + \right. \\ \label{SEMQ} &&- \theta^\beta \theta^\alpha \partial_{[n} \tilde{B}_{m] \alpha \beta}^{(p+1)} (x) \Big) ( d \theta^\alpha )^{p} \delta^{(p + 1)} ( d \theta^\beta ) \ ; \\
	\nonumber A_3 &=& \sum_{p=0}^{\infty} dx^m   dx^n   dx^r \left( \tilde{A}_{[m n r] \alpha \beta}^{(p)} (x) + \theta^\beta \tilde{B}_{[m n r] \alpha \beta}^{(p)} (x) + \theta^\alpha \left( \partial_{[r} \tilde{A}_{m n] \alpha \beta}^{(p+1)} (x) + (p+3) \tilde{B}_{[m n r] \alpha \beta}^{(p+1)} (x) \right) + \right. \\ \label{SEMR} &&- \theta^\beta \theta^\alpha \partial_{[r} \tilde{B}_{m n] \alpha \beta}^{(p+1)} (x) \Big) ( d \theta^\alpha )^{p} \delta^{(p + 2)} ( d \theta^\beta ) \ .
\end{eqnarray}
It is now a matter of rearranging all the terms correctly; for example, consider an expression like
\begin{equation}
	d \left( \tilde{A}_{\alpha \beta}^{(p)} \theta^\alpha (d \theta^\alpha )^p \delta (d \theta^\beta )^{(p)} \right) = dx^m \theta^\alpha \partial_m \tilde{A}_{\alpha \beta}^{(p)} (d \theta^\alpha )^p \delta (d \theta^\beta )^{(p)} + \tilde{A}_{\alpha \beta}^{(p)} (d \theta^\alpha )^{p+1} \delta (d \theta^\beta )^{(p)} \ ,
\end{equation}
where we have used the fact that $\tilde{A}_{\alpha \beta}^{(p)}$ is even. This means that the first term in (\ref{SEMO}) and the third term in (\ref{SEMP}) arrange in a $d-$exact term. This means that we can omit them in order to get the right cohomological field. Let us now consider an expression like
\begin{align}\label{SEMS}
	\nonumber d \left( - \theta^\beta \theta^\alpha \tilde{B}_{\alpha \beta}^{(p)} ( d \theta^\alpha )^p \delta^{(p)} (d \theta^\beta ) \right) = & p \theta^\alpha \tilde{B}_{\alpha \beta}^{(p)} ( d \theta^\alpha )^p \delta^{(p-1)} (d \theta^\beta ) + \theta^\beta \tilde{B}_{\alpha \beta}^{(p)} ( d \theta^\alpha )^{p+1} \delta^{(p)} (d \theta^\beta ) + \\
	& - dx^m \theta^\beta \theta^\alpha \partial_m \tilde{B}_{\alpha \beta}^{(p)} ( d \theta^\alpha )^p \delta^{(p)} (d \theta^\beta ) \ .
\end{align}
This means that we can arrange the second and third terms from (\ref{SEMO}) and the last term from (\ref{SEMP}) in a $d$-exact term. Observe that in order to arrange the terms correctly, we have to shift the first term in (\ref{SEMS}) $p \to p+1$ and the expression is valid for $p=0$ as well.

Let us now consider an expression like
\begin{equation}\label{SEMT}
	d \left( \theta^\alpha \tilde{A}_{m \alpha \beta}^{(p)} ( d \theta^\alpha )^{p-1} \delta^{(p)} ( d \theta^\beta ) \right) = \tilde{A}_{m \alpha \beta}^{(p)} ( d \theta^\alpha )^{p} \delta^{(p)} ( d \theta^\beta ) + dx^n \theta^\alpha \partial_{[n} \tilde{A}_{m] \alpha \beta}^{(p)} ( d \theta^\alpha )^{p-1} \delta^{(p)} ( d \theta^\beta ) \ ;
\end{equation}
this means that we can arrange the first term in (\ref{SEMP}) and the third term in (\ref{SEMQ}) as a $d$-exact term. Even in this case we have to make a shift on the second term of (\ref{SEMT}). Moreover the previous statement is not valid for the $p=0$ term; in that case we have to consider an expression like
\begin{align}
	\nonumber d \left( - \theta^\beta \tilde{A}_{m \alpha \beta}^{(0)} \delta^{(1)} ( d \theta^\beta ) \right) = & - d \theta^\beta \tilde{A}_{m \alpha \beta}^{(0)} \delta^{(1)} ( d \theta^\beta ) - dx^n \theta^\beta \partial_{[n} \tilde{A}_{m] \alpha \beta}^{(0)} \delta^{(1)} ( d \theta^\beta ) = \\
	= & \tilde{A}_{m \alpha \beta}^{(0)} \delta^{(0)} ( d \theta^\beta ) - dx^n \theta^\beta \partial_{[n} \tilde{A}_{m] \alpha \beta}^{(0)} \delta^{(1)} ( d \theta^\beta ) \ ,
\end{align}
which is exactly the separated $p=0$ term.

Let us now consider an expression like
\begin{align}
	\nonumber d \left( - \theta^\beta \theta^\alpha \tilde{B}_{m \alpha \beta}^{(p)} ( d \theta^\alpha)^{p-1} \delta^{(p)} ( d \theta^\beta ) \right) = & \theta^\alpha p \tilde{B}_{m \alpha \beta}^{(p)} ( d \theta^\alpha)^{p-1} \delta^{(p-1)} ( d \theta^\beta ) + \theta^\beta \tilde{B}_{m \alpha \beta}^{(p)} ( d \theta^\alpha)^{p} \delta^{(p)} ( d \theta^\beta ) + \\
	& - dx^n \theta^\beta \theta^\alpha \partial_{[n} \tilde{B}_{m] \alpha \beta}^{(p)} ( d \theta^\alpha)^{p-1} \delta^{(p)} ( d \theta^\beta ) \ .
\end{align}
This allows us to fix as a $d$-exact term the second and fourth terms from (\ref{SEMP}) together with the last term from (\ref{SEMQ}), but we have a fundamental observation to do: the previous relation allows us to fix the described terms after a shift $p \to p+1$. Notice that the term $dx^m \theta^\beta \tilde{B}_{m \alpha \beta}^{(0)} (x) \delta^{(0)} ( d \theta^\beta )$ remains free.

Let us now consider an expression like
\begin{equation}\label{SEMU}
	d \left( \tilde{A}_{[m n] \alpha \beta}^{(p)} \theta^\alpha (d \theta^\alpha )^{p-1} \delta ( d \theta^\beta )^{(p+1)} \right) = dx^r \theta^\alpha \partial_{[r} \tilde{A}_{m n] \alpha \beta}^{(p)} (d \theta^\alpha )^{p-1} \delta ( d \theta^\beta )^{(p+1)} + \tilde{A}_{[m n] \alpha \beta}^{(p)} (d \theta^\alpha )^{p} \delta ( d \theta^\beta )^{(p+1)} \ ;
\end{equation}
this means that we can arrange the first term in (\ref{SEMQ}) and the third term in (\ref{SEMR}) as a $d$-exact term. As we previously noticed, we have to make a shift on the first term of (\ref{SEMU}). Moreover the previous statement is not valid for the $p=0$ term; in that case we have to consider an expression like
\begin{equation}
	d \left( - \frac{1}{2} \theta^\beta \tilde{A}_{[m n] \alpha \beta}^{(0)} \delta^{(2)} ( d \theta^\beta ) \right) = \tilde{A}_{[m n] \alpha \beta}^{(0)} \delta^{(1)} ( d \theta^\beta ) - \frac{1}{2} dx^r \theta^\beta \partial_{[r} \tilde{A}_{m n] \alpha \beta}^{(0)} \delta^{(2)} ( d \theta^\beta ) \ ,
\end{equation}
which is exactly the separated $p=0$ term.

Let us now consider an expression like
\begin{align}
	\nonumber d \left( - \theta^\beta \theta^\alpha \tilde{B}_{[m n] \alpha \beta}^{(p)} ( d \theta^\alpha )^{p-1} \delta^{(p+1)} ( d \theta^\beta ) \right) = & (p+1) \theta^\alpha \tilde{B}_{[m n] \alpha \beta}^{(p)} ( d \theta^\alpha )^{p-1} \delta^{(p)} ( d \theta^\beta ) + \theta^\beta \tilde{B}_{[m n] \alpha \beta}^{(p)} ( d \theta^\alpha )^{p} \delta^{(p+1)} ( d \theta^\beta ) + \\ & - dx^r \theta^\beta \theta^\alpha \partial_{[r} \tilde{B}_{m n] \alpha \beta}^{(p)} ( d \theta^\alpha )^{p-1} \delta^{(p+1)} ( d \theta^\beta ) \ .
\end{align}
This expression allows us to fix as $d$-exact the second and fourth terms from (\ref{SEMQ}) together with the last of (\ref{SEMR}). Let us now consider the expression
\begin{equation}
	d \left( - \frac{1}{p+3} \theta^\beta \tilde{A}_{[m n r] \alpha \beta}^{(p)} (d \theta^\alpha )^{p} \delta^{(p + 3)} ( d \theta^{\beta} ) \right) = \tilde{A}_{[m n r] \alpha \beta}^{(p)} (d \theta^\alpha )^{p} \delta^{(p + 2)} ( d \theta^{\beta} ) \ ,
\end{equation}
i.e. the first term in (\ref{SEMR}) is trivially $d$-exact. Finally, let us consider the expression
\begin{equation}
	d \left( - \theta^\beta \theta^\alpha \tilde{B}_{[m n r] \alpha \beta}^{(p)} (d \theta^\alpha )^{p} \delta^{(p + 3)} ( d \theta^{\beta} ) \right) = (p+3) \theta^\alpha \tilde{B}_{[m n r] \alpha \beta}^{(p)} (d \theta^\alpha )^{p} \delta^{(p + 2)} ( d \theta^{\beta} ) + \theta^\beta \tilde{B}_{[m n r] \alpha \beta}^{(p)} (d \theta^\alpha )^{p+1} \delta^{(p + 3)} ( d \theta^{\beta} ) \ ;
\end{equation}
this allows as to arrange the second and third terms of (\ref{SEMR}) as $d$-exact. Even in this case it is necessary to do a shift $p \to p+1$.

Thus we have found that, modulo $d-$exact terms, a pseudoform $A^{1|1}$ of the Chern-Simons Lagrangian which is a representative of the cohomology class is
\begin{equation}
	A^{1|1} = dx^m \theta^\beta \tilde{B}_{m \alpha \beta}^{(0)} (x) \delta^{(0)} ( d \theta^\beta ) \ ,
\end{equation}
and the relative equation of motion is
\begin{equation}
	\partial_{[n} \tilde{B}_{m] \alpha \beta}^{(0)} (x) = 0 \ .
\end{equation}
Moreover we have obtained that the free Super Chern-Simons action with a general $A^{1|1}$ pseudoform leads to
\begin{equation}
	A^{1|1} = A^{1|0} \wedge \mathbb{Y}^{0|1} \ , \ \text{ s.t. } \mathbb{Y}^{0|1} = \theta^\beta \delta ( d \theta^\beta ) + d \Omega^{-1|1} \ .
\end{equation}

\subsection{The Interaction Term}

In this subsection we determine the explicit expression for the interaction term announced in subsection 4.8. In order to do so, we recall that a general $(1|1)$-pseudoform in $\mathcal{SM}^{(3|2)}$ is expanded as seen in \eqref{TLAAA} $\div$ \eqref{TLAAD}.
Let us apply the operator $\Theta (\iota_v)$ to these expressions:
\begin{align}
	A_0 : \ \  \Theta (\iota_v) A_0 = \sum_{p=0}^{\infty} A_{\alpha \beta}^{(p)} \Theta (\iota_v) ( d \theta^\alpha )^{p+1} \delta^{(p)} ( d \theta^\beta ) \ ,
\end{align}
since the operator $\Theta (\iota_v)$ acts only on the $d \theta \delta$ parts; we can now use \eqref{T2} in order to get
\begin{equation} \label{APF}
	\Theta (\iota_v) A_0 = - \sum_{p=0}^{\infty} A_{\alpha \beta}^{(p)} i (-1)^p p! \left[ \left( \frac{v^\alpha}{v^\beta} \right)^{p+1} - \left( \frac{d \theta^\alpha}{d \theta^\beta} \right)^{p+1} \right] \ .
\end{equation}
For the action of $\Theta$ on $A_1 , A_2 , A_3$ we can make use of \eqref{T1}:
\begin{align}
	\label{APG} A_1 &: \ \  \Theta (\iota_v) A_1 = \sum_{p=0}^{\infty} dx^m A_{m \alpha \beta}^{(p)} i (-1)^p p! \frac{(d \theta^\alpha)^p}{( d \theta^\beta )^{p+1}} \ ; \\
	\label{APH} A_2 &: \ \  \Theta (\iota_v) A_2 = \sum_{p=0}^{\infty} dx^m   dx^n A_{[m n] \alpha \beta}^{(p)} i (-1)^{p+1} (p+1)! \frac{(d \theta^\alpha)^p}{( d \theta^\beta )^{p+2}} \ ; \\
	\label{API} A_3 &: \ \  \Theta (\iota_v) A_3 = \sum_{p=0}^{\infty} dx^m   dx^n   dx^r A_{[m n r] \alpha \beta}^{(p)} i (-1)^p (p+2)! \frac{(d \theta^\alpha)^p}{( d \theta^\beta )^{p+3}} \ .
\end{align}
Now we want to find the expression for $A \wedge A$; due to the relation $d \theta^\alpha \delta ( d \theta^\alpha ) = 0 = d \theta^\beta \delta ( d \theta^\beta )$, we can expand $A \wedge A$ explicitly as
\begin{equation} \label{APJ}
	A \wedge A = A_0 \wedge A_2 + A_1 \wedge A_1 + A_2 \wedge A_0 \ .
\end{equation}
We can evaluate the terms of \eqref{APJ} by matching the power of $d \theta$ and the derivation order of $\delta$ as usual:
\begin{align}
	A_0 \wedge A_2 &= - \sum_{p=0}^\infty dx^m \wedge dx^n p! (p+1)! A_{\alpha \beta}^{(p)} A_{[m n] \beta \alpha}^{(p)} \delta ( d \theta^\beta ) \delta ( d \theta^\alpha ) \ , \\
	A_2 \wedge A_0 &= - \sum_{p=0}^\infty dx^m \wedge dx^n p! (p+1)! A_{[m n] \alpha \beta}^{(p)} A_{\beta \alpha}^{(p)} \delta ( d \theta^\beta ) \delta ( d \theta^\alpha ) \ ; \\
	A_1 \wedge A_1 &= - \sum_{p=0}^\infty dx^m \wedge dx^n p! p! A_{[m \alpha \beta}^{(p)} A_{n] \beta \alpha}^{(p)} \delta ( d \theta^\beta ) \delta ( d \theta^\alpha ) \ ,
\end{align}
Therefore $A \wedge A$ reads
\begin{equation} \label{APK}
	A \wedge A = - \sum_{p=0}^\infty dx^m \wedge dx^n p! p! \left[ (p+1) \left[ A_{[m n] \alpha \beta}^{(p)} , A_{\beta \alpha}^{(p)} \right] + A_{[m \alpha \beta}^{(p)} A_{n] \beta \alpha}^{(p)} \right] \delta ( d \theta^\beta ) \delta ( d \theta^\alpha ) \ .
\end{equation}
In order to evaluate the action of the operator $\Theta$ on \eqref{APK}, we can use \eqref{T4}:
\begin{equation} \label{APL}
	\Theta (\iota_v) A \wedge A = \sum_{p=0}^\infty dx^m \wedge dx^n p! p! \left[ (p+1) \left[ A_{[m n] \alpha \beta}^{(p)} , A_{\beta \alpha}^{(p)} \right] + A_{[m \alpha \beta}^{(p)} A_{n] \beta \alpha}^{(p)} \right] \frac{i v^\alpha}{ d \theta^\alpha} \delta \left( \epsilon^{\alpha \beta} v \cdot d \theta \right) \ .
\end{equation}
Now that we have evaluated the action of $\Theta$ on all the pieces appearing in the product $m_2$, we have to apply the exterior derivative $d$ to those expressions in order to obtain a formula for the first part of the anticommutator defining the operator $Z_v$. Therefore, \eqref{APF} $\div$ \eqref{API} become
\begin{align}
	\nonumber d \Theta (\iota_v) A_0 =& - \sum_{p=0}^{\infty} i (-1)^p p! \left[ dx^m \partial_m A_{\alpha \beta}^{(p)}  \left[ \left( \frac{v^\alpha}{v^\beta} \right)^{p+1} - \left( \frac{d \theta^\alpha}{d \theta^\beta} \right)^{p+1} \right] + \right. \\
	\label{APM} & \left. + \partial_\alpha A_{\alpha \beta}^{(p)}  \left[ d \theta^\alpha \left( \frac{v^\alpha}{v^\beta} \right)^{p+1} - d \theta^\alpha \left( \frac{d \theta^\alpha}{d \theta^\beta} \right)^{p+1} \right] + \partial_\beta A_{\alpha \beta}^{(p)}  \left[ d \theta^\beta \left( \frac{v^\alpha}{v^\beta} \right)^{p+1} - d \theta^\beta \left( \frac{d \theta^\alpha}{d \theta^\beta} \right)^{p+1} \right] \right] \ ; \\
	\label{APN} d \Theta (\iota_v) A_1 =& - \sum_{p=0}^{\infty} i (-1)^p p! \left[ dx^m \wedge dx^n \partial_{[n} A_{m] \alpha \beta}^{(p)} \frac{(d \theta^\alpha)^p}{( d \theta^\beta )^{p+1}} + dx^m \partial_\alpha A_{m \alpha \beta}^{(p)} \left( \frac{d \theta^\alpha}{d \theta^\beta} \right)^{p+1} + dx^m \partial_\beta A_{m \alpha \beta}^{(p)} \left( \frac{d \theta^\alpha}{d \theta^\beta} \right)^{p} \right] \ , \\
	\nonumber d \Theta (\iota_v) A_2 =& \sum_{p=0}^{\infty} i (-1)^{p+1} (p+1)! \left[ dx^m \wedge dx^n \wedge dx^r \partial_{[r} A_{m n] \alpha \beta}^{(p)} \frac{(d \theta^\alpha)^p}{( d \theta^\beta )^{p+2}} + \right. \\
	\label{APO} & \left. + dx^m \wedge dx^n \partial_\alpha A_{[m n] \alpha \beta}^{(p)} \frac{(d \theta^\alpha)^{p+1}}{( d \theta^\beta )^{p+2}} + dx^m \wedge dx^n \partial_\beta A_{[m n] \alpha \beta}^{(p)} \frac{(d \theta^\alpha)^{p}}{( d \theta^\beta )^{p+1}} \right] \ ; \\
	\label{APP} d \Theta (\iota_v) A_3 =& - \sum_{p=0}^{\infty} dx^m \wedge dx^n \wedge dx^r i (-1)^p (p+2)! \left[ \partial_\alpha A_{[m n r] \alpha \beta}^{(p)} \frac{(d \theta^\alpha)^{p+1}}{( d \theta^\beta )^{p+3}} + \partial_\beta A_{[m n r] \alpha \beta}^{(p)} \frac{(d \theta^\alpha)^{p}}{( d \theta^\beta )^{p+2}} \right] \ ,
\end{align}

We have now to evaluate the action of $\Theta$ on $dA$ as well; since the procedure is analogous to the one already described (i.e. we have to make extensive use of \eqref{T1}, \eqref{T2} and \eqref{T3}) we give the results directly:

\begin{align}
	\nonumber \Theta ( \iota_v ) dA_0 = & - \sum_{p=0}^{\infty} \left[ - dx^m \left( \partial_m A_{\alpha \beta}^{(p)} \right) i (-1)^p p! \left[ \left( \frac{v^\alpha}{v^\beta} \right)^{p+1} - \left( \frac{d \theta^\alpha}{d \theta^\beta} \right)^{p+1} \right] + \right. \\
	\nonumber & \left. - \left( \partial_\alpha A_{\alpha \beta}^{(p)} \right) i (-1)^p p! \left[ (p+2) d \theta^\alpha \left( \frac{v^\alpha}{v^\beta} \right)^{p+1} - (p+1) d \theta^\beta \left( \frac{v^\alpha}{v^\beta} \right)^{p+2} - d \theta^\alpha \left( \frac{d \theta^\alpha}{d \theta^\beta} \right)^{p+1} \right] + \right. \\
	\label{APQ} & \left. + p \left( \partial_\beta A_{\alpha \beta}^{(p)} \right) i (-1)^{p-1} (p-1)! \left[ (p+1) d \theta^\alpha \left( \frac{v^\alpha}{v^\beta} \right)^{p} - p d \theta^\beta \left( \frac{v^\alpha}{v^\beta} \right)^{p+1} - d \theta^\alpha \left( \frac{d \theta^\alpha}{d \theta^\beta} \right)^{p} \right] \right] \ . \\
	\nonumber \Theta ( \iota_v ) dA_1 = & \sum_{p=0}^{\infty} \left[ dx^m \wedge dx^n \left( \partial_{[n} A_{m] \alpha \beta}^{(p)} \right) i (-1)^p p! \frac{(d \theta^\alpha)^p}{( d \theta^\beta )^{p+1}} + \right. \\
	\nonumber & \left. - dx^m \left( \partial_\alpha A_{m \alpha \beta}^{(p)} \right) i (-1)^p p! \left[ \left( \frac{v^\alpha}{v^\beta} \right)^{p+1} - \left( \frac{d \theta^\alpha}{d \theta^\beta} \right)^{p+1} \right] + \right. \\
	\label{APR} & \left. + p dx^m \left( \partial_\beta A_{m \alpha \beta}^{(p)} \right) i (-1)^{p-1} (p-1)! \left[ \left( \frac{v^\alpha}{v^\beta} \right)^{p} - \left( \frac{d \theta^\alpha}{d \theta^\beta} \right)^{p} \right] \right] \ . \\
	\nonumber \Theta ( \iota_v ) dA_2 = & - \sum_{p=0}^{\infty} \left[ dx^m \wedge dx^n \wedge dx^r \left( \partial_{[r} A_{m n] \alpha \beta}^{(p)} \right) i (-1)^{p+1} (p+1)! \frac{(d \theta^\alpha)^p}{( d \theta^\beta )^{p+2}} + \right. \\
	\nonumber & \left. + dx^m \wedge dx^n \left( \partial_\alpha A_{[m n] \alpha \beta}^{(p)} \right) i (-1)^{p+1} (p+1)! \frac{(d \theta^\alpha)^{p+1}}{( d \theta^\beta )^{p+2}} + \right. \\
	\label{APS} & \left. - (p+1) d x^m \wedge dx^n \left( \partial_\beta A_{[m n] \alpha \beta}^{(p)} \right) i (-1)^p p! \frac{(d \theta^\alpha)^p}{( d \theta^\beta )^{p+1}} \right] \ . \\
	\nonumber \Theta ( \iota_v ) dA_3 = & \sum_{p=0}^{\infty} \left[ dx^m \wedge dx^n \wedge dx^r \left( \partial_\alpha A_{[m n r] \alpha \beta}^{(p)} \right) i (-1)^{p+2} (p+2)! \frac{(d \theta^\alpha)^{p+1}}{( d \theta^\beta )^{p+3}} + \right. \\
	\label{APT} & \left. - (p+2) dx^m \wedge dx^n \wedge dx^r \left( \partial_\beta A_{[m n r] \alpha \beta}^{(p)} \right) i (-1)^{p+1} (p+1)! \frac{(d \theta^\alpha)^p}{( d \theta^\beta )^{p+2}} \right] \ .
\end{align}
The action of the operator $Z_v$ is now defined by $\displaystyle i Z_v = \left\lbrace d , \Theta(\iota_v) \right\rbrace $, so we need to sum \eqref{APM} $\div$ \eqref{APP} with \eqref{APQ} $\div$ \eqref{APT}. The results are
\begin{align}
	i Z_v A_0 = & \sum_{p=0}^\infty i (-1)^{p+1} (p+1)! \frac{1}{v^\beta} \left( \frac{v^\alpha}{v^\beta} \right)^p \epsilon^{\alpha \beta} v \cdot d \theta \left( \frac{v^\alpha}{v^\beta} \partial_\alpha A_{\alpha \beta}^{(p)} + \partial_\beta A_{\alpha \beta}^{(p)} \right) \ ; \\
	i Z_v A_1 = & \sum_{p=0}^\infty i (-1)^{p+1} p! dx^m \left( \frac{v^\alpha}{v^\beta} \right)^p \left( \partial_\alpha A_{m \alpha \beta}^{(p)} \frac{v^\alpha}{v^\beta} + \partial_\beta A_{m \alpha \beta}^{(p)} \right) \ ; \\
	i Z_v A_2 = & 0 \ ; \\
	i Z_v A_3 = & 0 \ .
\end{align}
The action of $Z_v$ on $A \wedge A$ is evaluated in the same way. Omitting the tedious algebraic manipulations, the result is
\begin{align}
	\nonumber i Z_v \left( A \wedge A \right) = & \sum_{p=0}^\infty \left[ dx^m \wedge dx^n p! p! \left[ (p+1) \partial_\alpha \left[ A_{[m n] \alpha \beta}^{(p)} , A_{\beta \alpha}^{(p)} \right] + \partial_\alpha \left( A_{[m \alpha \beta}^{(p)} A_{n] \beta \alpha}^{(p)} \right) \right] i v^\alpha \delta \left( \epsilon^{\alpha \beta} v \cdot d \theta \right) + \right. \\
	& \left. + dx^m \wedge dx^n p! p! \left[ (p+1) \partial_\beta \left[ A_{[m n] \alpha \beta}^{(p)} , A_{\beta \alpha}^{(p)} \right] + \partial_\beta \left( A_{[m \alpha \beta}^{(p)} A_{n] \beta \alpha}^{(p)} \right) \right] i v^\beta \delta \left( \epsilon^{\alpha \beta} v \cdot d \theta \right) \right] \ .
\end{align}


We can now apply these results to \eqref{ITB}. Since the "$\wedge$" product is associative, it is much more convenient to evaluate the second term as
\begin{equation*}
	\left( A \wedge A \right) \wedge Z_D A \ ;
\end{equation*}
indeed, as we have seen in the previous calculations, the $Z_v A$ term is made only of two parts, $Z_v A_0$ and $Z_D A_1$. Since the $Z_v A_0$ part is proportional to $d \theta$ and the $A \wedge A$ term is proportional to $\delta ( d \theta^\beta ) \delta ( d \theta^\alpha )$, it follows that their product is automatically annihilated. Therefore it follows that the interaction term is constructed from the two terms:
\begin{align}
	\nonumber \left( A \wedge A \right) \wedge Z_v ( A_1 ) = & - \left\lbrace \sum_{p=0}^\infty dx^m \wedge dx^n p! p! \left[ (p+1) \left[ A_{[m n] \alpha \beta}^{(p)} , A_{\beta \alpha}^{(p)} \right] + A_{[m \alpha \beta}^{(p)} A_{n] \beta \alpha}^{(p)} \right] \delta ( d \theta^\beta ) \delta ( d \theta^\alpha ) \right\rbrace \wedge \\
	& \wedge \left\lbrace \sum_{q=0}^\infty (-1)^{q+1} q! dx^r \left( \frac{v^\mu}{v^\nu} \right)^q \left( \partial_\mu A_{r \mu \nu}^{(q)} \frac{v^\mu}{v^\nu} + \partial_\nu A_{r \mu \nu}^{(q)} \right) \right\rbrace \ ; \\
	\nonumber A_1 \wedge Z_v \left( A \wedge A \right) = & \left\lbrace \sum_{p=0}^{\infty} dx^m A_{m \mu \nu}^{(p)} ( d \theta^\mu )^{p} \delta^{(p)} ( d \theta^\nu ) \right\rbrace \wedge \left\lbrace \sum_{q=0}^\infty \left[ dx^n \wedge dx^r q! q! \left[ (q+1) \partial_\alpha \left[ A_{[n r] \alpha \beta}^{(q)} , A_{\beta \alpha}^{(q)} \right] + \right. \right. \right. \\
	\nonumber & \left. \left. \left. + \partial_\alpha \left( A_{[n \alpha \beta}^{(q)} A_{r] \beta \alpha}^{(q)} \right) \right] v^\alpha \delta ( \epsilon^{\alpha \beta} v \cdot d \theta ) + dx^n \wedge dx^r q! q! \left[ (q+1) \partial_\beta \left[ A_{[n r] \alpha \beta}^{(q)} , A_{\beta \alpha}^{(q)} \right] + \right. \right. \right. \\
	& \left. \left. \left. + \partial_\beta \left( A_{[n \alpha \beta}^{(q)} A_{r] \beta \alpha}^{(q)} \right) \right] v^\beta \delta ( \epsilon^{\alpha \beta} v \cdot d \theta ) \right] \right\rbrace \ .
\end{align}
The second term contains a delta that we can recast as
\begin{equation}
	\delta ( \epsilon^{\alpha \beta} v \cdot d \theta ) = - \frac{1}{v^\beta} \delta \left( d \theta^\alpha - \frac{v^\alpha}{v^\beta} d \theta^\beta \right) = \frac{1}{v^\alpha} \delta \left( d \theta^\beta - \frac{v^\beta}{v^\alpha} d \theta^\alpha \right)  \ ;
\end{equation}
this implies that we can use $\displaystyle d \theta^\alpha = \frac{v^\alpha}{v^\beta} d \theta^\beta $ and that $\displaystyle \delta ( d \theta^\beta ) \delta \left( d \theta^\alpha - \frac{v^\alpha}{v^\beta} d \theta^\beta \right) = \delta ( d \theta^\beta ) \delta ( d \theta^\alpha ),$ since $\delta ( d \theta^\beta )$ has \virgolette support only in $d \theta^\beta = 0$". The same argument holds for $\alpha \leftrightarrow \beta$. By making use of these manipulations and by neglecting the terms that arrange as total fermionic derivative we get the final expression for the interaction term:
\begin{align}
	\nonumber A \wedge Z_v \left( A \wedge A \right) + 2 A \wedge A \wedge Z_v A = & \sum_{p,q=0}^\infty (-1)^p p! q! q! dx^m \wedge dx^n \wedge dx^r \delta^2 \left( d \theta \right) \cdot \\
	\nonumber & 3 \left\lbrace (q+1) \left[ A_{m n \alpha \beta}^{(q)} , A_{\beta \alpha}^{(q)} \right] + A_{m \alpha \beta}^{(q)} A_{n \beta \alpha}^{(q)} \right\rbrace \cdot \\
	\label{APPIT} & \left\lbrace \left( \frac{v^\alpha}{v^\beta} \right)^p \left( \frac{v^\alpha}{v^\beta} \partial_\alpha + \partial_\beta \right) A_{r \alpha \beta}^{(p)} + \left( \frac{v^\beta}{v^\alpha} \right)^p \left( \frac{v^\beta}{v^\alpha} \partial_\beta + \partial_\alpha \right) A_{r \beta \alpha}^{(p)} \right\rbrace \ .
\end{align}

\end{document}